\documentclass[fleqn,usenatbib]{mnras} 

\usepackage{newtxtext,newtxmath}

\usepackage[T1]{fontenc}
\usepackage[normalem]{ulem}
\usepackage{xcolor}

\usepackage{graphicx}	
\usepackage{amsmath}	
\usepackage{multicol}
\usepackage{caption}
\usepackage{subcaption}
\usepackage{orcidlink}
\usepackage[acronym]{glossaries}

\DeclareRobustCommand{\VAN}[3]{#2}
\let\VANthebibliography\thebibliography
\def\thebibliography{\DeclareRobustCommand{\VAN}[3]{##3}\VANthebibliography}

\newcommand{\chunliang}[1]{\textbf{\textcolor{violet}{[Chunliang: #1]}}}

\newcommand{\cmt}[1]{}
\newcommand{\updated}[1]{#1} 
\newcommand{\Msun}{M$_{\odot}$}
\newcommand{\msun}{M$_{\odot}$}
\newcommand{\Rsun}{R$_{\odot}$}
\newcommand{\Lsun}{L$_{\odot}$}

\newcommand{\phant}{{\sc phantom}}

\newcommand{\mesa}{{\sc mesa}}

\makeglossaries
\newacronym{AGB}{AGB}{asymptotic giant branch}
\newacronym{CE}{CE}{common envelope}
\newacronym{EoS}{EoS}{equation of state}
\newacronym[\glsshortpluralkey={LRNe}, \glslongpluralkey={luminous red novae}]{LRN}{LRN}{luminous red nova}
\newacronym{LTE}{LTE}{local thermodynamic equilibrium}
\newacronym{RLOF}{RLOF}{Roche lobe overflow}
\newacronym{SED}{SED}{spectral energy distribution}
\newacronym{SPH}{SPH}{smoothed particle hydrodynamics}

\title[Common envelope lightcurves]{Dust Formation in Common Envelope Binary Interactions --- III. Lightcurves}

\author[Mu et al.]{Chunliang Mu\orcidlink{0000-0003-1848-6507}$^{1,2}$ \thanks{E-mail: chunliang.mu@students.mq.edu.au},
Orsola De Marco\orcidlink{0000-0002-1126-869X}$^{1,2}$,
Luis C. Bermúdez-Bustamante\orcidlink{0000-0002-3629-6259 }$^{1,2}$
\newauthor
Ryosuke Hirai$^{3,4,5}$,
Daniel J. Price\orcidlink{0000-0002-4716-4235}$^{3,6}$,
Lionel Siess\orcidlink{0000-0001-6008-1103}$^7$,
Miguel González-Bolívar\orcidlink{0000-0002-5939-9269 }$^{2,8,9}$,
\newauthor
Mike Y. M. Lau\orcidlink{0000-0002-6592-2036}$^{10,11}$,
and
Nadejda Blagorodnova\orcidlink{0000-0003-0901-1606}$^{12,13,14}$
\\
$^1$School of Mathematical and Physical Sciences, Macquarie University, Sydney NSW 2109, Australia\\
$^2$Astrophysics and Space Technologies Research Centre, Macquarie University, Sydney NSW 2109, Australia\\
$^3$School of Physics and Astronomy, Monash University, Clayton, Victoria 3800, Australia\\
$^4$OzGrav: The ARC Centre of Excellence for Gravitational Wave Discovery, Australia\\
$^5$Astrophysical Big Bang Laboratory (ABBL), RIKEN Pioneering Research Institute (PRI), 2-1 Hirosawa, Wako, Saitama 351-0198, Japan\\
$^6$IPAG, Univ. Grenoble Alpes, CNRS, 38000 Grenoble, France\\
$^7$Institut d’Astronomie et d’Astrophysique, Université Libre de Bruxelles, CP 226, 1050 Brussels, Belgium\\
$^8$The ARC Centre of Excellence for All Sky Astrophysics in 3 Dimensions\\
$^9$Australian Astronomical Optics, Macquarie University, Sydney, NSW, Australia\\
$^{10}$Zentrum f\"ur Astronomie der Universit\"at Heidelberg, Astronomisches Rechen-Institut, M\"onchhofstr. 12-14, 69120 Heidelberg, Germany\\
$^{11}$Heidelberger Institut f\"{u}r Theoretische Studien, Schloss-Wolfsbrunnenweg 35, 69118 Heidelberg, Germany\\
$^{12}$Institut de Ciències del Cosmos (ICCUB), Universitat de Barcelona (UB), c. Martí i Franquès, 1, 08028 Barcelona, Spain\\
$^{13}$Departament de Física Quàntica i Astrofísica (FQA), Universitat de Barcelona (UB), c. Martí i Franquès, 1, 08028 Barcelona, Spain\\
$^{14}$Institut d’Estudis Espacials de Catalunya (IEEC), Edifici RDIT, Campus UPC, 08860 Castelldefels (Barcelona), Spain
}

\date{Accepted XXX. Received YYY; in original form ZZZ}

\pubyear{2026}

\begin{document}
\label{firstpage}
\pagerange{\pageref{firstpage}--\pageref{lastpage}}
\maketitle

\begin{abstract}
Luminous red novae are transient events thought to arise from common envelope binary interactions.
In this paper, we perform post-processing light-curve calculations for two, 3D hydrodynamic simulations of common envelope events between 1.7~M$_\odot$ and 3.7~M$_\odot$ asymptotic giant branch stars and a 0.6~M$_\odot$ compact companion, \updated{simulated for 44 years under adiabatic conditions, with} dust nucleation \updated{included}.
In both our simulations, we observe a bright, hot peak lasting 3--5 years, primarily due to \updated{photospheric expansion} before and during inspiral. Additional peaks \updated{appear for different viewing angles due to the  interaction's asymmetry.}
Dust forms about 1--3 years after the beginning of the simulated interaction, and shortly afterwards, followed by a partial recovery and a plateau with a \updated{colour} temperature of $\sim$400~K. The dust photosphere reaches a size of $\sim$250~au by the end of the simulations, \updated{but would become optically thin at visible wavelengths in 2000--6000 years,} revealing an inner, warmer photosphere.
\updated{Our results are most reliable around the time of peak and the immediate light decline; outside this window, they remain largely qualitative due to two well-quantified but substantial uncertainties---insufficient surface resolution in the first 1--2 years and the adiabatic assumption in the later years.}
We finally contextualise our simulations within a group of observed luminous red nova transients, drawing particular attention to the outburst of OGLE-2002-BLG-360 and AT~2025abao, which \updated{are the only two luminous red novae to date understood to be the interaction of a companion with low-to-intermediate-mass giant stars.}
\end{abstract}


\cmt{
Note: To remove comments like this one, remove the \#1 in
\textbackslash newcommand\{\textbackslash cmt\}[1]\{\textbackslash textcolor\{gray\}\{\#1\}\}
}

\section{Introduction}
\label{sec:introduction}

Red novae and \glspl{LRN} are stellar transients arising from low- and high-mass binary interactions, respectively \citep{Pastorello2019b,KaminskiBlagorodnova2026arXiv}. These events occur when two non-degenerate stars---or a growing subgiant and a compact companion---undergo a \gls{CE} interaction \citep{Ivanova2013}, culminating in either a stellar merger or a tight binary system. While detections in our own Galaxy are rare---V~838~Mon \citep{Munari2002}, V~1309~Sco \citep{Tylenda2011}, V~4332~Sgr \citep{Martini1999}, OGLE-2002-BLG-360 \citep{Tylenda2013}---the advent of dedicated transient surveys has led to a surge in extragalactic LRN events, resulting in a total of $\sim$30 studied objects \citep[see sample in][]{KaminskiBlagorodnova2026arXiv}. With the Vera Rubin Telescope now operational, we anticipate that its population will dramatically increase during the Rubin's 10-year Legacy Survey of Space and Time \citep[LSST, see][]{Howitt2020}, tracing CE evolution for a broad range of binary configurations and remnants. Characterizing these inherently three-dimensional interactions requires 3D hydrodynamic models capable of predicting lightcurves and physical observables, ultimately enabling the ``fingerprinting'' of the entire interaction process.

Three-dimensional hydrodynamic simulations of the \gls{CE} binary interaction have come a long way \citep{Roepke2023}, but the prediction of the lightcurve has been plagued with a number of challenges \citep{Galaviz2017}. Recently \citet{Hatfull2025a} predicted the lightcurve of the \gls{LRN} V~1309~Sco, which represented the merger of a low mass subgiant 1.5\Msun\ star with a 0.16\Msun\ companion. They used the SPH code {\sc starsmasher} \citep{Gaburov2010} and post-processed the light to provide observable (photospheric) properties of the outburst, as well as multi-wavelength lightcurves in the photometric Johnson-Cousins system. Since most \glspl{LRN} observations show the fast production of dust \citep[e.g.,][]{Nicholls2013,Blagorodnova2020}, \citet{Hatfull2025a} mimicked dust production, but treated it as an instantaneous process by using opacity tables that include both gas and dust opacities tabulated as a function of quantities like density and temperature, as an indication of when dust formation would be taking place. 

\citet{Bermudez2024a} have carried out \gls{CE} simulations of 1.7~\msun\ and 3.7~\msun\ \gls{AGB} stars interacting in a \gls{CE} with 0.6~\msun\ compact companions, including the nucleation of carbon-based dust. They demonstrated that dust takes a short, but non-zero time to form in the context of a CE interaction; the formation is very efficient and all the gas that can be converted into carbon grains does so in a matter of a couple of decades. Additionally, the increasing amount of dust rapidly obscures the gas photosphere. The timing of dust formation therefore leads to completely different  lightcurve properties and accurate predictions of the dust formation, properties and distribution is therefore paramount.

In this paper, we carry out a first step at bridging the gap between simulations and observations, by performing a calculation of the light emitted from the dusty \gls{CE} simulations of \citet{Bermudez2024a}.
The method 
is similar to what was used by \citet{Galaviz2017} and \citet{Hatfull2025a}, namely it carries out radiation transport along rays from the CE to the observer. The light contribution into the ray is then calculated with a blackbody approximation assuming \gls{LTE}. 


In Section~\ref{sec:post-processing}, we explain how we calculate the lightcurve, starting with the hydrodynamic models (Section~\ref{sec:hydro-models}), the computation of their grey luminosity (Section~\ref{sec:grey_luminosity}), and the explanation of the opacities adopted (Section~\ref{sec:opacity}).
We then present our results in Section~\ref{sec:results} starting with the lightcurve in Section~\ref{ssec:lightcurve}, a discussion of the properties of the photosphere in Section~\ref{sec:the_evolutuion_of_the_photospheric_radius}. In Section~\ref{sec:uncertainties} we discuss the various sources of uncertainty. We start with the SPH simulation resolution in Section~\ref{sec:uncertainties:SPH}; we continue with other issues stemming from the properties of the surface layers in Section~\ref{sec:uncertainties:SPHres}, carry out  a critique of the late time lightcurve in Section~\ref{sec:uncertainties:cooling} and a comparison with observations in Section~\ref{ssec:observations}. 
We finally present our conclusions in Section~\ref{sec:discussion}.

\section{Method}
\label{sec:post-processing}


\subsection{Hydrodynamic models}
\label{sec:hydro-models}

We calculate the lightcurve from four 3D hydrodynamic simulations performed with the smoothed particle hydrodynamics code {\sc phantom} \citep{Price2018}, fully described in \citet{Bermudez2024a} and \citet{GonzalezBolivar2022}. Two of the simulations are \gls{CE} interactions
between a 1.7~\Msun{}, 260~\Rsun{}, 5180~\Lsun{}, 3227~K, asymptotic giant branch star and a 0.6~\Msun\ companion, while the other two used a 3.7~\Msun{}, 330~\Rsun{}, $11\,900$~\Lsun{}, 3317~K, AGB star with the same companion mass. All simulations are started at an orbital separation such that the giant fills its Roche lobe (550 and 637~\Rsun, respectively).
\updated{The giants' cores and companions are point mass particles with a softening length of 2~\Rsun{}\footnote{
    We note that 2~\Rsun{} is the correct value for the companion point mass particle's softening length in the 1.7~\Msun{} simulation. The 2.5~\Rsun{} value reported by \cite{Bermudez2024a} is incorrect.
} for the 1.7~\Msun\ simulation and 8~\Rsun\ for the 3.7~\Msun\ simulation. The softening length size has a minimal impact on the inspiral as demonstrated by, e.g., \citet{Lau2022}.
}

Each pair of simulations computed the evolution of the CE interaction with and without dust nucleation. We employ a tabulated \gls{EoS} that includes the input of recombination energy into the gas, where it is assumed to be instantly thermalised \citep{Reichardt2020a}. The envelope of both stars is almost entirely unbound after a simulation time of $\sim$40~years. Both simulations are adiabatic with no radiation transport.

Dust forms approximately 1--3 years after the start of the simulations, before the in-spiral has completed at about 5 years.
As illustrated in figure~11 of \citet{Bermudez2024a}, in both simulations almost all dust that can form has formed by 30 to 40 years,
limited by the C/O=2.5 and a metallicity of $Z=0.02$. No dust destruction was implemented. The dust layer quickly becomes optically thick, so that the photosphere is at the outer edge of the dusty shell, expanding and cooling over four decades.  

\subsection{Calculation of the Grey Luminosity}
\label{sec:grey_luminosity}
In this section, we apply the radiative transfer equation to determine the system's bolometric luminosity at specific time intervals using a custum ray-tracing code -- see Data Availability below.
We define our coordinate system such that the observer is positioned at $z \rightarrow +\infty$, viewing the object (centred at $z=0$) along the negative $z$-axis. To calculate the surface brightness, we project a grid onto the object’s observable face and trace a single ray per pixel—parallel to the $z$-axis—to solve for the specific intensity, $I_i$ (see Figure~\ref{fig:diagram:ray-grid} in Appendix~\ref{app:light-calculation}, where we also provide additional details of the computation). From these values, the isotropic-equivalent luminosity, $L$, is derived as follows
\begin{equation}
    L \approx 4 \pi \sum_i I_i \Delta A_i,
    \label{eq:def:luminosity}
\end{equation}
where $\Delta A_i$ is the area of a pixel associated with the $i$-th ray. As rays are evenly spaced, $\Delta A_i = \Delta A$ are the same for all rays. 

To derive the specific intensity, $I_i$, for each ray, 
\updated{we use the hydrodynamic model grey opacities, $\kappa$, assuming they represent the total extinction. This includes  scattering out of the line of sight -- attenuation, while  scattering into the line of sight is under the assumption that the local incoming photon field is the same as the local Planck spectrum 
(more on  opacities in Section~\ref{sec:opacity})}. 
We then solve the radiative transfer equation, for a given ray
\begin{equation}
    \frac{dI_i}{d\tau_i} = I_i(\tau_i) - S_i(\tau_i),
    \label{eq:rad_trans}
\end{equation}
where $S_{i}$ is the source function along the ray, and $\tau_i$ is the optical depth at any given position $z$
\begin{equation}
    \tau_i(z) = \int_z^{+\infty} \kappa_i(z') \rho_i(z') dz',
    \label{eq:def:tau}
\end{equation}
where $\kappa_{i}$ is the wavelength-integrated opacity and $\rho_i$ is the density along the ray.
Ignoring any input of light from the far side of the object and any scattering into the beam, $I_i(\tau=\tau_{i, \mathrm{max}})$, 
the solution to Equation~\ref{eq:rad_trans} is therefore
\begin{equation}
    I_i(z=+\infty)
    = \int_{-\infty}^{+\infty} e^{-\tau_i(z')} S_i \kappa_i \rho_i dz'.
    \label{eq:radiative-transfer-solution-backwards}
\end{equation}
Here we note that $S \kappa \rho = S \frac{d\tau}{dz}$ is the source function integral per physical distance instead of per optical depth.

In the specific case of \gls{SPH} simulations, the specific intensity $I_i$ is ultimately determined by the many particles that intersect with the ray, $i$, within the particle's kernel cutoff radius, $2h$ \updated{(for a discussion of resolution see Section~\ref{sec:uncertainties:SPHres})}.
This is because all quantities on the ray are derived from the quantities from these neighbouring particles, as illustrated in Figure~\ref{fig:diagram:ray-vs-particles}, Appendix~\ref{app:light-calculation}.
More specifically, to discretize Equation~\ref{eq:radiative-transfer-solution-backwards} in \gls{SPH}, we use kernel interpolation (\citealt{Price2007}) to interpolate the product $S \kappa \rho$ from neighbouring particles
\begin{equation}
    I_i(z=+\infty) \approx
    \int_{-\infty}^{+\infty}
    e^{-\tau_i(z')} \sum_j \frac{m_j (S \kappa \rho)_j}{\rho_j h_j^3} w(q_{ij}(z')) dz',
    \label{eq:calc:kernel-interpolation:I_i}
\end{equation}
where $m_j, S_j, \kappa_j, \rho_j, h_j$ is the mass, source function, opacity, density and smoothing radius of the $j$-th particle;
$q_{ij}$ is the dimensionless distance from $j$-th particle to $i$-th ray, defined as
\begin{equation}
    q_{ij} (z) \equiv
    \frac{\sqrt{(x_i-x_j)^2+(y_i-y_j)^2+(z-z_j)^2}}{h_j};
    \label{eq:def:q_ij}
\end{equation}
and $w$ is the dimensionless kernel, derived  from the smoothing kernel, $W$, as
\begin{equation}
    w(q_{ij}(z)) \equiv h_j^3 W(q_{ij}(z), h_j).
\end{equation}
The value of $w(q_{ij}(z))$ goes to zero when $q_{ij}(z) \geq 2$ from a central peak of $\sim$0.318 in 3 dimensions \citep{Price2018}.
In other words, only neighbouring particles within $2 h_j$ from the ray are counted when interpolating the values on the ray. When kernel-interpolating quantities deriving from the product of primary variables, we first multiply the relevant variables at each particle's location, and then we kernel-interpolate the product to the ray. We discuss the merit of this procedure in Appendix~\ref{app:uncertainties:integration-order}.

To further stress that the \gls{SPH} particles are the actual source of information that drives the calculation of $L$,
we define the (wavelength and viewing-angle-dependent) effective cross-section area $A_{\mathrm{eff}, j}$ of each \gls{SPH} particle, $j$, to be
\begin{equation}
    A_{\mathrm{eff}, j} \equiv
    \frac{m_j \kappa_j}{h_j^3}
    \sum_i \Delta A_i 
    \int_{-\infty}^{+\infty}
    e^{-\tau_i(z')} w(q_{ij} (z')) dz'.
    \label{eq:def:Aeffj}
\end{equation}
By combining Equation~\ref{eq:def:luminosity}, \ref{eq:calc:kernel-interpolation:I_i}, and~\ref{eq:def:Aeffj}, we can now find the luminosity by
\begin{equation}
    L \approx
    4 \pi \sum_j S_j A_{\mathrm{eff}, j}.
    \label{eq:calc:luminosity}
\end{equation}

\updated{For the source function, $S_j$ for particle $j$, we adopt the Planck, blackbody function.} 

\begin{equation}
    S_j = \frac{\sigma_\mathrm{sb} T_j^4}{\pi},
    \label{eq:assumption:S_j}
\end{equation}
where $\sigma_\mathrm{sb}$ is the Stefan-Boltzmann constant\footnote{Using a full Planck spectrum and adding up intensities for each pixel at each frequency, rather than adding up pre-integrated blackbodies on a single grey ray, gives the same results. The reason is that we are using grey opacities, so all the terms in $A_{\rm eff}$ are wavelength independent, thus, simply changing the source function to be wavelength-dependent in Equation~\ref{eq:calc:luminosity} (and then integrating the result across wavelength) does not change the outcome.}.
\updated{Considering the fact that our opacity represents total extinction, including absorption and scattering out of and into the beam, this is like assuming that the photons scattered into the beam have a Planck distribution (assumption of \gls{LTE}). We further elaborate on the possible implications of this assumption in  Appendix~\ref{app:source}.}

 

For the optical depth in Equation~\ref{eq:def:Aeffj}, we use kernel interpolation (\citealt{Price2012a}) again to derive a discretised version of $\kappa_i(z') \rho_i(z')$ in Equation~\ref{eq:def:tau}, namely
$$
    \kappa_i(z') \rho_i(z') \approx \sum_j \frac{m_j (\kappa \rho)_j }{\rho_j h^3_j} w(q_{ij}(z')).
$$
Equation~\ref{eq:def:tau} now becomes
\begin{equation}
    \tau_i(z) \approx \sum_j \frac{m_j \kappa_j}{h^3_j} \int_{z}^{+\infty} w(q_{ij}(z')) dz',
    \label{eq:calc:tau_z}
\end{equation}
which can be used to calculate the optical depth at any given point.

The photosphere is defined as the place where $\tau = 2/3$ along rays that are parallel to the line of sight. From the locations of the photosphere along each ray, a mean photospheric radius can be defined as explained in Section~\ref{sec:the_evolutuion_of_the_photospheric_radius}.
The value of $\kappa, \tau$ and $T$ as a function of position along two specific rays are shown in Figure~\ref{fig:profile:2md-single-ray} and corresponding values are reported in Appendix~\ref{app:photosphere_table}.

\begin{figure*}
    \centering
    \includegraphics[width=0.49\linewidth]{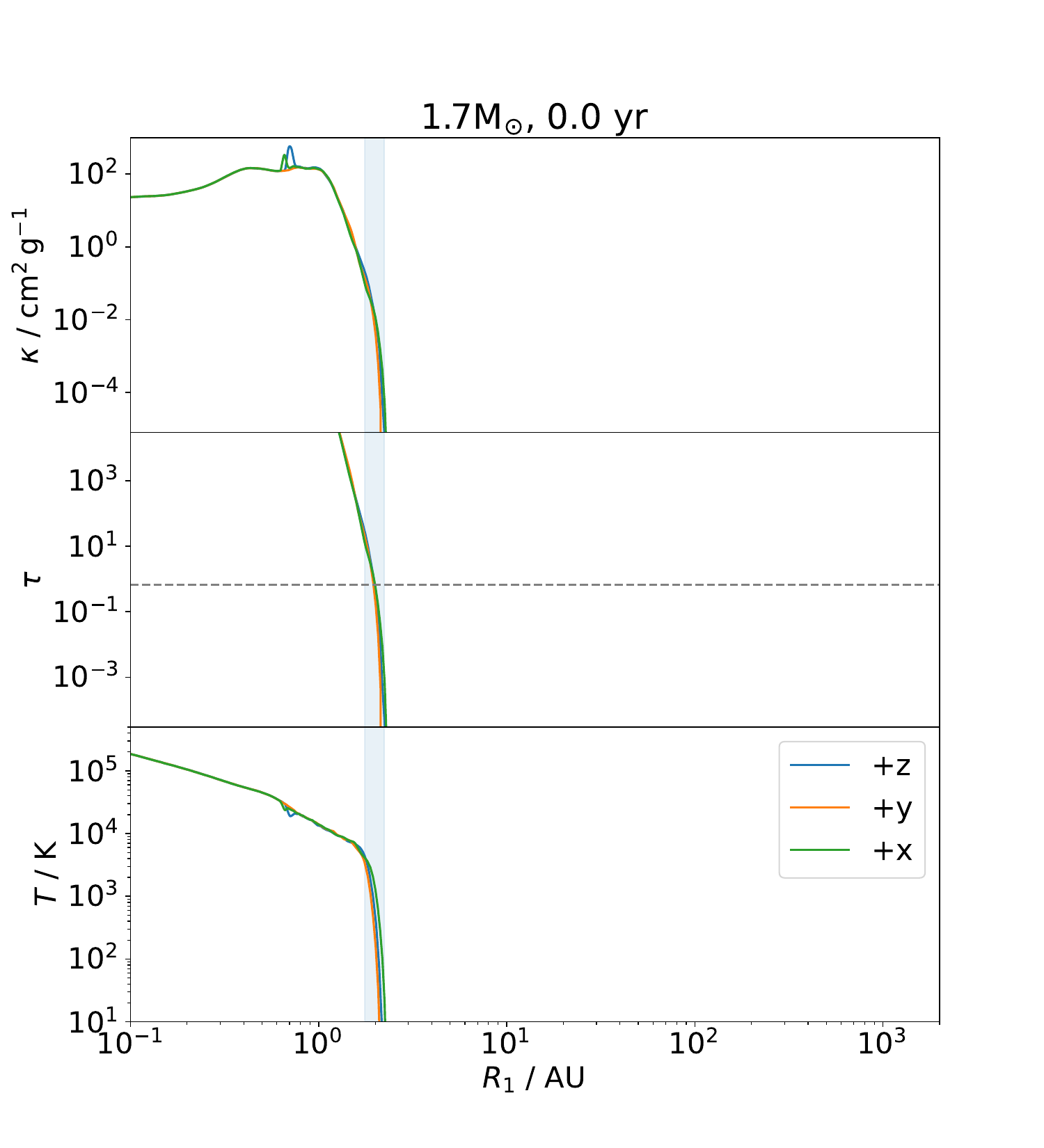}
    \includegraphics[width=0.49\linewidth]{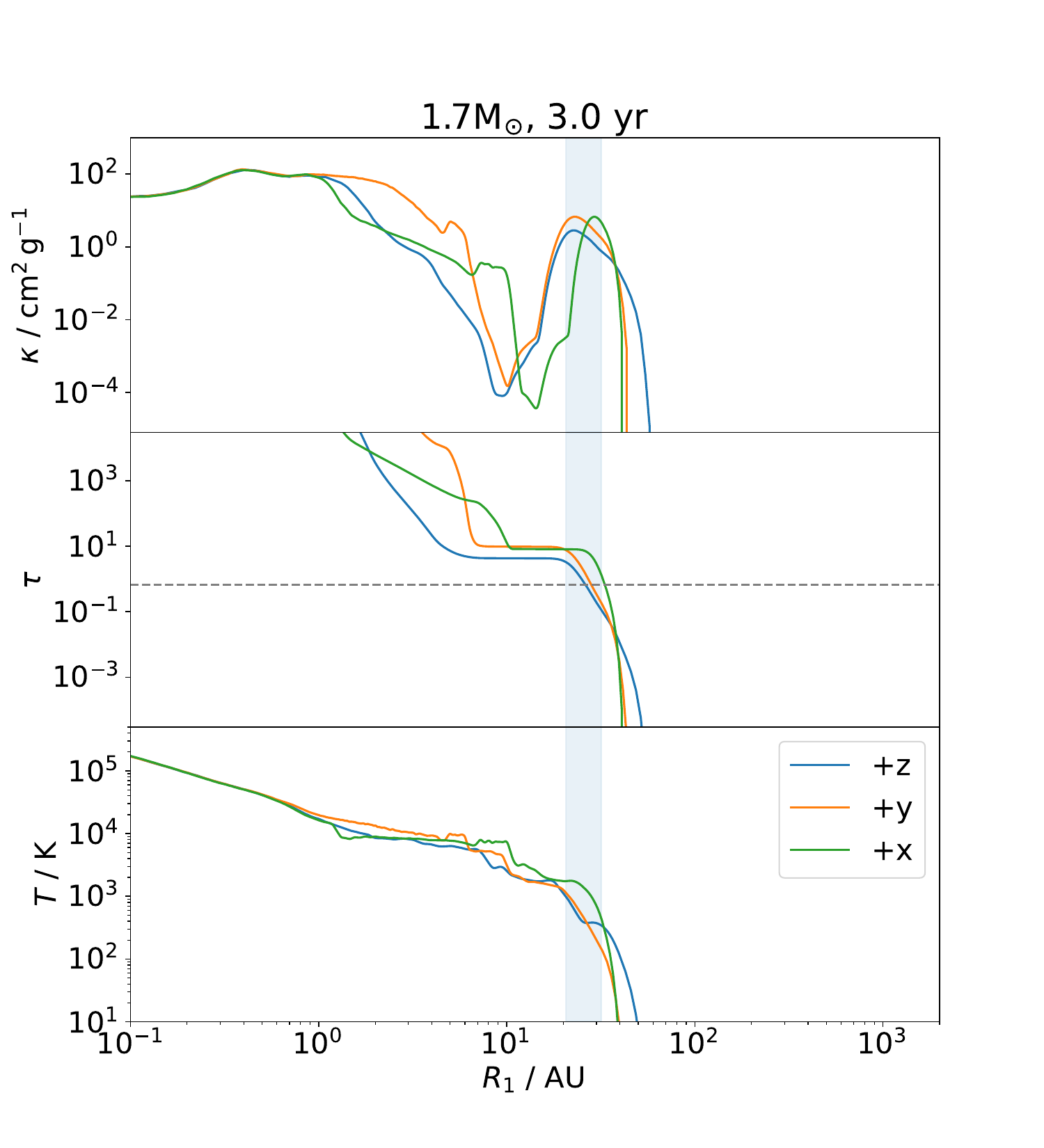}
    \includegraphics[width=0.49\linewidth]{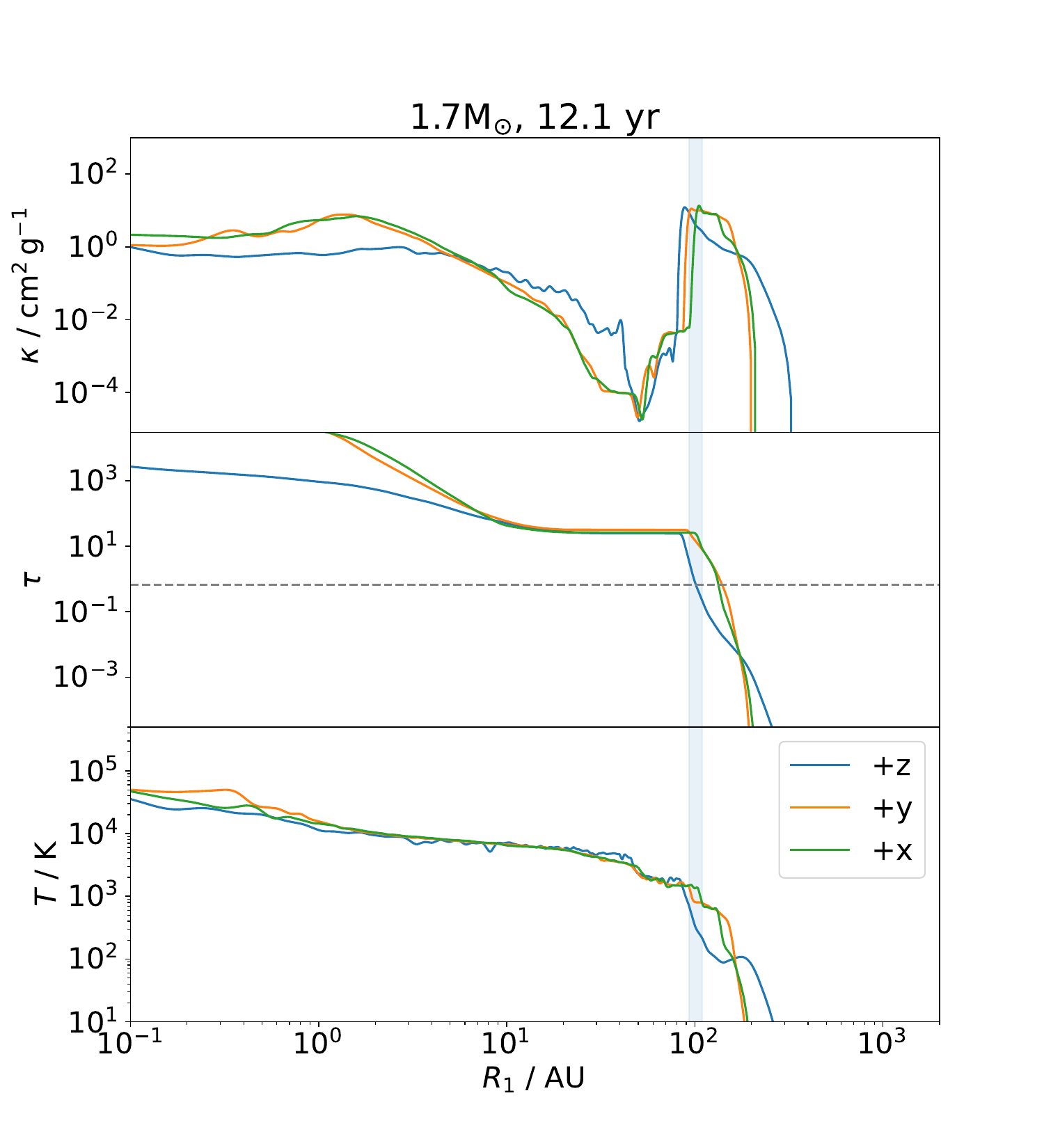}
    \includegraphics[width=0.49\linewidth]{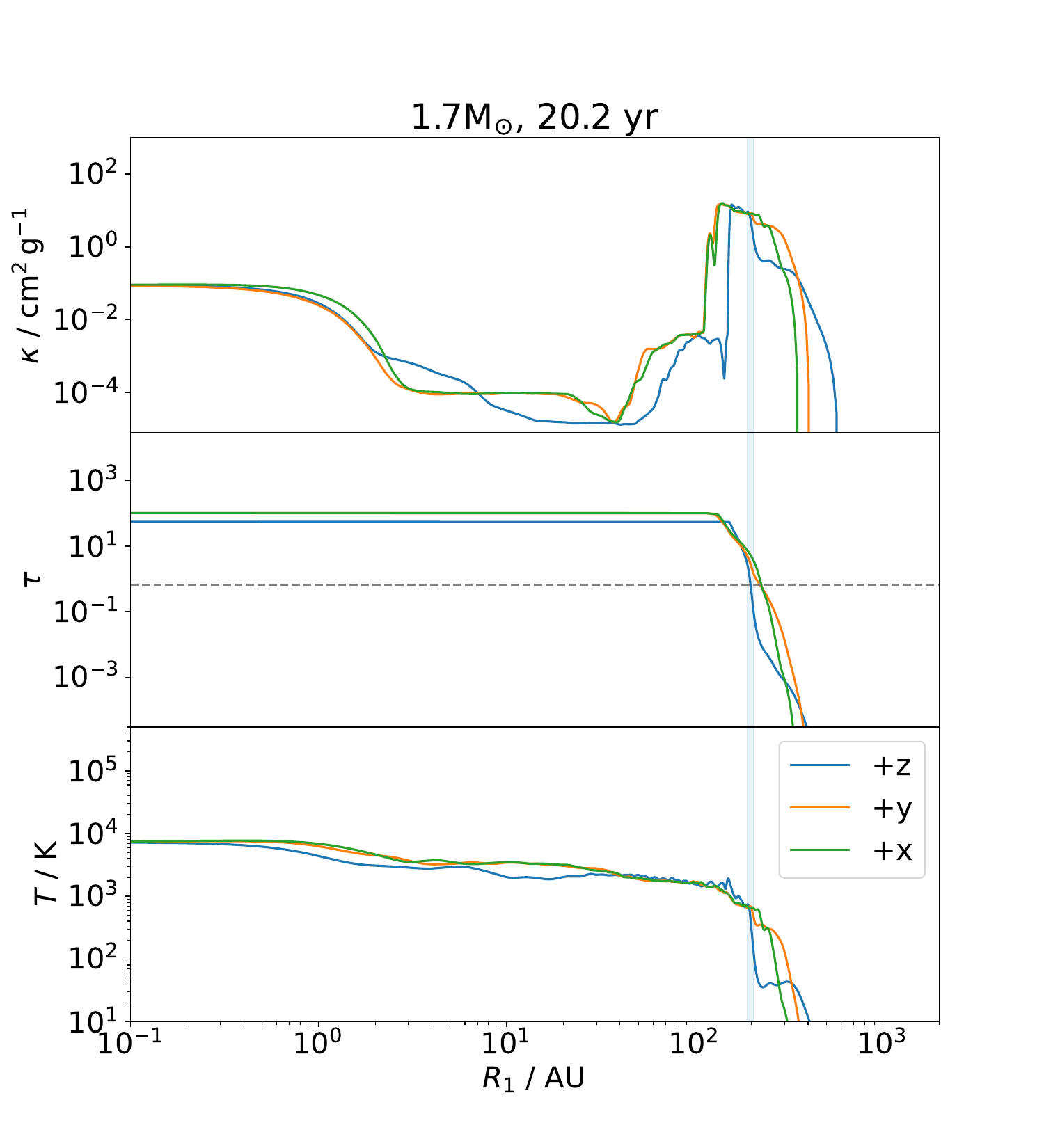}
    \caption{
    Opacity, $\kappa$, optical depth, $\tau$, and temperature, $T$, along one line-of-sight as viewed from the polar $z \rightarrow +\infty$ (blue) and the orbital $y \rightarrow +\infty$ (orange) and $x \rightarrow +\infty$ (green) directions, for the 1.7~\Msun{} simulation at $t=0$ (top left), $t=3~\mathrm{yr}$ (top right), $t=12~\mathrm{yr}$ (bottom left), and $t=20~\mathrm{yr}$ (bottom right). The \updated{horizontal} dashed line in the middle panels marks $\tau=2/3$.
    The photospheric location is marked with blue vertical shaded lines as seen from the $z \rightarrow +\infty$ direction, with the line width representing the local smoothing length at the photosphere, giving a sense of uncertainty of the photospheric $\kappa, \tau$, and $T$. Corresponding values are listed in Table~\ref{tab:ray_values} in Appendix~\ref{app:photosphere_table}. 
    Full movie version are at \url{https://zenodo.org/records/18239846}.} 
\label{fig:profile:2md-single-ray}
\end{figure*}

\subsection{Gas and Dust Opacities}
\label{sec:opacity}

The dust opacity, $\kappa_\mathrm{dust}$, for our carbon dust simulations is the same as that calculated by \citet{Bermudez2024a},
where the Planck mean opacity of the dust-gas mixture is calculated using Mie theory (see equation~7 of that paper).
Planck mean dust opacities are used instead of Rosseland mean opacities,
as they are more suitable for the optically thin outer regions, near the photosphere, where dust forms and light emits. \updated{These opacities include absorption and scattering processes. However, in the radiation transport calculation outlined in Section~\ref{sec:grey_luminosity} we have used these opacities as if they were absorption only.}
\cmt{ \label{where:c:is-mesa-opacity-planck} !See (C\ref{list:todo:c:is-mesa-opacity-planck}) }
\cmt{ \label{where:q:planck-or-rosseland} !See <Q\ref{list:todo:q:planck-or-rosseland}> }

\begin{figure}
    \centering
    \includegraphics[width=10cm]{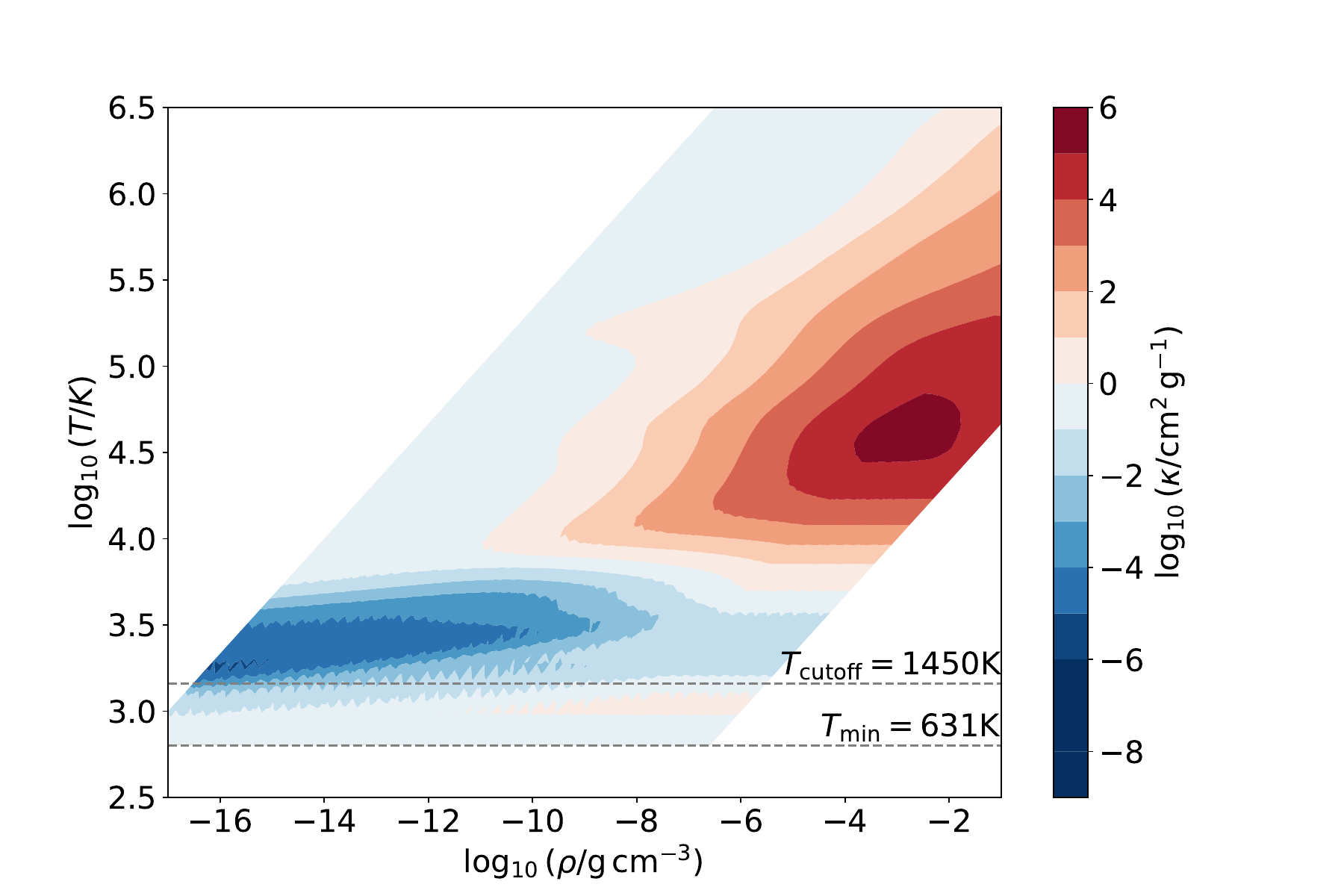}
    \caption{Tabulated \mesa{} opacity,
    $\kappa_\mathrm{MESA}$, as a function of temperature $T$ and density $\rho$.
    Grey cut-off line marks the approximate oxygen dust condensation temperature below which we override the opacity value with a constant gas opacity of $2 \times 10^{-4} \mathrm{cm}^2 \mathrm{g}^{-1}$.
    The table only covers regions with temperature above $631~\mathrm{K}$, although particles near the photosphere often have much lower temperatures. See text for details of how we handle this issue.}
    \label{fig:kappa-mesa-table}
\end{figure}

\begin{figure}
    \centering
    \includegraphics[width=9cm]{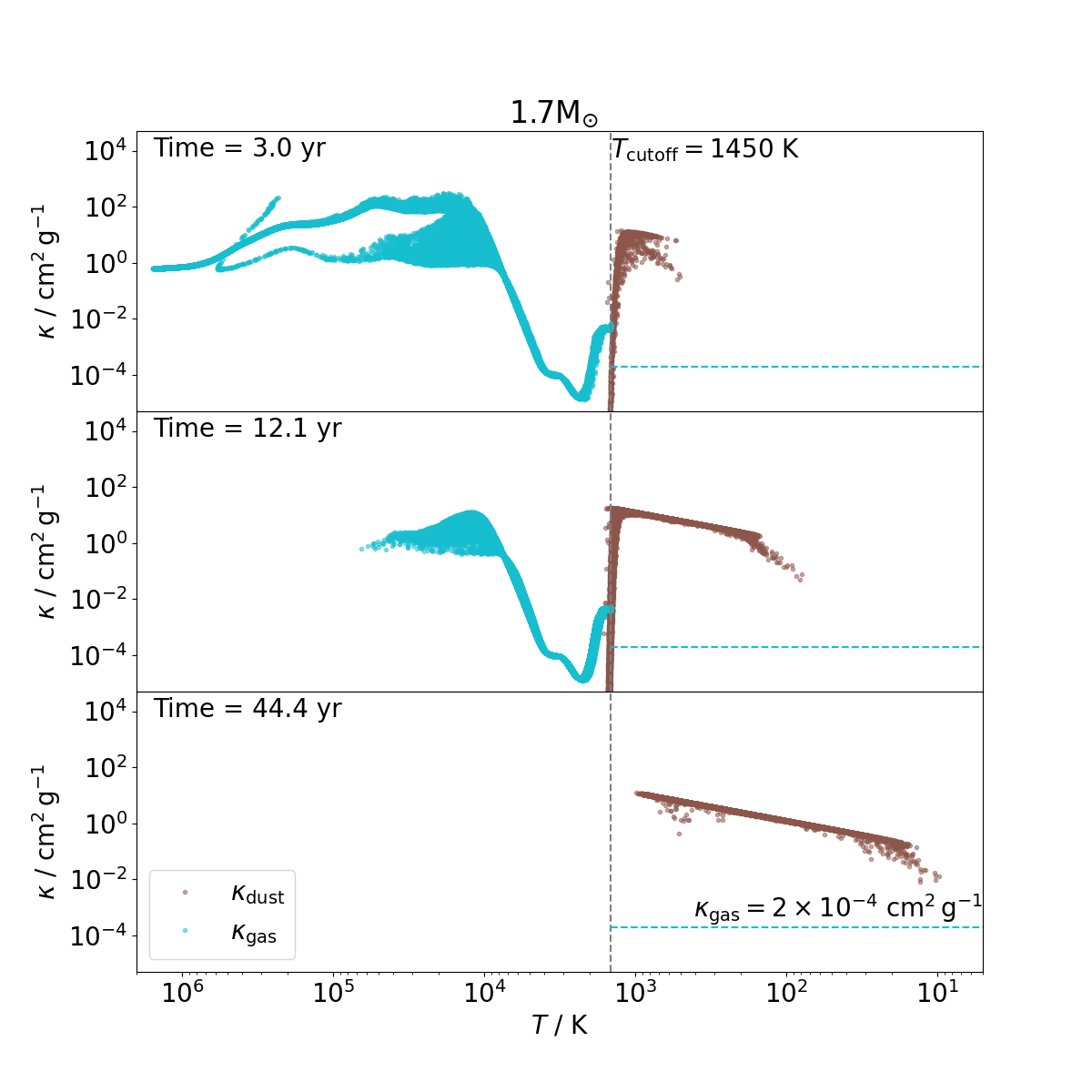}
    \caption{Dust (brown symbols to the left of $T_{\rm{cutoff}}$) and gas (cyan symbols to the right of $T_{\rm{cutoff}}$) opacities for all \gls{SPH} particles as a function of their temperature for the 1.7~\Msun{} dusty simulation at $t=3~\mathrm{yr}$ (top panel), $12~\mathrm{yr}$ (middle panel), and $44~\mathrm{yr}$  (bottom panel). When the SPH particle temperature is lower than $T_\mathrm{cutoff}$ (green dashed vertical line),
    $\kappa_\mathrm{gas} = 2 \times 10^{-4}$~cm$^2$~g$^{-1}$, 
    indicated by a blue dashed line. The total opacity $\kappa = \kappa_\mathrm{gas} + \kappa_\mathrm{dust}$ is the sum of the gas and dust opacities, with each dominating in different temperature regimes.
    The double valued distribution of gas opacities in the upper panel (cyan symbols) is due to the fact that opacities depend on density as well as temperature. This means that \gls{SPH} particles with the same temperature can have different opacities.
    }
    \label{fig:profile:2md-kappa}
\end{figure}

The gas opacity, $\kappa_\mathrm{gas}$, derives from the tabulated values adopted by the \mesa{} code, as implemented in {\sc phantom} by \citet{Reichardt2020a},
which combine a range of different opacity sources across a wide range of temperatures from $631~\mathrm{K}$ to $5\times10^8~\mathrm{K}$, 
as shown in Figure~\ref{fig:kappa-mesa-table}.\cmt{ \label{where:q:is-mesa-opacity-planck} !See <Q\ref{list:todo:q:is-mesa-opacity-planck}> }
We only adopt the \mesa\ gas opacity values for $T>1450$~K because for lower temperatures the tabulated opacities are those for oxygen-rich dust (\citealt{Allard2001a, Ferguson2005a}), while we implement our own carbon dust opacities as explained above. For $T<1450$~K, the gas opacity is assumed to be a low value compatible with recombined gas being transparent, or $\kappa_\mathrm{gas}=2 \times 10^{-4} \mathrm{cm}^2 \mathrm{g}^{-1}$.
The opacity discontinuity introduced by this temperature cutoff is negligible for our dusty simulations,
because at the temperature of the discontinuity, typically, the opacity from carbon dust ($\sim 10~\mathrm{cm}^2 \mathrm{g}^{-1}$) dominates the gas opacity, as seen in Figure~\ref{fig:profile:2md-kappa}.
The total opacity $\kappa$ is the sum of dust and gas opacities: $\kappa = \kappa_\mathrm{dust} + \kappa_\mathrm{gas}$.

\cite{Hatfull2025a} also assumed wavelength-independent opacities. They averaged two  opacity tables that extend to different temperature regions: that of \citet{Grevesse1998} in high temperature regions ($\log T > 3.75$), 
and those of \citet{Semenov2003} in low temperature regions ($\log T < 4$).
Similarly to us they set their gas opacity
to $10^{-4} \textrm{cm}^2 \textrm{g}^{-1}$ when $T \leq 3000$~K (instead of our lower $1450$~K) for no-dust scenarios. For dust-forming scenarios, they did not include dust nucleation, but instead represented the overall opacity with either the Planck mean opacity or the Rosseland mean opacity version of the \citet{Semenov2003} tables, for the full low temperature region. We only consider the Planck mean opacity because we do not include radiative cooling in our simulations, and the outgoing flux
is determined in the region around and above
the photosphere, which is optically thin when the resolution is good,
where Planck opacity applies. When the resolution is poor, the emerging luminosity is overestimated
anyway and it is determined only by the first particle encountered by the ray, such that changing opacity to Rosseland will make no difference.

\section{Results}
\label{sec:results}

\subsection{The lightcurve}
\label{ssec:lightcurve}

\begin{figure*}
    \centering
    \includegraphics[width=0.454\linewidth, trim={0 90 112 150}, clip]{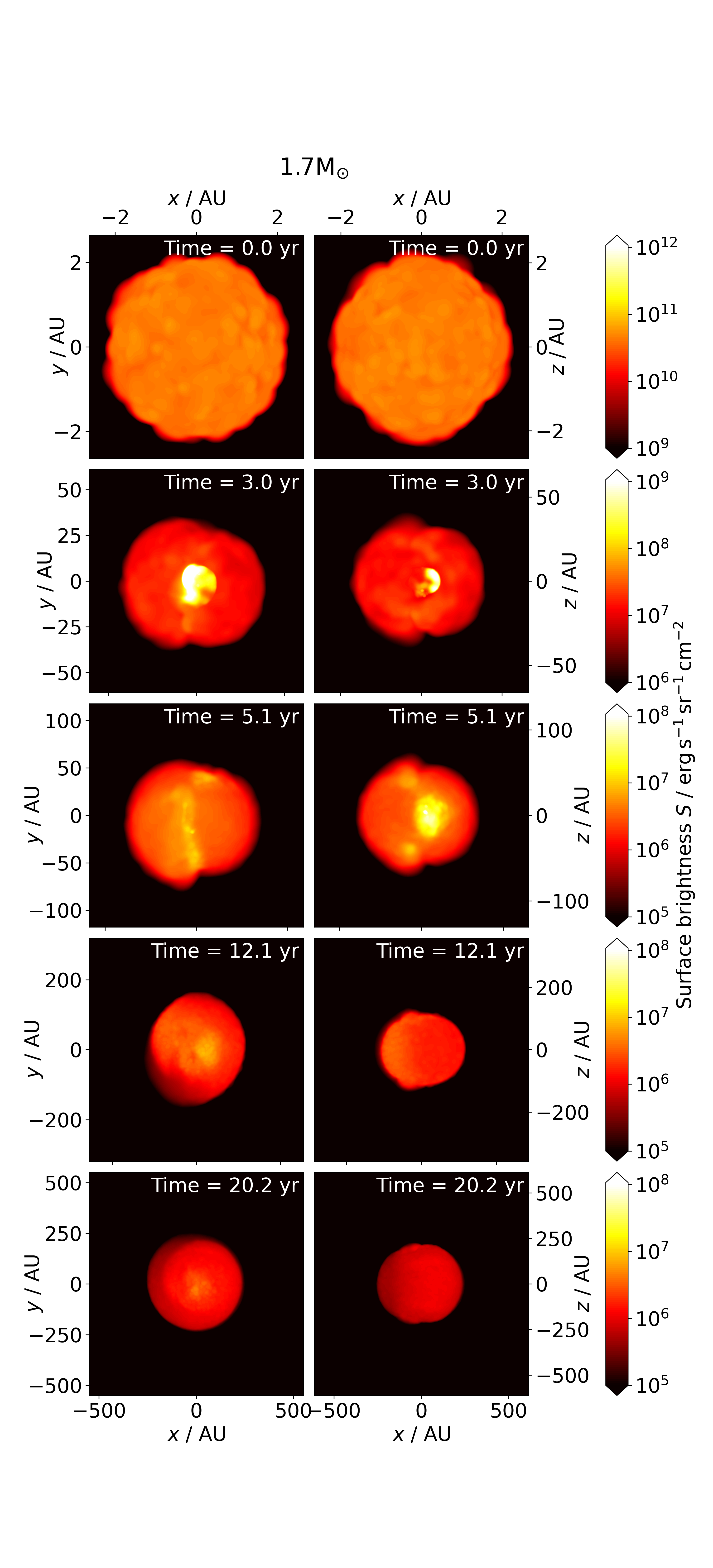}
    \includegraphics[width=0.454\linewidth, trim={0 90 112 150}, clip]{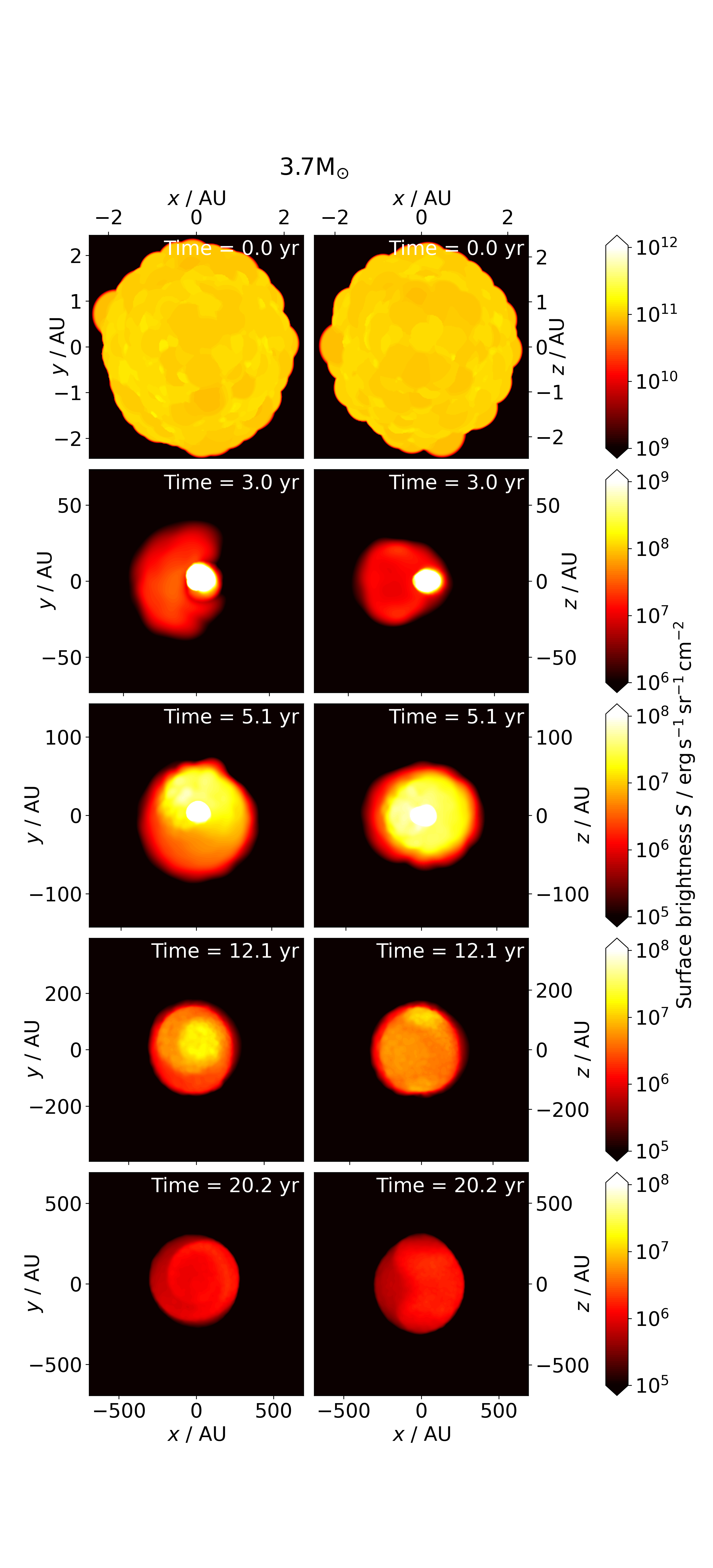}
    \includegraphics[width=0.082\linewidth, trim={610 90 0 150}, clip]{figures/image_4md_256x256.png}
    \caption{
    Surface brightness from each pixels for the 1.7~\Msun{} (left two columns) and 3.7~\Msun{} (right two columns) simulations, as viewed from the polar ($z \rightarrow +\infty$) direction (first and third columns) and the orbital ($y \rightarrow +\infty$) direction (second and fourth columns), at $t=0, 3, 5.1, 12.1, {\rm and} 20.2~\mathrm{yr}$ for the top, second, third, fourth and bottom rows, in each panel, respectively - {\it note the progressively increasing size of the object and variable colour scales}. The origin coincides with the primary's core. Movies corresponding to this still image can be found at \url{https://zenodo.org/records/18239846}
    for both 1.7\Msun{} (2md) and 3.7\Msun{} (4md) simulations, viewed from $z \rightarrow +\infty$ (xyz), $y \rightarrow +\infty$ (xzy) and $x \rightarrow +\infty$ (zyx).
    }
    \label{fig:lum-image:2md}
\end{figure*}

\begin{figure*}
    \centering
    \includegraphics[width=0.49\linewidth]{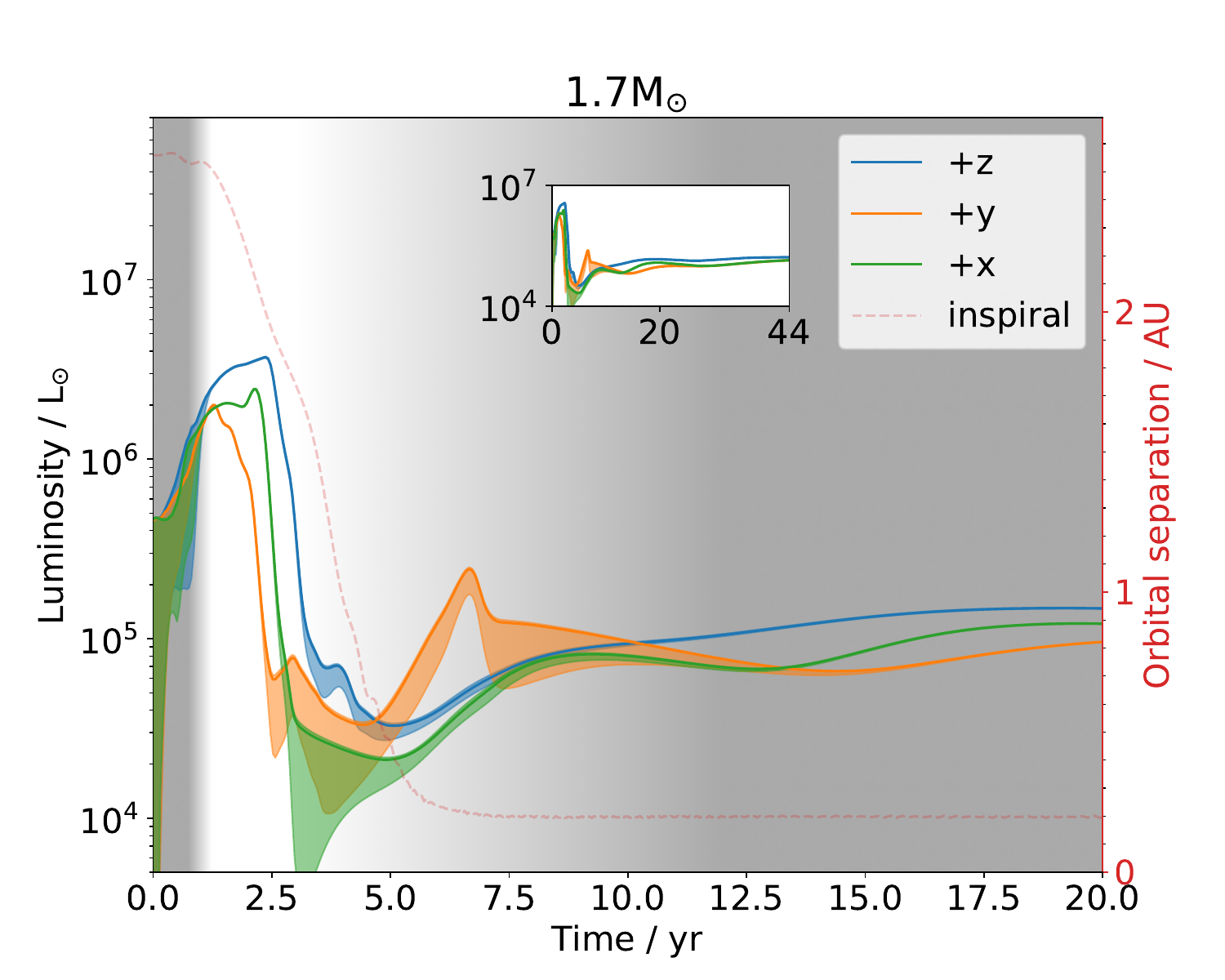}
    \includegraphics[width=0.49\linewidth]{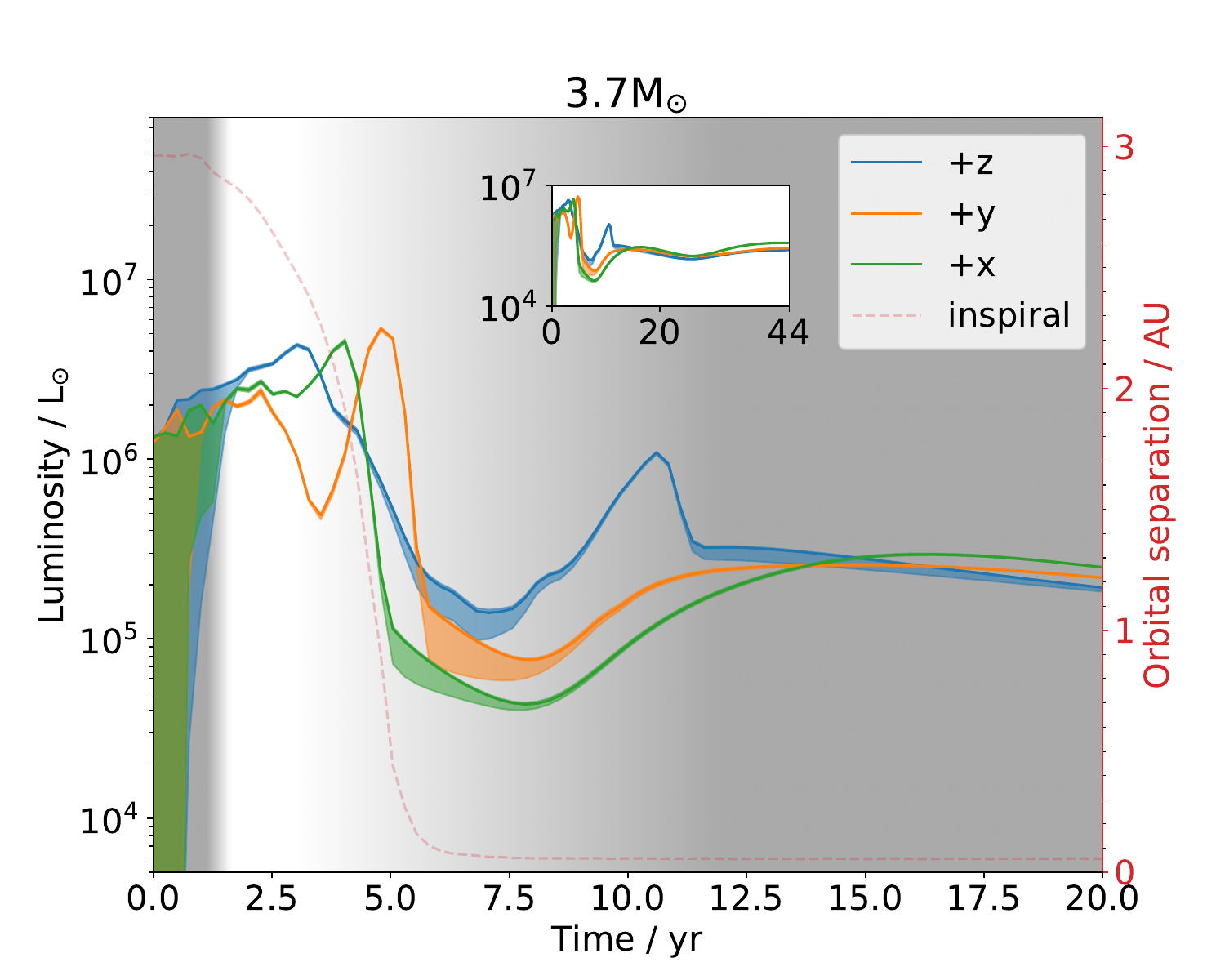}
    \includegraphics[width=0.49\linewidth]{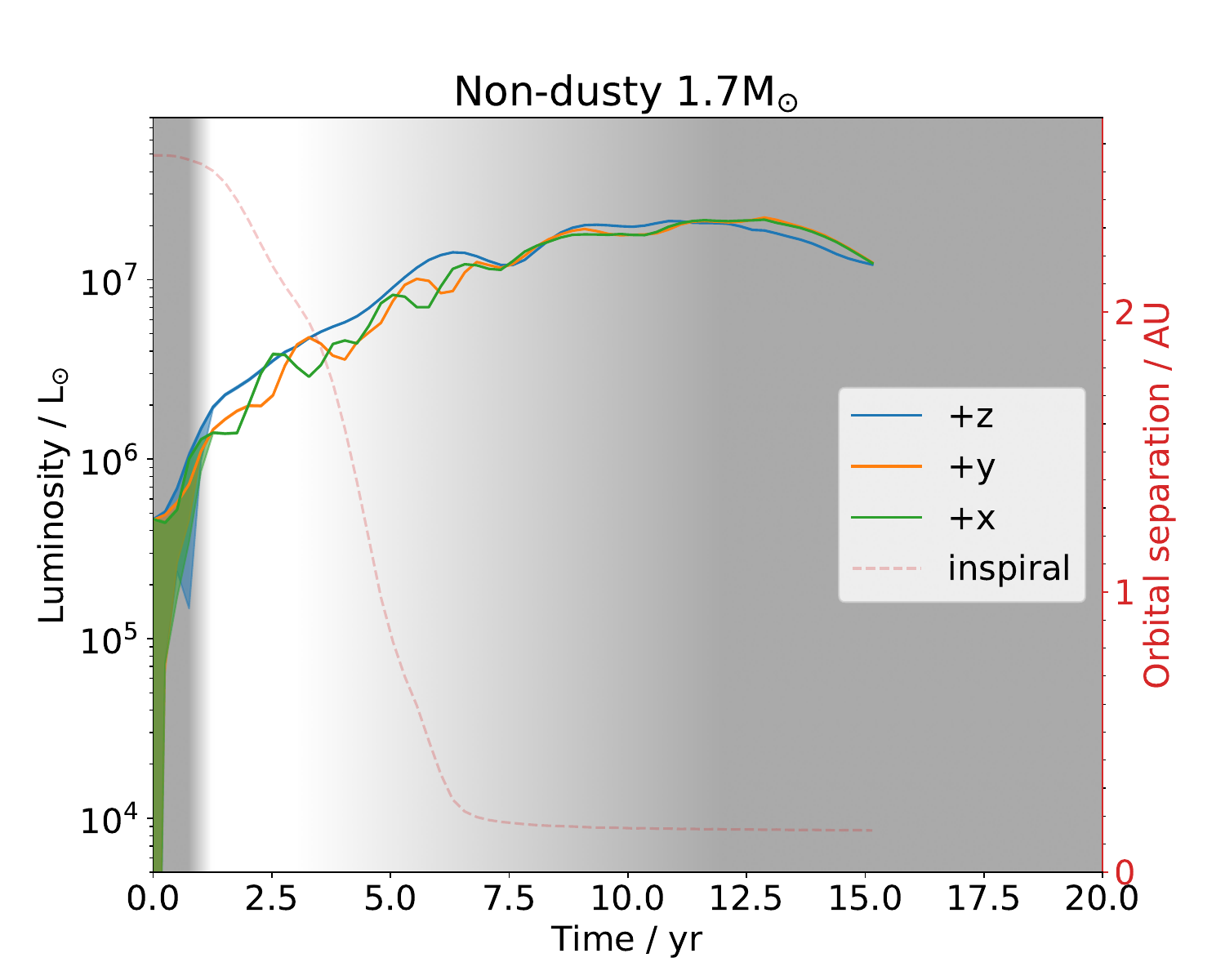}
    \includegraphics[width=0.49\linewidth]{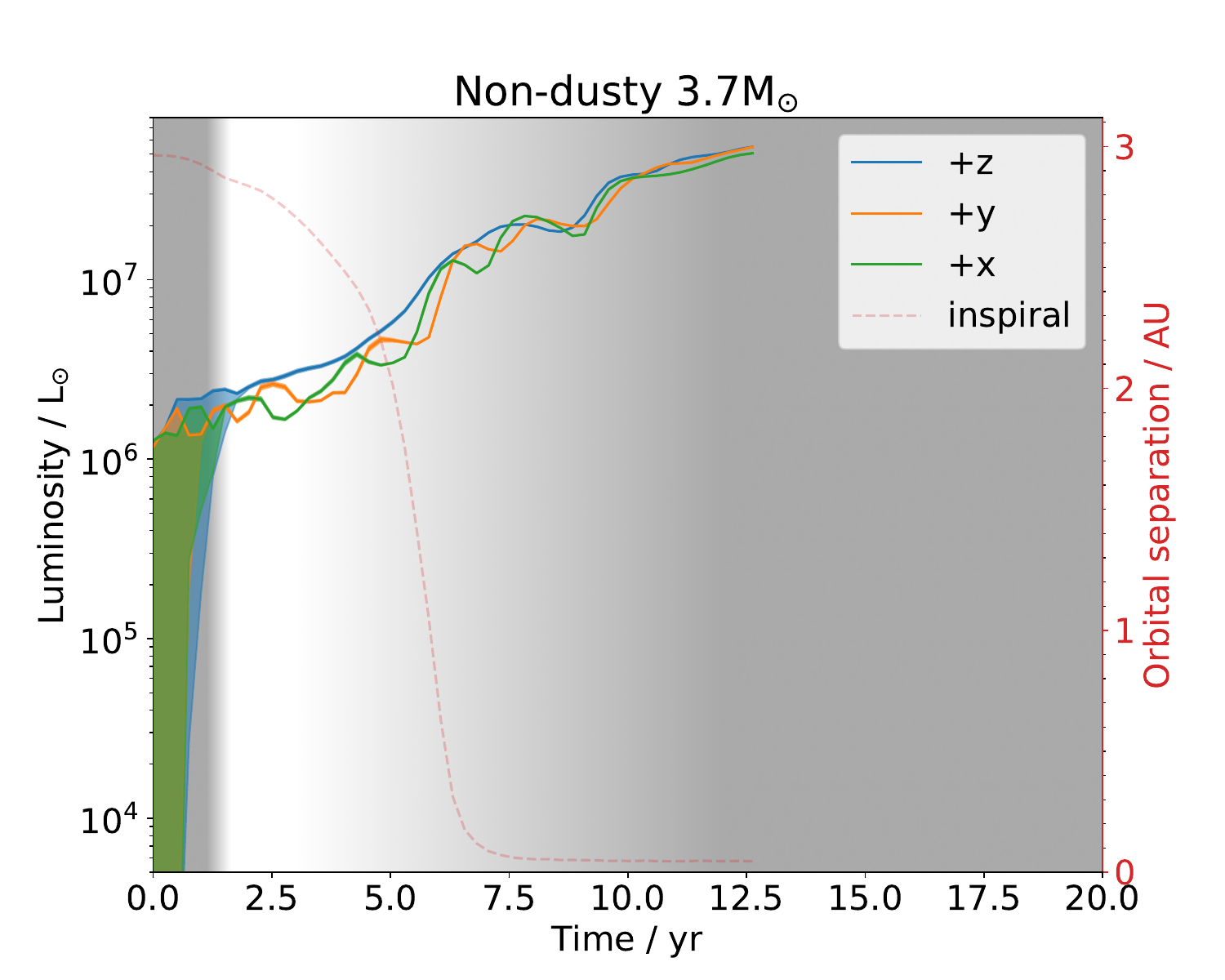}
    \caption{
    Bolometric luminosity lightcurve of the  of the  1.7~\Msun{} (left panels) and the 3.7~\Msun{} (right panels) models for dusty (top panels) and non-dusty (bottom panels) simulations, integrated using a grid of $256\times256$ rays. Blue, orange, and green curves are the results viewed from $+z$, $+y$, and $+x$ directions, respectively, with the semitransparent areas of the corresponding colours indicating the uncertainty due primarily to resolution (see Section~\ref{sec:uncertainties:surface-particles}).
    The light red dashed line is an illustration of the orbital separation between the donor and companion point masses, superimposed here for reader's convenience. \updated{The grey gradient background indicates where the lightcurve is more (light grey) or less (dark grey) trustworthy because of the combination of all uncertainties, not just resolution,  discussed in this paper.}
    \updated{The insets in the top panels show the extended 44-year evolution, whereas the main panels focus on the first 20 years.}
    }
    \label{fig:lc:both}
\end{figure*}

In Figure~\ref{fig:lum-image:2md} we present a total intensity 
image obtained by tiling $256 \times 256$ pixels (this resolution choice is justified in Appendix~\ref{app:convergence:ray-grid}).
The surface brightness at each pixel, $i$, is equal to the specific intensity, $I_i$.
An integrated luminosity is presented in Figure~\ref{fig:lc:both} for both simulations with and without dust.

\begin{figure}
    \centering
    \includegraphics[width=\linewidth, trim={10 40 80 90}, clip]{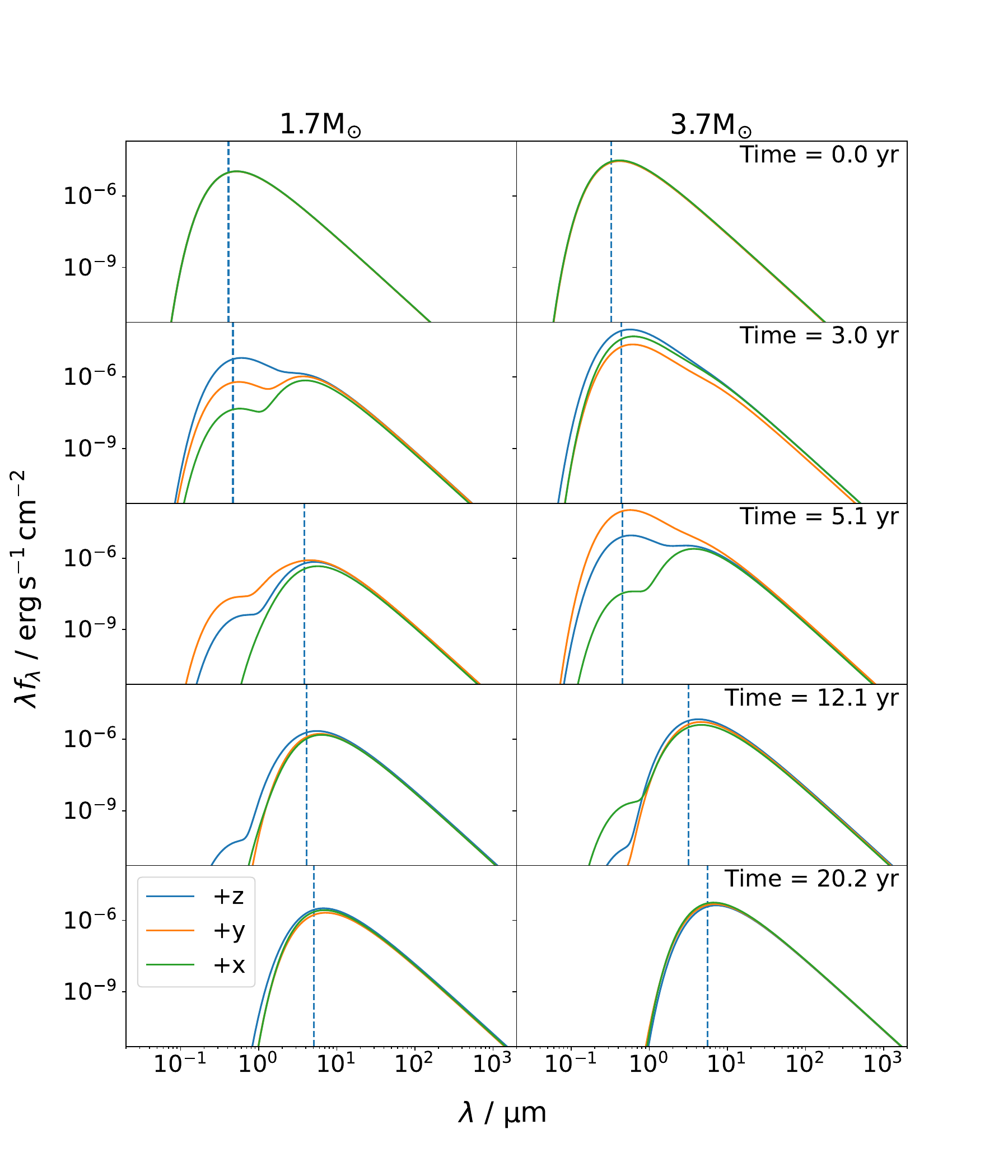} 
    \caption{Spectral flux density as a function of wavelength for the 1.7~\Msun{} (left column) and the 3.7~\Msun{} (right column) dusty simulations, integrated using a grid of 256$\times$256 rays, seen from $1~\mathrm{kpc}$.
    Blue, orange, and green are the three viewports, $+z$, $+x$ and $+y$, respectively.
    The highest peak of $f_\lambda$ (which is not the same as the peak of $\lambda f_\lambda$) is marked with a dashed vertical blue line.
    The colour reddens during the simulation,
    with a redder peak forming between 3~yr and 5.1~yr for the 1.7~\Msun{} and 3.7~\Msun{} simulations, caused by dust formation happening at that time \citep{Bermudez2024a}.
    }
    \label{fig:spec:both}
\end{figure}

In Figure~\ref{fig:spec:both}, we approximate the \gls{SED} by considering a wavelength dependent blackbody emissions from each particle, while still assuming grey opacity. In other words, we assume each particle has the following source function
$$
S_{\lambda, j} = \frac{2 h c^2}{\lambda^5 ( e^{\frac{hc}{\lambda k_B T}} - 1 ) },
$$
and use this expression in Equation~\ref{eq:calc:luminosity}. 

\begin{figure*}
    \centering
    \includegraphics[width=0.49\linewidth]{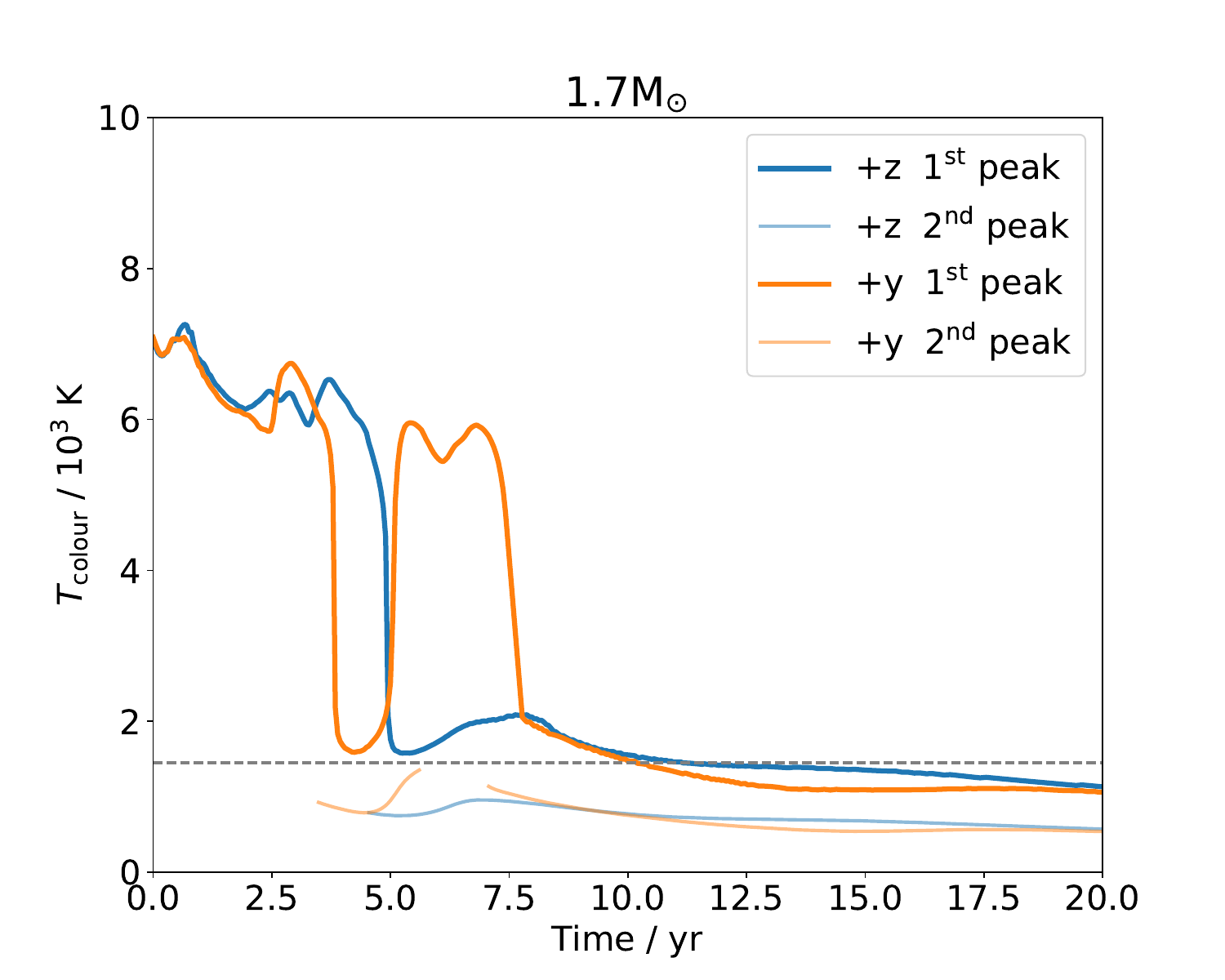}
    \includegraphics[width=0.49\linewidth]{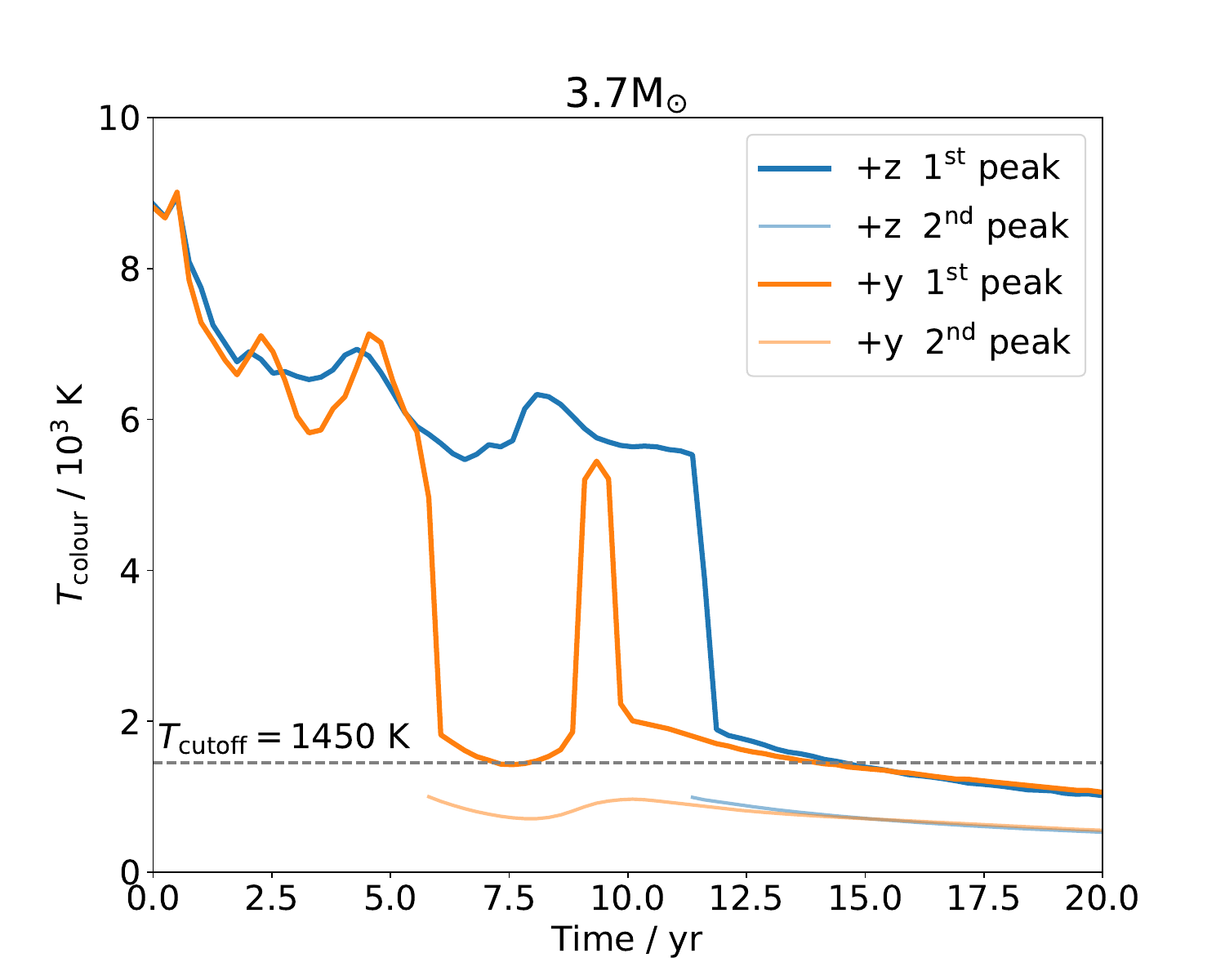}
    \caption{
    Evolution of the \updated{colour temperature, $T_{{\rm colour}}$ (as would be derived by an observer from the \gls{SED})},  for the 1.7~\Msun{} model (left panel) and the 3.7~\Msun{} model (right panel). The thick lines indicate the temperature of the first, hot peak in the \gls{SED} for two viewing directions. The presence of a corresponding thin line at a given moment in time indicates the presence of a second, cooler peak in the \gls{SED} (evident in Figure~\ref{fig:spec:both}).
    The \updated{horizontal}, dashed line marks the cut-off temperature $T_\mathrm{cutoff}$ below which dust can condensate (see Figure~\ref{fig:profile:2md-kappa}).
    }
    \label{fig:Teff}
\end{figure*}

In Figure~\ref{fig:Teff} we derive the mean \updated{colour temperature of the model, $T_\mathrm{colour}$, determined by first fitting the \gls{SED} with a 2-component blackbody spectrum model, and selecting the temperature corresponding to the brighter peak.
After dust starts to form, a second cooler component appears in the \gls{SED}, resulting in the sudden drop of $T_\mathrm{colour}$ when the cool peak becomes stronger than the hotter \gls{SED} component peak from the stellar gaseous photosphere. The momentary comeback of the hot peak in $T_\mathrm{colour}$
} around $\sim$6 yr for the $+y$ direction in the 1.7~\msun\ model is caused by the hot gas component shining through the dust and temporarily outshining the dust component again.

Next, we divide the lightcurve in Figure~\ref{fig:lc:both} into three phases and describe their characteristics, guided by the $+z$ viewport --- looking down the pole; blue curves in all figures.

\begin{itemize}
\item {\it The rise to the peak.} The exact reason for the luminosity increase that we call the ``outburst" in LRNe has been a source of debate. \citet{MacLeod2017} interpreted the peak and plateau emission of M31 LRN 2015 as different phases of a common envelope ejection, the first comprised by the release of a small amount of fast material, while the latter by the slower release of recombination energy from a larger amount of expanding material. \citet{Metzger2017} attributed it instead to emission from an ejected and cooling shell, which shortly thereafter impacts and shocks previous, toroidal mass loss. 

In our simulations, the bolometric luminosity increases to a peak value of $\sim4\times 10^6$~\Lsun\ (Figure~\ref{fig:lc:both}) and reaches temperatures of $\sim 6500$~K (Figure~\ref{fig:Teff}), at 2.5 and 3 years after the start of the simulations, for the 1.7 and 3.7~\msun\ models, respectively. A strong expansion to 15--30~au (geometric radius, solid line in Figure~\ref{fig:Rph}) or 7~au (observed radius; dot-dashed line in the same Figure) powers the light rise. 

\updated{The timescale and brightness of the lightcurve rise is driven by the inspiral which takes place between 1 and 6 years and is expected to be well reproduced. Timescale and brightness of the early rise are {\it also}  affected by (i) resolution (Section~\ref{sec:uncertainties:SPHres}), (ii) the insertion of the companion in the computational domain and (iii) the surface instability of the stellar model \citep[e.g.,][]{GonzalezBolivar2022}. As discussed in Appendix~\ref{app:uncertainties:initial-years}, however, the uncertainty due to surface instability is not a major problem, seemingly not affecting the peak luminosity and having a relatively modest impact only later on. }


The side viewports (indicated by the green and orange lines in all figures)  reveal that the 1.7~\Msun\ model peaks earlier, whereas the 3.7~\Msun\ model exhibits a delayed peak.
Furthermore, the more massive model shows a broader and notably more irregular peak preceding the decline phase discussed below, characterized by multiple sub-peaks and troughs. By 5.1 years, the surface brightness maps (Figure \ref{fig:lum-image:2md} and the accompanying movie) show that the 1.7~\Msun\ model is already dimming due to dust obscuration (discussed in detail below). In contrast, the 3.7~\Msun\ model still exposes its bright core, particularly from the polar perspective; this occurs because dust formation is delayed in this higher-mass, lower-$q$ ($q = M_2 / M_1$) simulation.


\updated{
We note that, at time zero, the models are more luminous and hotter than the corresponding time-zero {\sc mesa} profile. This offset arises from the limited resolution at the stellar surface, as we explain and quantify in Section~\ref{sec:uncertainties:SPHres} and Appendix~\ref{app:uncertainties:initial-years}. Consequently, the light-curve plot shows large initial uncertainties (shaded regions in Figure~\ref{fig:lc:both}; the uncertainty definition is given in Section~\ref{sec:uncertainties:surface-particles}).
}

\item {\it The decline to the trough.} This phase goes from the peak to the trough (again using the pole-on view as reference), which starts at $\sim 2.5$~yrs, for both models, at which point the bolometric light declines over 2.5 -- 5 years for the two models  (Figure~\ref{fig:lc:both}). While the photosphere is still expanding in both models (from $\sim$20 to 70~\Rsun\ for the 1.7~\msun\ model and from 20 to 100~\Rsun\ for the 3.7~\msun\ model), the dominant peak in the \gls{SED} for the 1.7~\msun\ model goes from thousands of degrees Kelvin to hundreds, as the dust enshrouds the object and dominates its colour. The 3.7~\msun\ model (pole view), on the other hand, is quite different: first of all, the light decline is not as severe as it is for the lower mass model, because the \gls{SED} remains dominated by a (cooling) stellar photosphere. However, the other two (side) viewports display more variability during the peak, and decline later than the pole view light, but more abruptly and to a deeper \updated{trough}, as the dust forms a little later, and preferentially on the equator. In all cases, the light decline is that expected for an expanding, adiabatically cooling photosphere, where the only variable is when, where and how much dust forms.

\updated{
Given the adiabatic assumption, the lack of radiative transport and the lack of dust destruction in our simulations, the dust in the newly formed dust shell cannot be destroyed by the numerous hot photons produced by the inner gas layers. Thus, the exact lifetime of the dusty shell and the exact brightness and width of the trough in the lightcurve will not be accurate; however, the qualitative claim of a trough in luminosity following the peak should survive this limitation.
}

\item {\it The rise to the plateau.} After the trough, the light recovers to a plateau value that is approximately flat (see Figure~\ref{fig:lc:both} insert, where the full evolution to 44 years is seen). The 1.7~\msun\ model rises rapidly and then more slowly to the end of the simulation. The 3.7~\msun\ model rises to a new peak and then rapidly declines (the peak has a width of$\sim2$~yrs) to a plateau, which then continues to decline gently. As is clear from the movie associated with Figure~\ref{fig:lum-image:2md}, the 3D nature of the structure displays an irregular surface that can reveal outer or inner layers making the light temporarily brighter or dimmer. The luminosity of the plateau, at $1-2\times 10^5$~\Lsun\ for both models is too large, due to the lack of any radiation/cooling, as explained in Section~\ref{sec:uncertainties:cooling}.


In Figure~\ref{fig:lc:both} we also show the lightcurve obtained for the two simulations carried out with no dust nucleation (and carefully compared to the dusty ones in \citet{Bermudez2024a}).
The peak and trough behaviour lasting a few years from the start of  \gls{RLOF} observed in the dusty simulations is not present in the non-dusty cases.
There, the \gls{CE} gradually brightens and only much later dims (after about 11 years in the case of the 1.7~\msun\ model, while for the 3.7~\msun\ model no dimming is observed within the 12.5~yrs of the simulation), as the photosphere expands and adiabatically cools, with a decrease of the \updated{colour} temperature.

\end{itemize}

\subsection{Properties of the photospheric surface}
\label{sec:the_evolutuion_of_the_photospheric_radius}

\begin{figure*}
    \centering
     \includegraphics[width=0.49\linewidth]{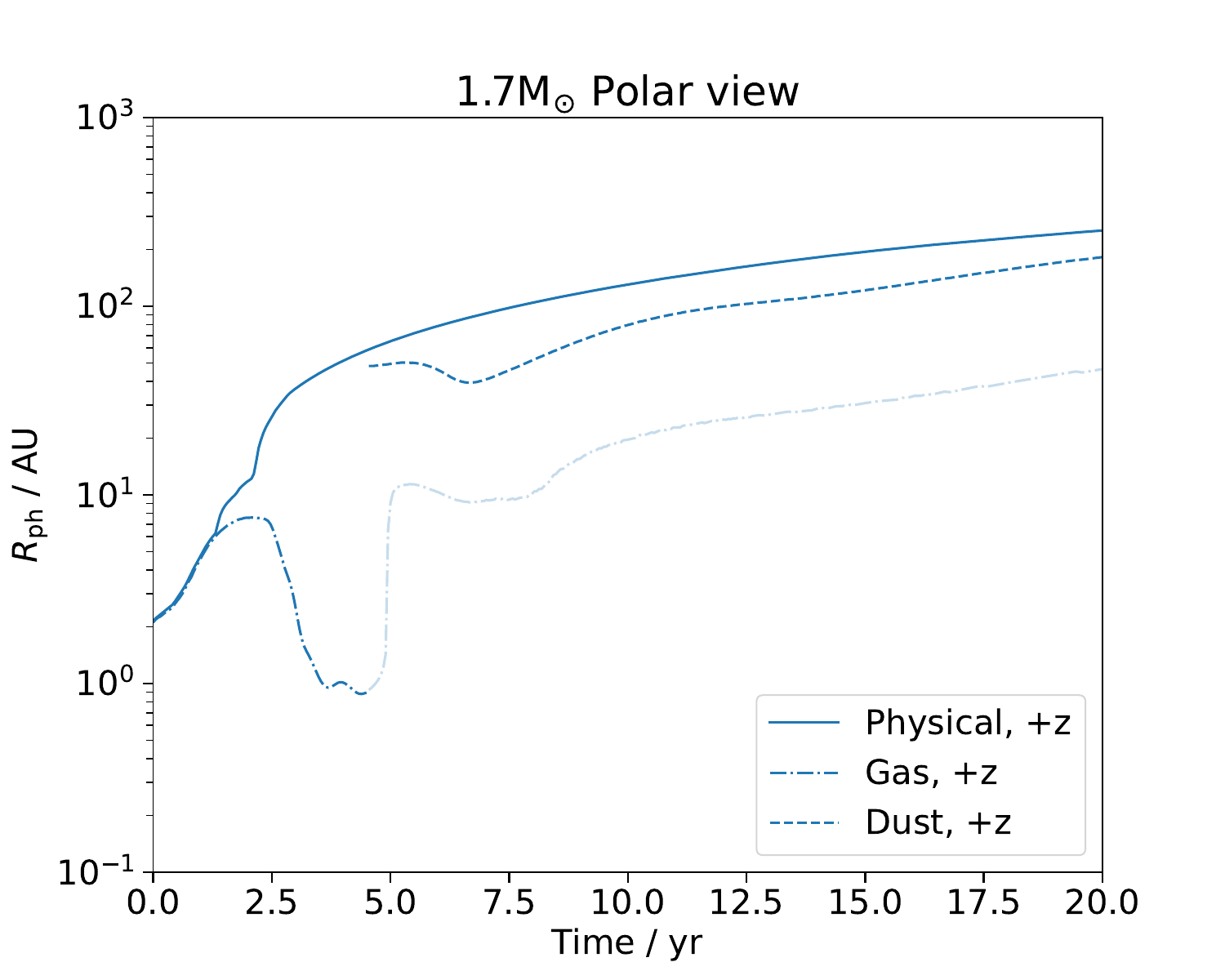}
    \includegraphics[width=0.49\linewidth]{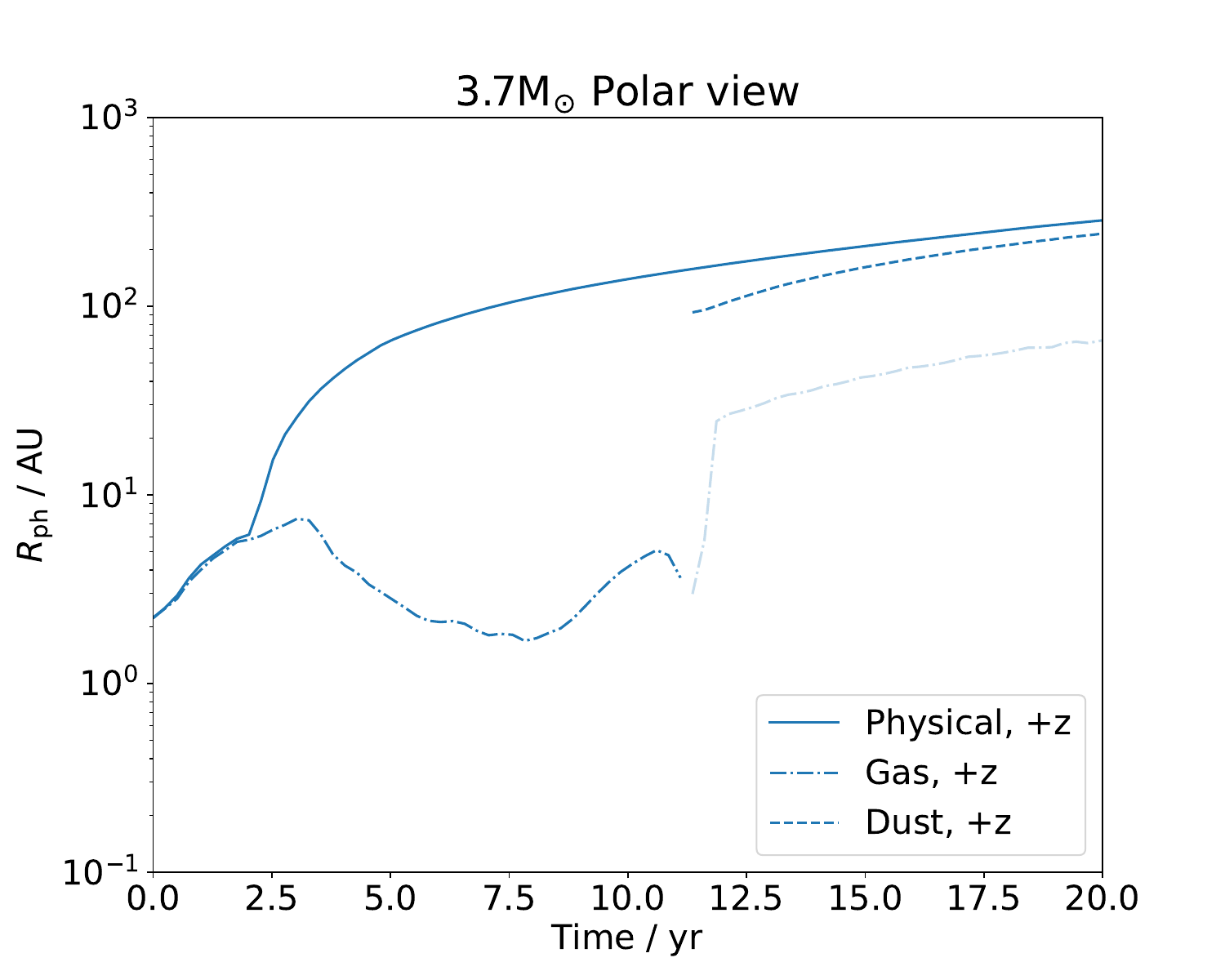}
    \includegraphics[width=0.49\linewidth]{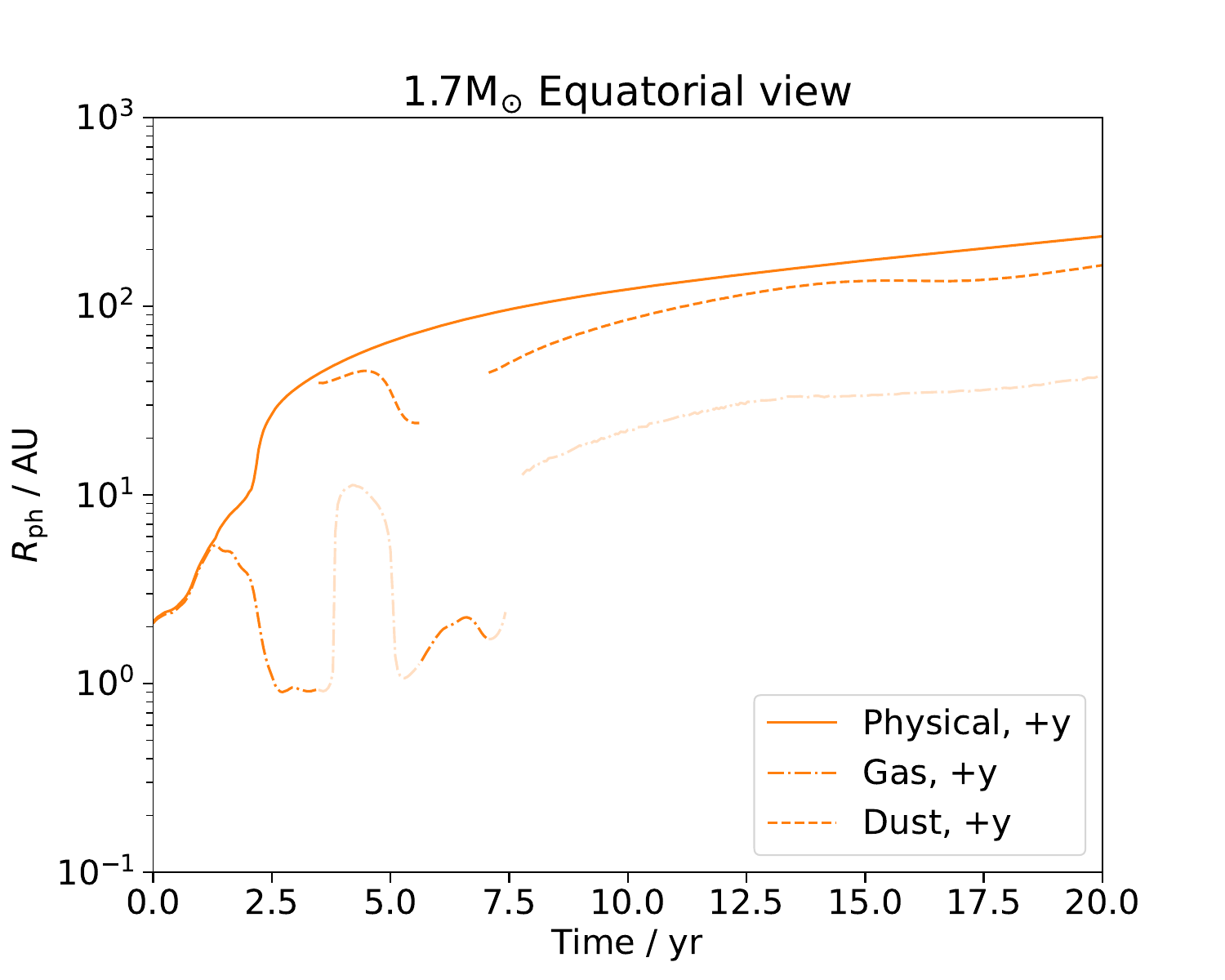}
    \includegraphics[width=0.49\linewidth]{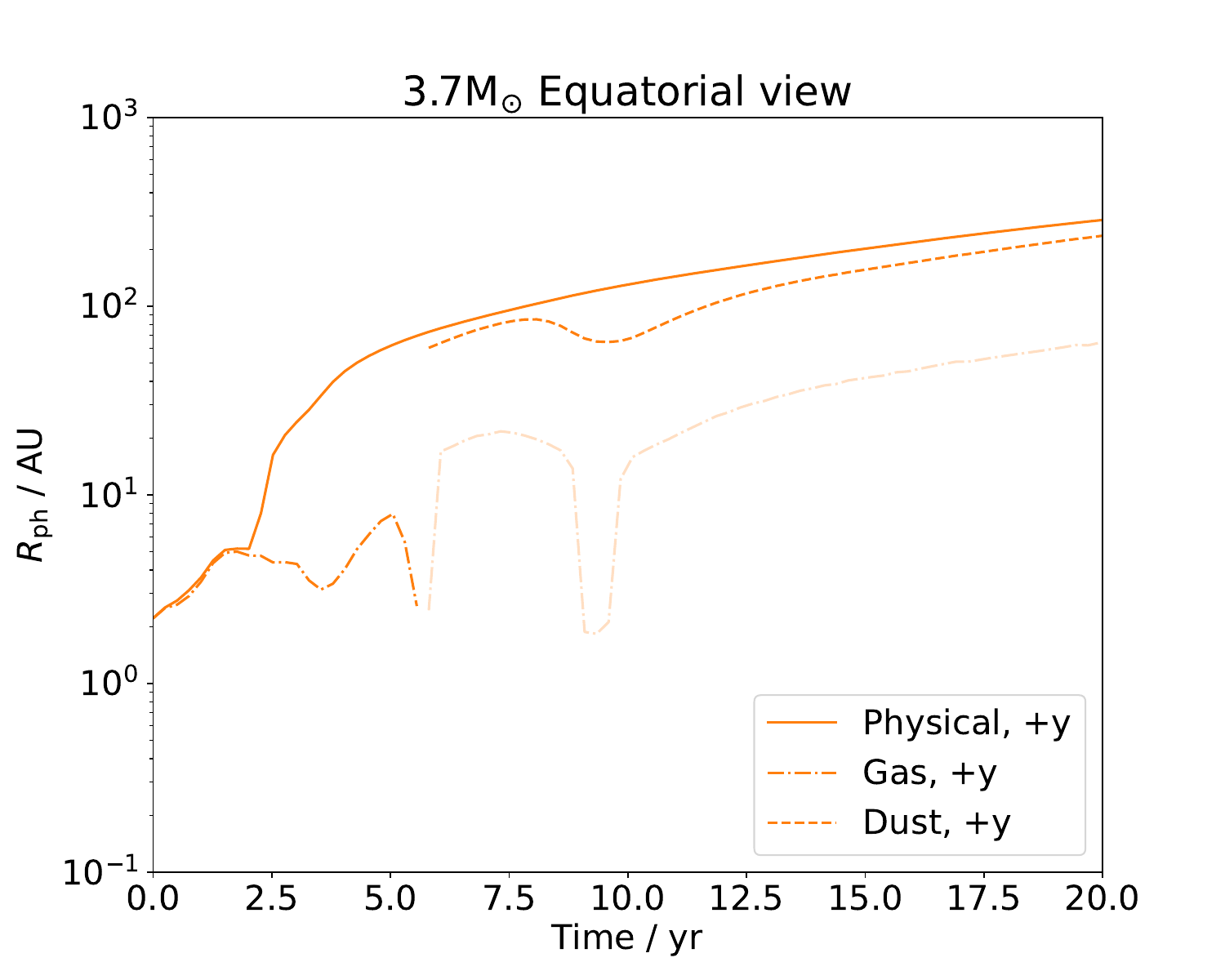}
    \caption{
    Photosphere radius for the 1.7~\Msun{} (left panels) and 3.7~\Msun{} (right panels) simulations. Solid lines: estimates are by the ``geometric" method, using Equations~\ref{eq:def:Aph} and~\ref{eq:def:Rph}. Dot-dash and dashed lines: estimates are by the ``observer's" method -  \updated{$R_\mathrm{ph, obs} = \sqrt{L / 4 \pi \sigma T_{{\rm colour}}^4}$, where $T_{{\rm colour}}$} is the \updated{colour} temperature obtained by fitting the gas (dot-dashed) and dust (dashed) peak of the \gls{SED} as in Figure~\ref{fig:Teff}. Faint dot-dashed lines show the size of the gas photosphere seen through an optically-thin dust photosphere. Blue (top panels) and orange (bottom panels) lines are for measurements in the $z \rightarrow +\infty$ (polar) and $y\rightarrow +\infty$ (orbital) directions, respectively.}
    \label{fig:Rph}
\end{figure*}

We calculate the size of the photosphere using two methods. The first method is by using the Stefan-Boltzmann relation with the calculated bolometric luminosity and temperature from the peak(s) of the \gls{SED}. Observers fit \gls{SED} measurements using one or more Blackbody  curves, depending on what they observe. Often, in \glspl{LRN} two peaks in the \gls{SED} indicate that the stellar (gas) photosphere is visible behind a shell of optically thin dust. From fitting both peaks the temperatures and sizes of both can be extracted. At late times it is often the case that the dust becomes optically thick and only one, redder \gls{SED} peak remains visible, from which the size of the expanding dust cocoon can be deduced. This is also the case in the simulations, where a sample of the \gls{SED} at different times can be seen in Figure~\ref{fig:spec:both}. The first method therefore uses exactly the observers' method. 

The second method finds the size of the photosphere by counting the number pixels for a given viewing angle, where $\tau > \frac{2}{3}$
\begin{equation}
    A_\mathrm{ph} \equiv
    \sum_{ i \in \{ i | \tau_i > \frac{2}{3} \} } \Delta A_i.
    \label{eq:def:Aph}
\end{equation}
The photospheric size is hence simply
\begin{equation}
    R_\mathrm{ph} = \sqrt{\frac{A_\mathrm{ph}}{\pi}},
    \label{eq:def:Rph}
\end{equation}
assuming a circular photosphere as projected on the plane of the sky\footnote{Note that this method is distinct from the one adopted by \citet{Bermudez2024a}, where the photosphere size was simply the distance between the core of the star to the photosphere integrated along one ray only in the polar or equatorial direction.}
.

The value of $R_\mathrm{ph}$ using the two methods is plotted in Figure~\ref{fig:Rph}. Using the ``geometric" method to measure the photospheric radius, we see a near-constant expansion rate during the 40 years (in Figure~\ref{fig:Rph} we only plot the first 20 years). Both objects' photospheres grow to about 250~au over the simulation time of 44 years. Using a simple linear regression fit, for the 1.7~\Msun{} simulation after 5 years, the photosphere is expanding at almost 60~km~s$^{-1}$ in all three directions.
For the 3.7~\Msun{} simulation after 5 years, the photosphere is expanding at $\sim$70~km~s$^{-1}$ for the three directions.

Tracking the photospheric radius using the ``observers" method reveals a more complex evolution.  In Figure~\ref{fig:Rph}, the dot-dashed and dashed lines track the radius measurements of the first (warmer) and second (hotter) \gls{SED} peaks, respectively. When the cooler, dust-dominated \gls{SED} peak becomes stronger than the hotter, gaseous peak, the dashed line transitions to semi-transparent, signalling that the gaseous photosphere is being viewed through an increasingly optically thick layer of dust. Because the orbiting companion ejects chaotic, expanding, and eventually dusty plumes, the side (+y) view appears significantly more complex than the polar view. By roughly 20 years, the gaseous photosphere is entirely obscured by the dusty photosphere. Ultimately, while the ``geometric" photosphere consistently outlines the object's outermost boundary on the plane of the sky, the ``observer" photosphere represents the temperature components directed toward the viewer. These two definitions align early on, while the expanding cocoon is symmetric and dust free.


\subsection{The dispersion of the dust shell}
\label{sec:dispersion-of-the-dust-shell}

After the opaque dust shell has fully enshrouded the object, it experiences free expansion that will reduce its density,  decreasing its optical depth,
which eventually will render it transparent even to  optical light. Here we linearly extrapolate our results to investigate when and how this will happen.

\begin{figure*}
    \centering
    \includegraphics[width=0.49\linewidth]{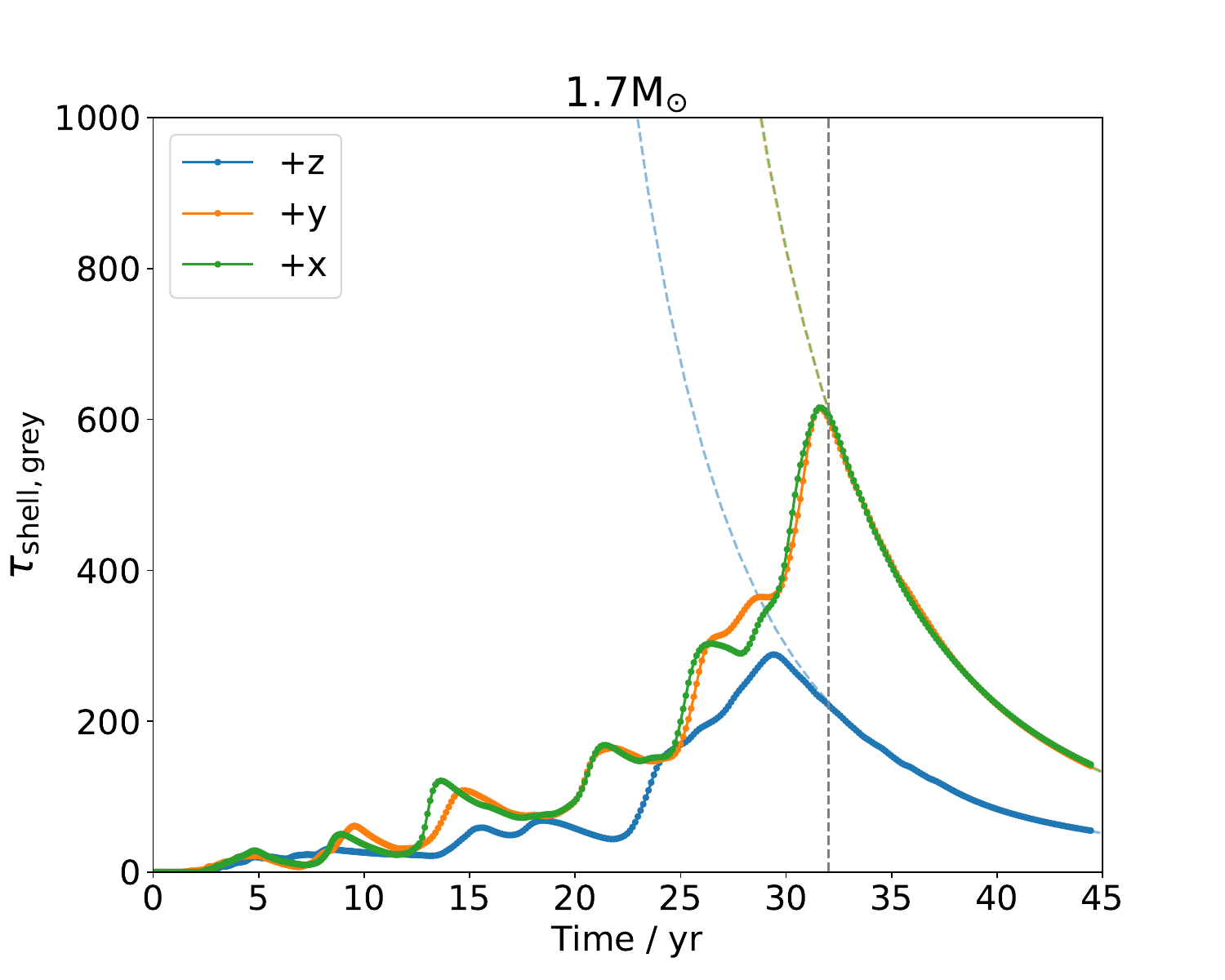}
    \includegraphics[width=0.49\linewidth]{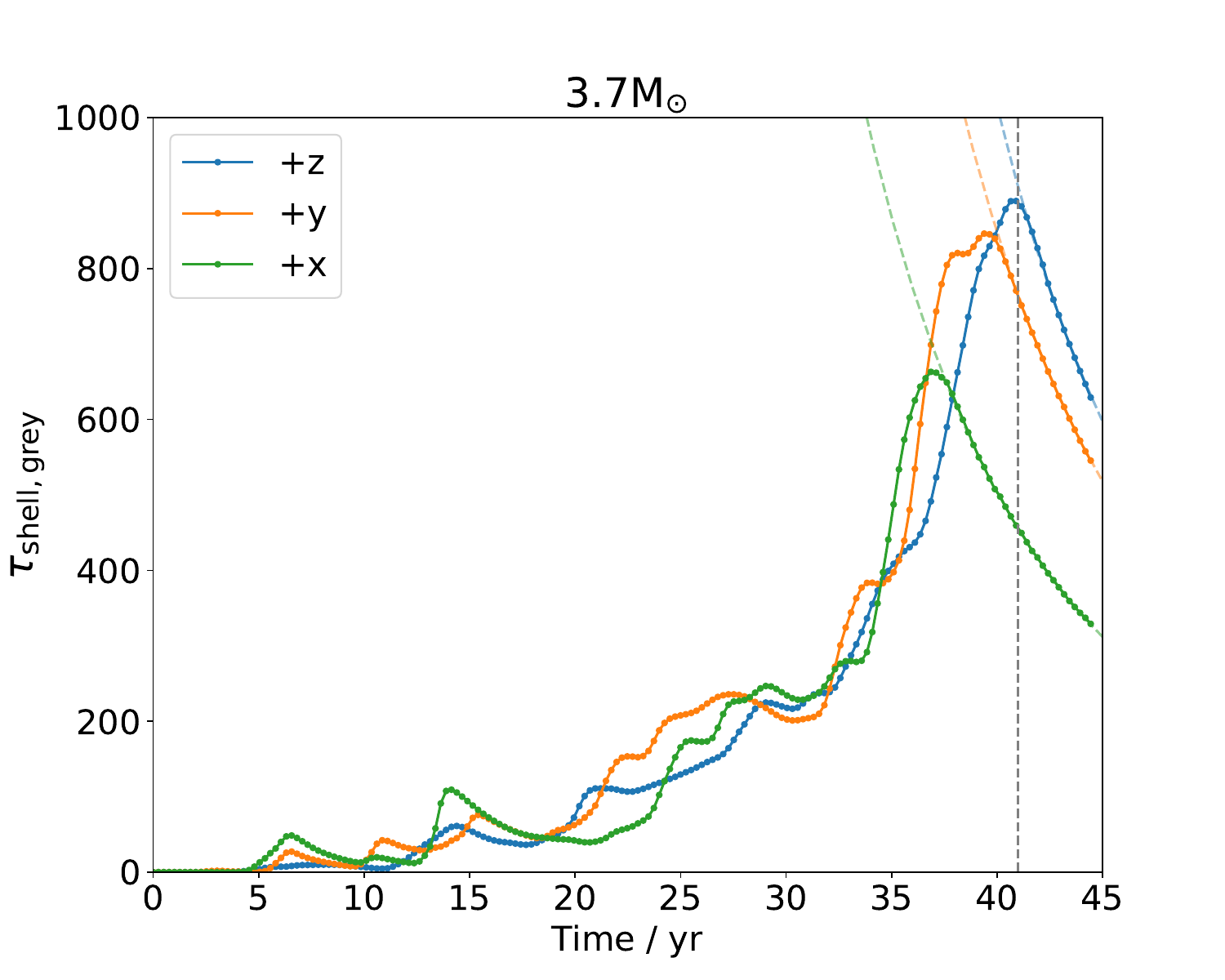}
    \caption{Grey optical depth of the dust shell, $\tau_\mathrm{shell,grey}$, for the 1.7~\Msun{} (left panel) and the 3.7~\Msun{} (right panel) simulations, observed from the three viewports.
    $\tau_\mathrm{shell, grey}$ starts to drop after the first few decades, as expected by the free expansion model of the dust shell.
    Curves in the corresponding colours (dashed lines) 
    use Equation~\ref{eq:calc:dust-dispersion:tau} to fit data on the right of the grey vertical line ($t_{\rm stop}$, when dust stops forming) and extrapolate the time when the dust becomes transparent (see Table~\ref{tab:dust-dispersion}).
    }
    \label{fig:tau-dust}
\end{figure*}

Using only the {\it grey} dust opacity, which dominates the total opacity, $\kappa_\mathrm{d, grey}$, we can obtain the optical depth of the dust shell, $\tau_\mathrm{\rm shell, grey} (t)$, at given time, $t$, using Equation~\ref{eq:def:tau}.
Using $\rho (r, t) = dm / (4 \pi r^2(m, t) dr)$,
we can express Equation~\ref{eq:def:tau} in mass coordinates
\begin{equation}
    \tau_\mathrm{\rm shell, grey}(t) = \int_0^M \frac{\kappa_\mathrm{{\rm d, grey}}}{4\pi r^2} dm.
\end{equation}
We assume that the dust opacity scales linearly with the dust/gas equilibrium temperature, $T$, in the later stage of the simulation as demonstrated by \citet{Bermudez2024a} --- ignoring radiative losses as well as the fact that dust and gas temperature may decouple, giving
\begin{equation}
    \kappa_\mathrm{d, grey}(T)
    = \kappa_\mathrm{d, 0} \frac{T}{T_0}
    = \kappa_\mathrm{d, 0} \frac{r_0^{2}}{r^{2}},
    \label{eq:calc:kappa-d-per-T}
\end{equation}
where $\kappa_\mathrm{d,0}$, $T_0$ and $r_0$ are constants, representing the quantities at the time when the flow becomes homologous \citep{Iaconi2019a}. 
Where for the second equality we have used the fact that temperature, volume and radius have the following relation for adiabatic changes
\begin{equation}
    T \propto V^{1-\gamma} \propto r^{3 - 3 \gamma},
\end{equation}
where $\gamma = 5/3$ for gas, and the volume dependency on $r^3$ as the shell is expanding homologously.
Substituting we obtain
\begin{equation}
    \tau_\mathrm{shell, grey}(t) = \int_0^M \frac{\kappa_\mathrm{d, 0} r_0^{2}}{4\pi r^{4}} dm.
\end{equation}
Finally, because of homologous expansion, we can assume
$r(m, t) = v_0(m) (t-t_0)$ do we can derive
\begin{equation}
    \tau_\mathrm{shell, grey}(t) = K (t - t_0)^{-4},
    \label{eq:calc:dust-dispersion:tau}
\end{equation}
where $K \equiv \int_0^M \frac{\kappa_\mathrm{d, 0} r_0^{2}}{4\pi v_0^{4}}dm$, where $M$ is the mass of the dusty shell. In Figure~\ref{fig:tau-dust} we fit the dust shell's optical depth, $\tau_\mathrm{shell, grey}(t)$, as a function of time to forecast when it will drop below $2/3$ (a time we call $t_{\tau_{10\mu m}}$, noting that this grey optical depth derives from a Planck-weighted opacity, which is skewed to the infrared). The fitting parameters, $K$ and $t_0$ in Equation~\ref{eq:calc:dust-dispersion:tau}, are listed in Table~\ref{tab:dust-dispersion}.
\updated{Under this extrapolation}
$\tau_{10\mu m}=120 - 160$ years from $t_0$ for the 1.7~\Msun{} simulation, and 
$200 - 230$ years for the 3.7~\Msun{} simulation. 

For the optical depth in the {\it visible} bands we have,

\begin{equation}
        \tau_{\rm shell, V}(t)
        = \int_0^M \frac{\kappa_{\rm d, V}}{4\pi r^{2}} dm
        \propto (t - t_0)^{-2},
\end{equation}
where $\kappa_{\rm d, V}$, the opacity in visible band, is approximately constant as temperature decreases (\citealt{Stagg1993}; instead of displaying a linear correlation with temperature like in Equation~\ref{eq:calc:kappa-d-per-T}\footnote{The main point of interest in \citet{Stagg1993} is their figure~3, which shows that the refractive indices used to estimate the opacity in the Mie theory are relatively constant with temperature, albeit for relatively high temperatures $> 273$K, showing (indirectly) the relative constancy of C-rich dust opacity with temperature.}). As a result, the visible optical depth drops far slower than the optical depth in the infrared.



\updated{
    Assuming that the shell has 20 to 300 times more optical depth at visible wavelengths (e.g., 5500~\AA) than at IR wavelengths (e.g., 10~$\mu$m; and assuming that the Plank mean opacity $\kappa_{\rm d, grey}$ is the same as $\kappa_{10\mu m}$), and using a $\kappa \propto \lambda^{-1}$ \cite[as is appropriate for visible wavelengths and large grains, see e.g.][]{Cardelli1989,Mathis1977}, 
    the dusty shell will become optically thin to visible radiation after $\sim$2\,400 years for the 1.7~\Msun{} simulation, and after 4\,700 years for the 3.7~\Msun{} simulation (see Table~\ref{tab:dust-dispersion}).
}

\begin{table}
\centering
\begin{tabular}{ rccccc }
    \hline
    Case & $t_0$ & $\log_{10}{(\mathrm{K / yr^4})}$ & $t_\mathrm{stop}$ & $t_\mathrm{\tau_{10\mu m}}$ & $t_{\tau_{5500\AA}}$ \\
    & (yr) & & (yr) & (yr) & (yr) \\

    \hline
                $+z$ & $ 2.85 \pm 0.12 $ & $ 8.212 \pm 0.006 $ & 32 & $130$ & 2400\\ 
    1.7~\Msun{} $+y$ & $ 4.03 \pm 0.06 $ & $ 8.574 \pm 0.004 $ & 32 & $160$ & 3800 \\ 
                $+x$ & $ 4.24 \pm 0.05 $ & $ 8.562 \pm 0.003 $ & 32 & $160$ & 3800\\ 
                $+z$ & $ 4.78 \pm 0.50 $ & $ 9.194 \pm 0.023 $ & 41 & $220$ & 6000\\ 
    3.7~\Msun{} $+y$ & $ 1.91 \pm 0.13 $ & $ 9.252 \pm 0.006 $ & 41 & $230$ & 5900\\ 
                $+x$ & $ 0.72 \pm 0.23 $ & $ 9.079 \pm 0.010 $ & 41 & $210$ & 4700\\ 
    \hline
\end{tabular}
\caption{
    \updated{
        Fitting parameters of Equation~\ref{eq:calc:dust-dispersion:tau}, $K$ and $t_0$; the time when dust stops forming, $t_\mathrm{stop}$, deduced from the peak of the optical depth plot in Figure~\ref{fig:tau-dust} - which coincides with the actual time of dust formation cessation observed by \citet{Bermudez2024a}, and the predicted times, $t_{\tau_\mathrm{10\mu m}} \equiv t(\tau_\mathrm{10\mu m} = 2/3)$ and
        $t_{\tau_\text{5500\AA}} \equiv t(\tau_\text{5500\AA} = 2/3)$,
        when the dust shell becomes transparent to infrared and visible radiation, respectively, and the gas' photosphere becomes visible again,  for the 1.7~\Msun{} and 3.7~\Msun{} simulations and three viewports (observer located at $z \rightarrow +\infty$, $y \rightarrow +\infty$, or $x \rightarrow +\infty$).
    }
}
\label{tab:dust-dispersion}
\end{table}

In both models dust formation {\it starts} in an equatorial belt (figures 3, 4 and 13 in \citealt{Bermudez2024a}) with the 1.7~\msun\ model starting somewhat earlier than the more massive model. In Figure~\ref{fig:tau-dust} we see that dust formation also stops first for the 1.7~\msun\ model. Dust stops  forming first and disperses first in the equatorial direction for the 3.7~\msun\ model, similarly to how it started. The opposite is the case for the 1.7~\msun\ model, where dust formation stops first and disperses first in the {\it polar} directions.

These behaviours are related to the lower envelope mass of the 1.7~\msun\ model as well as its larger mass ratio which, as we saw in \citet{Bermudez2024a}, leads to a great deal more asymmetry in the dust shell and ejecta, such as larger radial velocities in the polar direction.

\section{Discussion}
\label{sec:uncertainties}

Here, we discuss the uncertainty deriving from finite simulation resolution, particularly at the surface, the adiabatic assumption, as well as what observations can tell us about our simulation efforts so far.




\subsection{SPH simulation resolution test}
\label{sec:uncertainties:SPH}

\begin{figure}
    \centering
    \includegraphics[width=\linewidth]{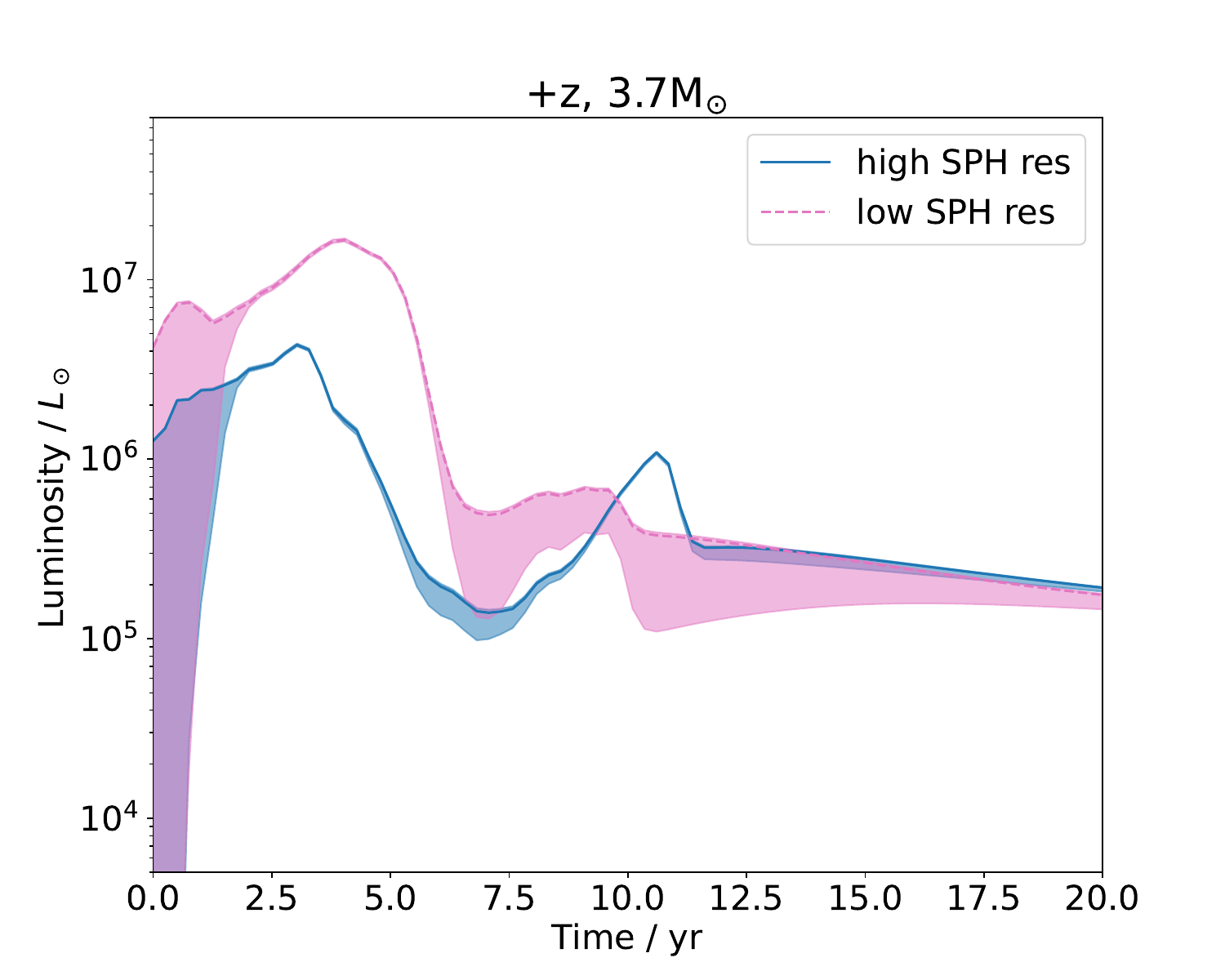}
    \caption{
    Lightcurves of the 3.7~\Msun{} dusty simulations as viewed from the polar ($+z$) direction, comparing  high resolution (1.3M particles, blue curve) and low resolution (200k particles, pink curve). The shaded areas are the uncertainties,  as computed following the method explained in Section~\ref{sec:uncertainties:SPHres} and Appendix~\ref{app:upper-lower-bounds}.
    }
    \label{fig:lc:SPHres}
\end{figure}
To determine the effect of resolution (i.e., the total number of \gls{SPH} particles) on the resulting lightcurve, we conducted the  same analysis described in Section~\ref{ssec:lightcurve} but using the 3.7~\Msun{} simulation carried out with 200\,000 particles by \cite{Bermudez2024a}; cf. the 1.3 million SPH particles of our production simulations. The two simulations have a number of \gls{SPH} particles that differ by a factor of $\sim$ 7, just shy of the factor of 8 ($2^3$) that achieves a factor of 2 in linear resolution. 

The results presented by \updated{\citet{Bermudez2024a}} for exactly the same two simulations, were not significantly impacted by their different resolutions, as demonstrated by them - see, e.g., their figure A1. However, in Figure~\ref{fig:lc:SPHres} we show the two lightcurves that result from the two simulations with different resolutions, showing a significant difference, with the higher resolution simulation overall dimmer by a factor of 2-3\updated{, and 3.3 times dimmer at $t=0$ specifically---}closer to the input value from the 1D stellar structure.
The approximate shape of the luminosity peak and trough are similar, but the specific timing and brightness of the peak, the shape of the descent to the trough, and even the presence of another secondary peak are not converged. 

The lightcurve uncertainty (shaded areas; explained in Section~\ref{sec:uncertainties:SPHres} and Appendix~\ref{app:upper-lower-bounds}) is smaller in the high-resolution (blue curve), as expected. However, the shaded areas do not fully account for the difference between the two curves. This suggests that the error is not only due to the low surface resolution. In Appendix~\ref{app:uncertainties:initial-years} we demonstrate that the imperfect reproduction of the surface layer, i.e., the specific position of the relatively few \gls{SPH} particles at the surface of the star,  do impact the light calculation at some level, while it does not impact other simulation results, such as final orbital separation, unbound mass or dust mass, as already demonstrated by \citet{GonzalezBolivar2022} and \citet{Bermudez2024a}. This problem will be greatly ameliorated by higher surface resolution.

\citet{Hatfull2021a} carried out a resolution test with the number of \gls{SPH} particles changed by a factor of three (99\,955 vs. 299\,929). They remarked on some notable differences between the two lightcurves generated with the two resolutions. Ultimately a convergence test is extremely difficult to execute given prohibitive computational times\updated{: convergence can only be tested by running three simulations each with approximately 8 times more \gls{SPH} particles than the previous one (doubling the linear resolution), which is computationally infeasible.}

\subsection{Quantification of the uncertainty from low resolution at the surface}
\label{sec:uncertainties:SPHres}
\label{sec:uncertainties:surface-particles}

In this Section, we first find a way to formally quantify the accuracy of the lightcurve and then discuss a way to \updated{quantify the uncertainty} for the lightcurve itself, which is then shown as shaded areas in Figures~\ref{fig:lc:both}, \ref{fig:lc:ABmag}, \ref{fig:lc:SPHres} and \ref{fig:lc:xmdx}.

At the beginning of the simulation, the gas distribution is entirely optically thick. The photosphere is therefore located somewhere within one smoothing length ($25-30$~\Rsun{}) of the outermost SPH particle encountered by the ray.

More importantly, at this location, the temperature gradient is typically large, so that the error on the photospheric temperature is even larger, with a concomitant large error on the luminosity on a per-ray basis (Figure~\ref{fig:profile:2md-single-ray}). 
As the common envelope gas expands, the outer layer’s density decreases and the gas may become optically thin (at least before dust forms). This means that the photosphere recedes into the gas distribution, the local smoothing length decreases, which increases the precision with which we can determine the photosphere's location and temperature. 

To first approximation, we can think of the precision with which we can estimate the location and properties of the photosphere by estimating the number of particles, $N_\mathrm{res}$, used for the luminosity integration in Equation~\ref{eq:calc:luminosity}. This can be determined by considering each ray. Instead of integrating across all rays to determine the contribution to $L$ that each particle $j$ makes  (as we did in Equation~\ref{eq:calc:luminosity}),
we first separate the contribution to the luminosity from each particle $j$ and ray $i$ ($L_{ij} \equiv 4 \pi S_{j} A_{\mathrm{eff}, ij}$). We then select the particle that contributes the most to the luminosity of that pixel. Finally, we calculate the ratio of that particle's contribution to the total brightness of that pixel. In other words, we find the fraction of the highest contribution to the total effective pixel area $A_{\mathrm{eff}, i} \equiv \sum_j{A_{\mathrm{eff}, ij}}$ for a given ray $i$
\begin{equation}
    f_{\mathrm{contr}, i} \equiv \frac{\max_j{A_{\mathrm{eff}, ij}}}{\sum_j{A_{\mathrm{eff}, ij}}} 
    ,
    \label{eq:def:F_contr}
\end{equation}
where max$_j$ is the maximum area $A_{{\rm eff},ij}$ contribution among all particles, $j$, for a given ray, $i$. This shows the extent to which one particle dominates the integration of a particular ray.
The lower $f_{\mathrm{contr}, i}$, the higher the resolution, because it denotes that information for that ray was obtained from a large number of particles.

\begin{figure}
    \centering
    \includegraphics[width=\linewidth]{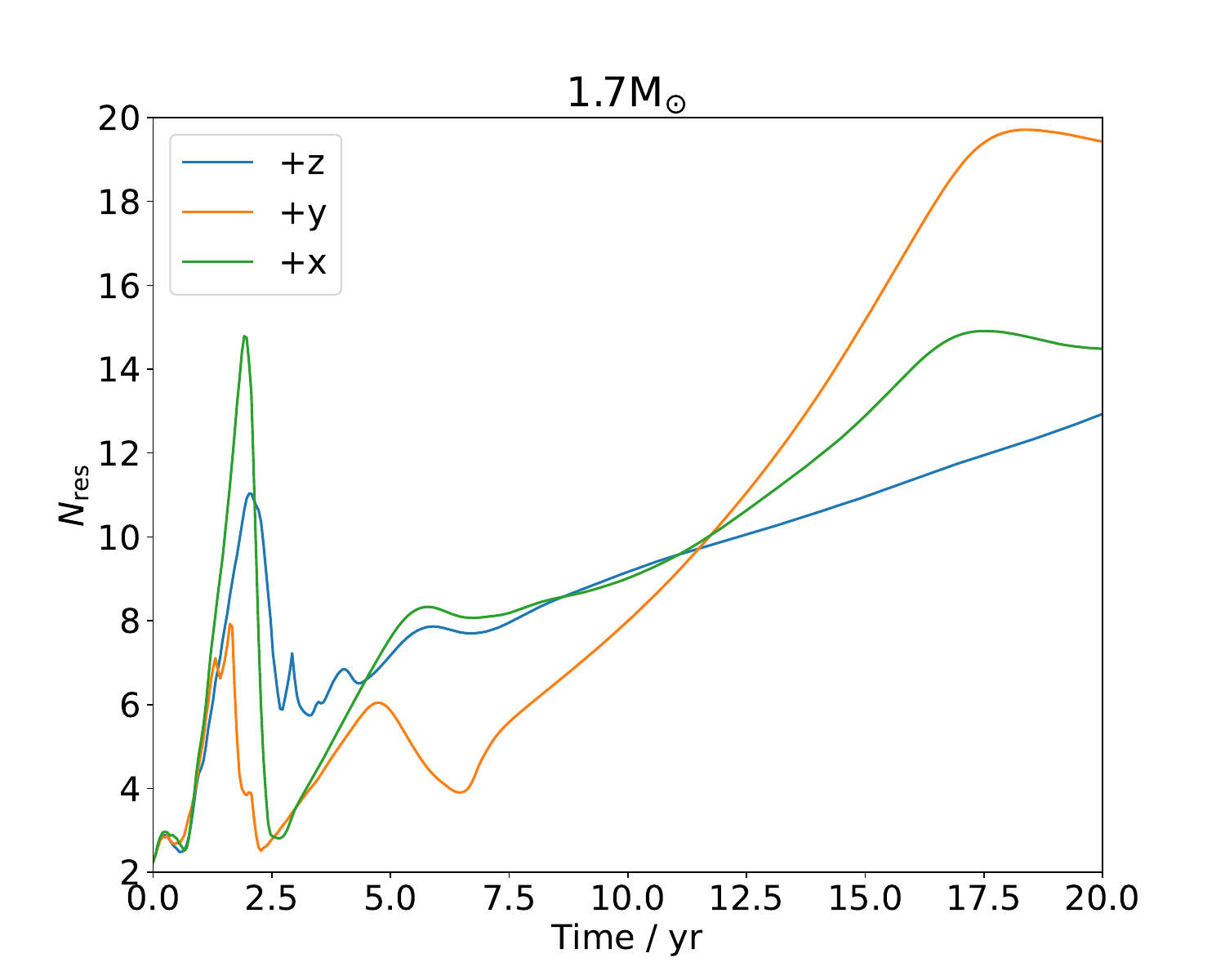}
    \includegraphics[width=\linewidth]{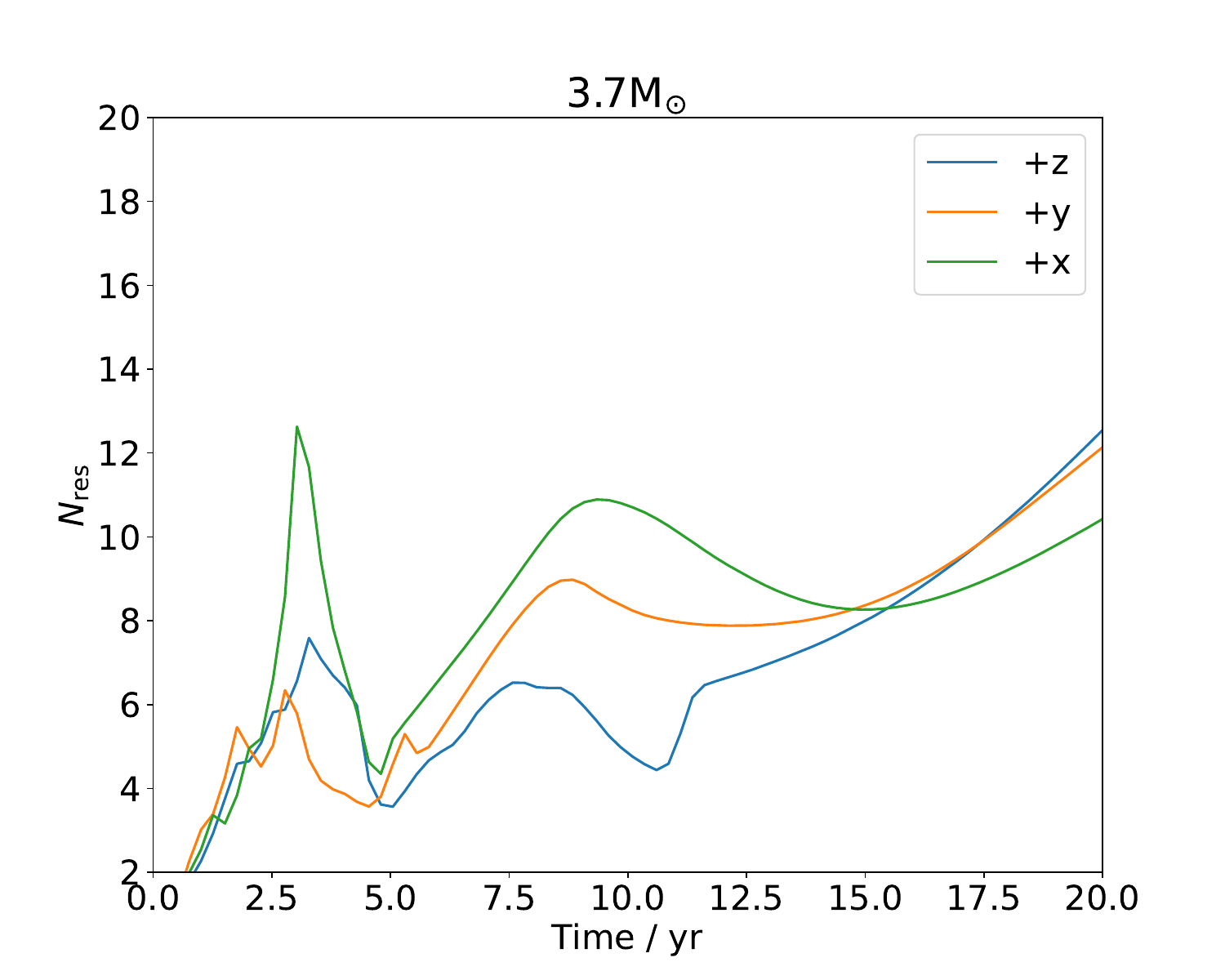}
    \caption{
    An estimate of the precision of the luminosity calculation, based
    on the number of particles, $N_\mathrm{res}$, used in the calculation of the specific intensity for each ray, for the 1.7~\Msun{} simulation (the 3.7~\Msun{} simulation is similar).
    Blue, orange, and green lines are averaged over a grid of $256 \times 256$ rays observed from the three three directions.
    }
    \label{fig:lc:Nres}
\end{figure}

The inverse of this fraction ($N_{\mathrm{res}, i}
    \equiv f_{\mathrm{contr}, i}^{-1}$) provides an estimate of the number of particles used to resolve the specific intensity per pixel $i$.
Averaging over pixels, using their luminosity contribution, $L_i$, as weights, we can estimate the \gls{SPH} resolution of our calculated luminosity, $L$, in terms of the number of particles per ray
\begin{equation}
    N_\mathrm{res}
    \equiv \frac{
    \sum_i{N_{\mathrm{res}, i} L_i}
    }{
    \sum_i{ L_i}
    },
\end{equation}
where $L_i \equiv 4 \pi \sum_j S_j A_{\mathrm{eff}, ij}$.
The quantity $f_{\mathrm{contr}, i}$ is plotted in Appendix~\ref{app:sph-resolution}, similarly to how the surface brightness was plotted in Figure~\ref{fig:lum-image:2md}, for easy comparison, while in Figure~\ref{fig:lc:Nres} we plot $N_{\rm res}$ as a function of time.
Note that by definition $N_\mathrm{res}$ cannot be lower than unity.
The larger the number, the better the resolution, the higher the confidence we have on the luminosity output.
In Figure~\ref{fig:lc:Nres}, for both dusty simulations, the resolution begins at 1 to 2, indicating an extremely low resolution where one or two particles dominate the result for each ray;
however, after 2 to 4 years, $N_\mathrm{res} \sim 5$ growing to a value of $\sim 20$,
meaning that information for the specific intensity of each ray is sourced from multiple particles and thus modelled more accurately.

Our resolution criterion, $N_\mathrm{res}$, is similar to the \gls{SPH} ray tracing relative accuracy metric, $\eta$, defined in equation 27 of \citet{Hatfull2021a}, but distinct from that work in that ours measures the resolution by the actual contribution to the luminosity by each particle for each ray, instead of by how much the number of neighbouring particles is lower than the average. Our method has the advantage that it summarises the result into a single number for each ray, instead of for each point on the ray; our method also has the advantage that it highlights the influence on the particles while ignoring those particles with very low optical depth and temperature (which don't matter to the final luminosity, even if they have a very low number of neighbours).
Our criterion is also similar to that of \citet{Price2024a}, where pixels with particles of optical depth $d\tau > 1/3$ present before $\tau$ reaches their photospheric value of 1 are flagged as bad pixels;
once again, our method provides a continuous assessment of photospheric resolution instead of a boolean judgement, and it is weighted by the particle's specific intensity, which matters directly to the luminosity error estimation.


An alternative way to quantify the surface uncertainty, which also allows us to calculate \updated{corresponding error margins} on the value of the calculated luminosity \updated{(noting that this is not the only source of error)}, uses the inherent spherical symmetry of each \gls{SPH} particle's smoothing kernel, which, when the SPH particle is located at the photospheric surface, misrepresents the steepness of the gradients in all quantities (e.g., density, pressure, temperature).

When the information on the light comes from one or a few optically thick surface particles, the kernel interpolation that assumes Gaussian profiles, overestimates the source function on the observer side of the surface layer, because the smoothing length of surface particles extends $h$ in all directions.  However, at these surface locations we have a {\it prior}, in that we know that all quantities have a sharp negative gradient. 

To quantify this problem, we start by treating the current lightcurve as an \updated{approximate} upper limit of the true lightcurve, considering that the \gls{SPH} kernel interpolation at the first locations on each ray where quantities are non-zero would overestimate the specific intensity contribution. This is because the surface particles are statistically distributed at the surface. However, each ray encounters the outer half of each particle first, starting the integration of the source function systematically $2h$ towards the observer side of the SPH particle. This means that each ray picks up an extra contribution to the source function, making the luminosity too large.

We then locate the point inside the star below which the resolution is high enough to make the calculation of the source function and optical depth accurate, and outside of which the resolution is considered too low. This depth is defined as follows: a ray passing this point from the outside, has counted 14.5 contributing particles to the lightcurve calculations (this number is a quarter of 57.9, the average number of neighbours in our \phant{} simulations). A detailed explanation of how we select the boundary depth is provided in Appendix~\ref{app:upper-lower-bounds}.

To determine \updated{an approximate} lower limit to the luminosity, we assume all the particles' above the point do not contribute any specific intensity to the ray (i.e., we set their temperature to 0~K), while their optical depth, $\tau$, is left unchanged. In this way, we simulate the opposite extreme with the source function at zero, while the $\kappa \rho$ has the normal behaviour.
The results are the shaded areas shown in Figure~\ref{fig:lc:both}.

\updated{The upper and lower limits calculated thus, do not account for other sources of error.} A further source of systematic uncertainty resides in the imperfect reproduction of the star in 3D. The initial surface temperature and density profile at $t=0$ after relaxation in \phant{} is different from the prescription of \mesa{}. This has little to no effect on the hydrodynamics, but plays a role in the lightcurve calculation. Together with the low surface resolution, this contributes to the initial lightcurve unreliability. In Appendix~\ref{app:uncertainties:initial-years} we carry out several tests that quantify the error in the lightcurve, demonstrating that it reduces greatly after time zero, even if, interestingly, and only for the 1.7~\msun\ model, we see a change of 3-6 months in the $\sim$3-yr width of the luminosity peak.

\subsection{The lightcurve between 20 and 45 years and the adiabatic approximation}
\label{sec:uncertainties:cooling}

These simulations are adiabatic. As such, the luminosity towards the end of the simulation plateaus at $1-2\times 10^5$~\Lsun{}, a very high value that is at odds with observations. Typical extreme carbon stars' luminosities are $\sim$10\,000~\Lsun{} \citep[e.g.,][]{Marini2021}, although some extreme cases are found with luminosities as high as $\sim 80\,000$~\Lsun\ \citep[for MSX SMC 055;][noting this is not a carbon-rich case]{Groenewegen2018}. 

An object radiating 100\,000~\Lsun\ over 30 years (the approximate length of the plateau) radiates $\sim 4\times 10^{47}$~ergs. This value is an order of magnitude larger than any of the typical energies  describing the star and the interaction: the envelope's gravitational potential energy, $\sim -6\times 10^{46}$~ergs, is overcome by the delivered orbital energy, $GM_c M_2 / 2 a_f - GM_1 M_2 / 2 a_i \approx 10^{46}$~ergs, and some of the envelope's internal energy (particularly the recombination energy), $\approx 6 \times 10^{46}$~erg. We also note that the undisturbed star, with 5180~\Lsun\ would radiate $\approx 2 \times 10^{46}$~erg in 30 years. 

The point is that in Nature only $1-2\times10^{46}$ ergs would be available to radiate over that time, lower than implied by the deduced luminosity during the late plateau. If the surface were allowed to radiate in our simulations, the likely outcome would be a smaller, cooler, less luminous surface. This discrepancy is most obvious at late times. At earlier times (within the inspiral timescales of 5-10 years) the energetics are likely dominated by the orbital energy injection and subsequent delivery of recombination energy at depth powering an adiabatic expansion.

\subsection{Comparison with observations}
\label{ssec:observations}
\begin{figure*}
    \centering
    \includegraphics[width=0.49\linewidth]{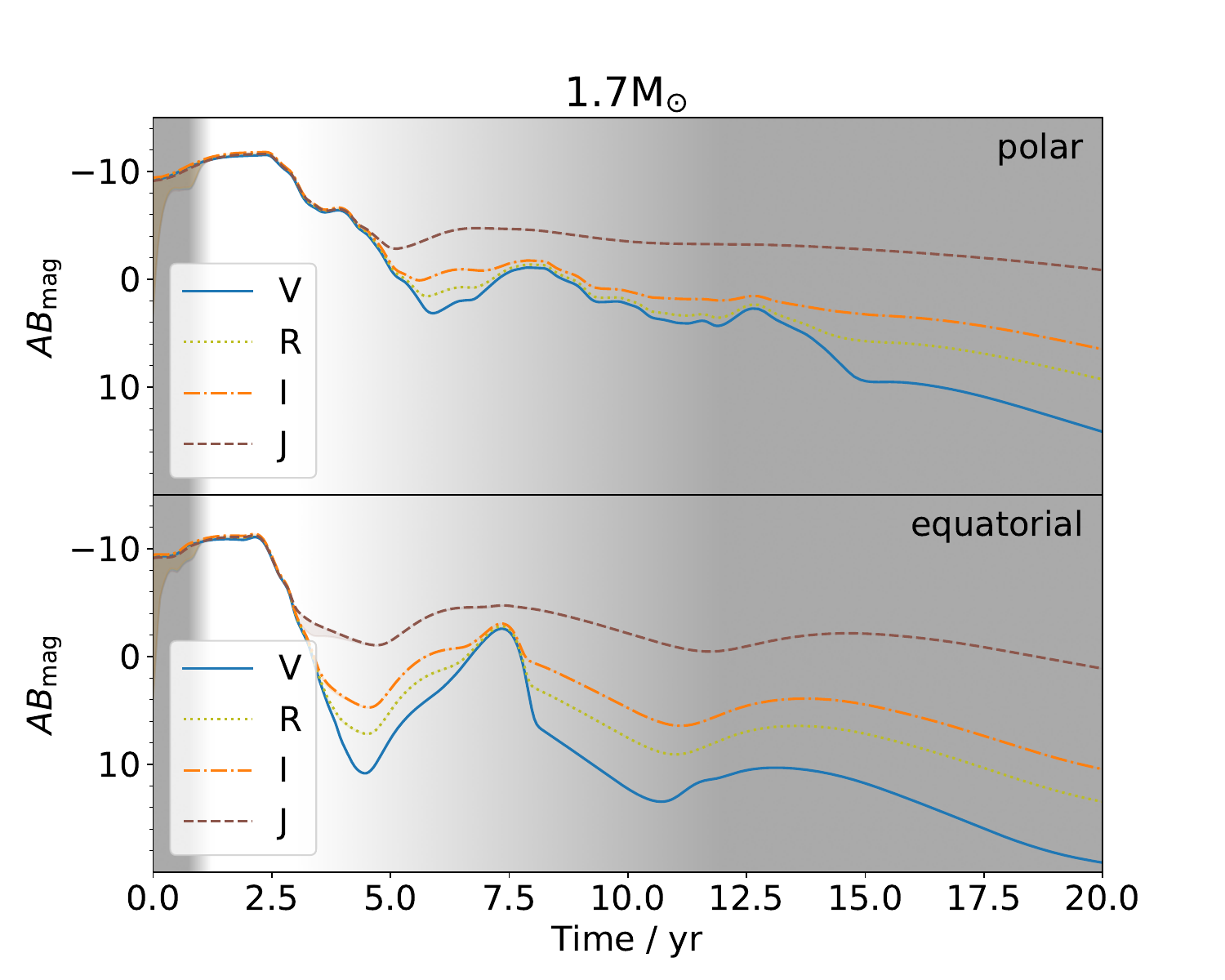}
    \includegraphics[width=0.49\linewidth]{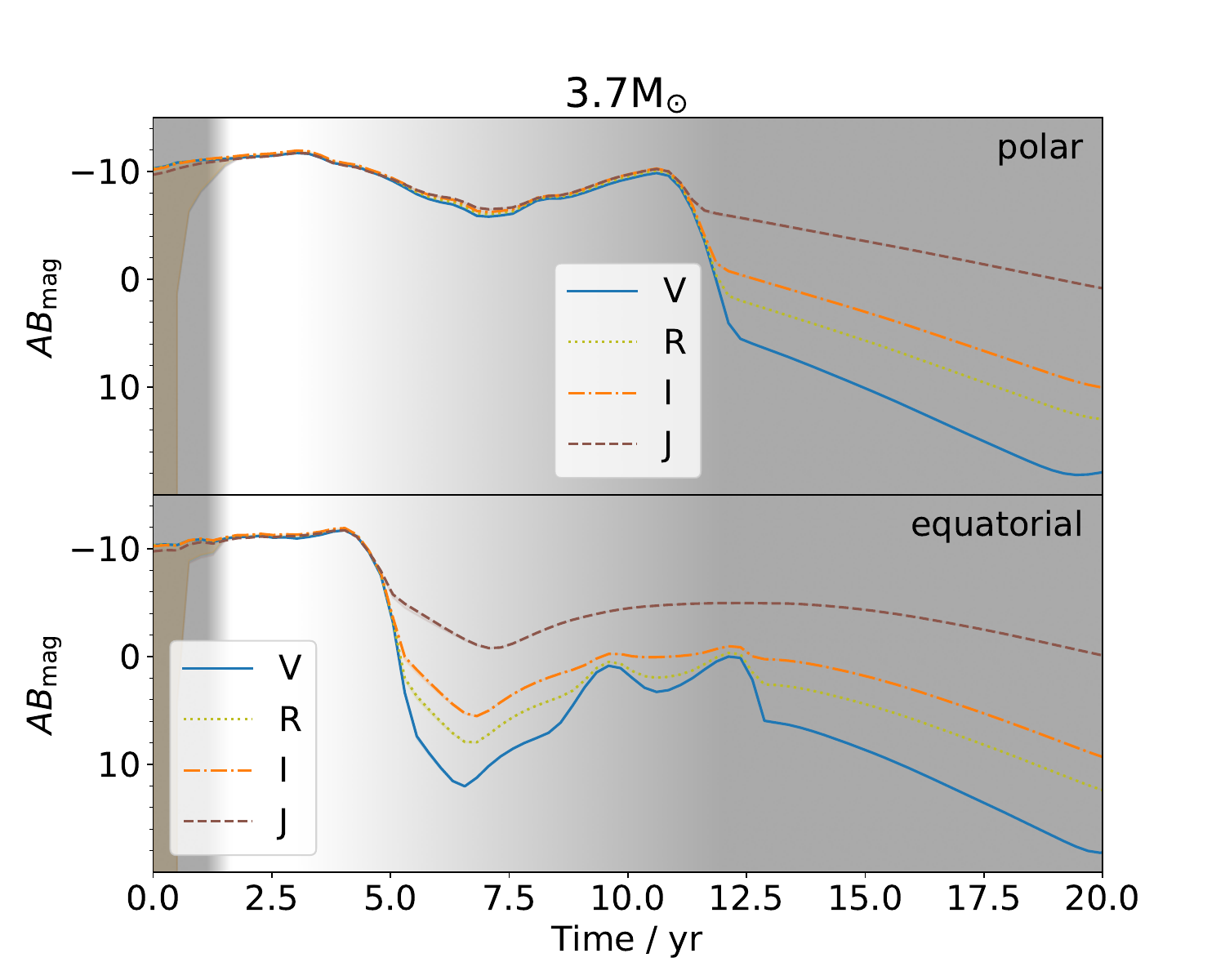}
    \caption{
    \updated{Approximate} absolute AB magnitude in Johnson V (solid), Cousins R (dotted), Cousins I (dash-dotted) and J bands (dashed) for the dusty 1.7~\Msun{} (left panels) and 3.7~\Msun{} (right panels) simulations \updated{using grey opacity}, as viewed from polar ($+z$, top panels) and equatorial ($+x$, bottom panels) directions. The shaded area are \updated{partial uncertainties stemming {\it only} from the low surface resolution}, the same as in Figure~\ref{fig:lc:both}---See Section~\ref{sec:uncertainties:SPHres} for more details. \updated{We stress however, that there are several other systematics that will need to be eliminated before the simulations are ready for a comparison with the observations.}
    }
    \label{fig:lc:ABmag}
\end{figure*}

\begin{table*}
    \centering
    \begin{tabular}{lcccc}
    \hline
    &Observations   & Observations & Our 1.7~\msun\  & Our 3.7~\msun\\
    & OGLE-2002  & AT~2025abao & simulation & simulation\\

    \hline
        &\multicolumn{4}{c}{Progenitor}\\
    \hline
    ZAMS mass (\msun)  &$<4$& $7\pm2$& 2 (1.7@AGB)& 4 (3.7@AGB) \\
    Radius  (\Rsun)   &30& 350&260&330\\ 
    Luminosity (\Lsun) &290 &10\,000&5180
    & 11\,900 \\
    \updated{$T_{\rm colour}$} (K) &4350 &3100&3227& 3317\\
    $M_2$ (\Msun) &Unknown&Unknown&0.6 ($q=0.35$)&0.6 ($q=0.16$)\\
    \hline
    &\multicolumn{4}{c}{At peak}\\
    \hline
    time (yr) &1.6$^a$&0.099$^b$&1-2.5$^c$ & 2.5-6$^d$ \\
    $R_{\rm phot}$ (au) &1.1$^e$& 4.1&$5-6$& $6-3$ \\
    $L$ (\Lsun) &8\,100$^f$ &$7.2\times10^5$&$2-4\times 10^6$& $2-4\times 10^6$\\
    \updated{$T_{\rm colour}$} (K) &3600 &5740&6000-7000& 6000-7000\\
    $v_{\rm exp}$ (km~s$^{-1}$) &--&450&86& 85 \\
    \hline
    &\multicolumn{4}{c}{At 8 years post outburst start} \\
    \hline
    $R_{\rm phot}$ (\updated{au}) &50 &--&50 (dust)& 2 (stellar) 50 (dust)\\
    $L$ (\Lsun) &8\,200 &--& $\sim 1\times10^5$& $30\,000-2\times10^5$\\
    $T_{\rm phot}$ (K) &550 &--&800-2000& 500-6000$^g$\\
    $v_{\rm exp}$ (km~s$^{-1}$) & 35 &--& 60& 68 \\
    $M_{\rm dust}$ &$>1.4\times 10^{-5}$ &--& $1\times10^{-4}$ & $1\times10^{-4}$\\
    \hline
    \multicolumn{5}{l}{$^a$Time of the second peak in April 2004 (MJD~53100), after outburst start on MJD~52500.}\\
    \multicolumn{5}{l}{$^b$36 days between the start of the rise selected by us to be at MJD~60950 to $g$ band maximum at }\\ \multicolumn{5}{l}{MJD~60986.3 \citep{Reguitti2026}.}\\
    \multicolumn{5}{l}{$^c$Peak time for both viewports.}\\
    \multicolumn{5}{l}{$^d$Peak time for equatorial viewport. First peak for polar viewport is $\sim$3yrs.}\\
    \multicolumn{5}{l}{$^e$Measured in October 2003 instead than April 2004 - see table 1 in \citet{Tylenda2013}. }\\
    \multicolumn{5}{l}{Both models' photospheric radii are by then dominated by dust, while in the observations the}\\
    \multicolumn{5}{l}{measured photosphere is that of the gas, where a larger, 23~au, 800~K dust shell is present but optically thin. }\\
    \multicolumn{5}{l}{$^f$Using the distance of 8.2~kpc of \citet{Tylenda2013}. Luminosity scales with the square of the distance,}\\ 
    \multicolumn{5}{l}{while radii scale as the distance. Note that \citet{Steinmetz2025} use a distance of 4.09~kpc.}\\
    \multicolumn{5}{l}{$^g$Viewing angle-dependant, with polar being hotter.}\\
    \end{tabular}
    \caption{\updated{Some observationally-derived quantities} of red transient OGLE-2002 (presumed to have been a RGB star; \citealt{Tylenda2013}) and AT 2025 abao (presumed to have been an AGB star; \citealt{Reguitti2026b}), \updated{listed alongside approximately equivalent data extracted from our simulations. While a meaningful comparison cannot be carried out at this time, after the sources of random and systematic uncertainty identified in this work have been reduced, a comparison like this one will become meaningful.}}
    \label{tab:observations}
\end{table*}

A class of astrophysical transients called LRNe have been identified as mergers of non-compact stars \citep[see ][for an observational review]{KaminskiBlagorodnova2026arXiv}. Typical lightcurves of such transients exhibit a peak and plateau or a multi-peak behaviour with a subsequent dimming, reddening, and quick dust production \citep{Kaminsi2010AA,Tylenda2016,Blagorodnova2020,Steinmetz2025,Karambelkar2026}. The transients identified in our Galaxy generally corresponded to lower mass donors, while extragalactic, brighter events likely correspond to more massive objects, as shown by the correlation between progenitor mass and peak brightness \citep[e.g.,][]{Kochanek2014,Blagorodnova2021}. Archival observations of pre-outburst progenitor systems also revealed that the majority of donor stars are subgiants, which are stars on the very early post-main-sequence, experiencing a fast, thermal-timescale, expansion phase. An often cited example is V~1309~Sco, for which observations were obtained during the entire pre-outburst and outburst phases such that this merger can reasonably be inferred to have been that of a low mass, 1.5~\msun, 4~\Rsun{} star with a 0.15~\msun\ companion \citep{Tylenda2011}. This object, along with several others such as V838~Mon (a more massive merger; \citealt{Bond2003}), took place on faster timescales compared to those with RGB/AGB progenitors, because of the shorter dynamical times of the more compact, dense envelopes. 

Among the observed red nova sample, only two examples are compatible with {\it bona fide} giant star progenitors, as opposed to subgiant or main-sequence progenitors. The red nova OGLE-2002-BLG-360 \citep[hereafter OGLE-2002;][]{Tylenda2013} was characterised by an outburst that took place over years, as opposed to days and weeks. It brightened by a factor of 30 (3.7 mag) in the $I$ band to a first peak, and then by another $\sim$1 mag by the third peak, for a total rise time of 2.2~yrs (where we take the beginning of the rise time to be on MJD~52500, corresponding to 14 August 2002; \citealt{Tylenda2013}). The object then steadily reddened and dimmed by $\sim$10 mag over the following 5 years, \updated{with the dust observed around this object most likely having an oxygen-based origin \citep{Steinmetz2025}.}

A second object, WNTR23bzdiq/WTP19aalzlk/AT~2025abao discovered in M~31, was initially presumed to represent a full common envelope ejection from a more massive AGB star \citep[a $350\pm50$~\Rsun{}, early AGB star, descending from a $7\pm2$~\Msun{}, main sequence progenitor;][]{Karambelkar2025}. The lightcurve activity was initially interpreted as a long and shallow outburst. However, in October 2025 the object experienced another outburst, showing a quick brightening over about a month of $\sim$5\,mag in the optical, to a peak bolometric luminosity of 870\,000~\Lsun{}\footnote{The $g$ band magnitude maximum quoted elsewhere corresponds to 720\,000~\Lsun, but the bolometric luminosity maximum took place 67~days later.}, analogous to other \gls{LRN}e \citep{Mikolajczyk2026,Reguitti2026b}. Hence, the initially observed slow brightening was, in fact, just pre-outburst activity preceding a red nova \citep{KaminskiBlagorodnova2026arXiv}. \citet{Reguitti2026b} analysed the lightcurve for 120 days past peak, revealing a cooling photosphere (from 5740~K at peak to 3600~K, 120 days later), a declining luminosity (down to 440~\Lsun\ at 120 days past peak) and an expanding photosphere (10~au at 70 days post peak, but reducing to 8.3~au by 120 days). 
\updated{The dust observed around this object is oxygen-based dust and likely has an AGB wind, rather than outburst, origin (Karambelkar, private communication).}

 In Table~\ref{tab:observations} we compare several parameters that were extracted from the observations of OGLE-2002 and AT~2025abao with equivalent parameters of our models, {\it noting that the models were not designed to fit these objects}, but these objects are, \updated{from an evolutionary point of view, the most similar to the models we have calculated and the comparison can therefore be very instructive even if at this time we are not looking for a match}.  Our two models have primary star (progenitor) masses that are in the ball park of those deduced for OGLE-2002 and AT~2025abao, but are more similar to the progenitor of AT~2025abao than the OGLE-2002's progenitor, which is likely a smaller RGB star. 
 
 In Figure~\ref{fig:lc:ABmag} we present \updated{an approximation of} the apparent magnitudes in the Johnson-Cousins $V$, $I$ \citep{Bessell1990a} and 2MASS $J$ bands \citep[][assuming a 1~kpc distance to the observer and no reddening]{Skrutskie2006} obtained by convolving \updated{grey-opacity-based-}$L_\lambda$ with filters' sensitivity curves. Spectra are computed using the {\sc pyphot} package~\citep{zenodopyphot}, using the filter table of {\sc ground\_johnson\_V} for the $V$ band, {\sc ground\_cousins\_I} for the $I$ band and {\sc 2mass\_J} for the $J$ band. We elect not to overlay the observations on these curves because the actual similarities are misleading given the misgivings we still have with our predictions. However, we notice that, particularly for AT~2025abao, which is the more similar progenitor to our more massive simulation, the peak $J$ band brightness of 12.6~mag, translates to an absolute $J$ band brightness of $M_J = -11.9$ (at a distance of 0.785~Mpc, $A_V=0.17$ adopted by \citet{Reguitti2026b} and $A_J/A_V=0.282$ of \citet{Cardelli1989}), a value that is indeed very similar to our simulation over the first few years. \updated{The oxygen-rich nature of this star's progenitor dust does not impact the comparison with our carbon-rich-dust model: the model has not yet nucleated sufficient dust to impact the light, and the characteristics are those of its gas photosphere, while the native dust in the observation is accounted for by its reddening.}
 
 The time of peak \updated{(which is not affected by dust formation)} in Table~\ref{tab:observations} is arbitrarily selected to correspond to the second of three peaks (at MJD~53\,100) in the lightcurve of \citet{Tylenda2013}, and it is with reference to the start of the outburst for the OGLE-2002 observation, which is selected by us as the time when the $I$ band brightness starts to increase (MJD~52\,500 in the lightcurve of \citet{Tylenda2013}). For AT~2025abao, we have used as reference the base of the rise at MJD60950 and $g$ band peak at MJD60986 - a short 36 days later. OGLE-2002 took a year and a half to rise to peak, while AT~2025abao had a rise that was quite a lot faster, only a month. For the models, time zero is defined as the time of the start of the simulations, and the timing of the luminosity peak is defined as the time elapsed form this moment until reaching maximum brightness. We note that while for the 1.7~\msun\ model this is quite well defined, and similar for all viewports, for the 3.7~\msun\ model it is more dependent on the viewport (the range is 2.5$-$5 yrs). This said we have little information on the actual rise as the error bar due to low surface resolution gives us little knowledge over the first year.
 
 At peak, the stellar/gas photospheric radius is of the order of 1~au for OGLE-2002, approximately 4~au for AT~2025abao, while for the simulations it is approximately 5~au, where we have used the "observer's" radius for the comparison (Section~\ref{sec:the_evolutuion_of_the_photospheric_radius}), and we note that at the selected time of peak, the dusty shell in the models has already formed, but it is still optically thin. 
 The OGLE-2002 luminosity peaks at a value of 8100~\Lsun{}, while for AT~2025abao it is much larger, 720\,000~\Lsun{}, and more similar to the simulations, with values ranging between 1 and $2\times 10^6$~\Lsun{}. Another LRN, M31-2015-LRN had a luminosity peak of 309\,000~\Lsun{} (Table~\ref{tab:LRNs}) and a progenitor mass of 3-5~\msun\ not dissimilar from our more massive model but likely more compact. \updated{We note if the rise to peak is powered by the dynamical inspiral, the gasseous photosphere radius is unlikely to be impacted by the type of dust present in the giant's envelope, because  the main source of expansion energy is the orbital energy of the companion. Hence the mismatch in dust type between the observations and simulations is not relevant.}

The next comparison is made approximately 8 years after the start of the outburst (or the start of the simulations), but only for OGLE-2002. At this time the photospheric radii of observation and 1.7~\msun\ simulation are dominated by dust and are similar (50~au). The 3.7~\msun\ model has a still dominating gas photosphere at 2~au, while a dust shell is already present with a radius of 50~au. Similarly, the photospheric (dust) temperatures are similar in the observation (550~K) and 1.7~\msun\ simulation ($\sim 800$~K for the emerging dust shell with an underlying $\sim 2000$~K gas photosphere), while for the 3.7~\msun\ simulation the dominance of dust or gas photosphere depends on the viewport: it is similar to the other model for the $+x$ and $+y$ views, but for the $+z$ view, the SED is still dominated at this time by the star ($T_{\rm phot}\sim 6000$~K), and the dust only fully obscures it at $\sim$12~yrs; Figure~\ref{fig:Teff}). The mass of the dust measured by \citet{Tylenda2013} is a lower limit ($>1.4\times10^{-5}$~\msun) due to its optically thick nature. In the simulations, the dust mass that has formed by 8 years is still relatively low but about to increase: it will reach the level of the measurement of \citet{Steinmetz2025} at $\sim$20~yrs after the start of the interaction \citep[][figure~11]{Bermudez2024a}. \updated{Here the dust type and the mismatch between the observations (oxygen-based) and simulations (carbon-based) may have an impact. The former is less opaque, possibly resulting in a different growth of the dusty photosphere and may result in a different dust mass  evolution.}

In Table~\ref{tab:LRNs} we list observationally-derived parameters for several additional red novae remnants that should be specifically pursued with simulations, such as M31-2015-LRN, which is similar to OGLE-2002, with a progenitor mass and radius of 3--5.5~\msun\ and 30--40~\Rsun{}, respectively \citep{MacLeod2017} and a typical optical lightcurve evolution over $\lesssim 1$~yr \citep{Williams2015,Blagorodnova2020}. Observationally-derived quantities such as these need to be correctly interpreted. During the optical outburst, when the visible optical depth is low, the observed SED reveals the outburst's photosphere, which then leads to a measurement of its photospheric radius, colour temperature, and bolometric luminosity. Several years after the optical peak, dust formation progressively obscures the central remnant, reprocessing the stellar luminosity into longer wavelengths. Hence, at late times, the dust radius, composition, density profile, and temperature define the photosphere, and the parameters of the merged remnant star can only be derived indirectly, via radiative transfer modelling of the dust. 

\updated{Finally, we should consider that our near and mid IR surveys have likely already observed populations of stars that are the progeny of the LRN, observed hundreds to thousands of years after outbursts. An example are the extremely red objects already mentioned in Paper~I \citep{Gruendl2008} and which have been hypothesized to derive from CE interactions \citep{DellAgli2021}. In order to carry out meaningful comparisons between these non-transient stars and our simulations predictions of observationally-derived quantities such as $\tau_{10}$, the optical depth at 10 micron, at a relevant late time are needed.}

\begin{table*}
    \centering
    \begin{tabular}{lcccccccccccl}
    \hline
               &Prog. &	Evol. &	Time$^a$	& $R_\star$&	$L_\star$ & \updated{$T_{\rm colour}^b$} &	$v_{\rm exp}$ 	&$R_{\rm dust}$ 	& $T_{\rm dust}$ &	$M_{\rm dust}$ 	&$\tau_{\rm V,dust}$&	Ref.\\
               &mass &stage& &&&&&&&&&\\
               & (\msun) &&(yr)&(au)&(\Lsun)&(K)&(km~s$^{-1}$)&(au)&(K)&(\Msun)&&\\
    \hline           
    OGLE-2002-&	$<$4	&RGB	&1$^c$&	1.1&	8100	&3600&--&	20	&	800$^d$	&5.5$\times 10^{-7}$&	2.5	&(1)\\
        -BLG-360      &&&8$^c$&	1.4	&	8200	&3200	&35&50&	550	&3.44$\times 10^{-3}$&	$>20$	&(1,2)\\
&&&	15-21$^c$	&0.70	&	1646$^e$&	3000&	226&1000$^f$&	50-200&	1.2$\times 10^{-2}$&	--&	(2)\\ \hline
    M31-2015-&	3--5& 	post-MS	& 0.12	& --&		3.09$\times 10^5$&	3390&	100	&23&1760	&1$\times 10^{-5}$&	0.7&	(3)\\
      -LRN          &	&	&	2.5	&--&	6\,900	&9780	&100	&38	&1110	&1$\times 10^{-2}$	&287&	(3)\\
                &&	&	10.0	& --&		4620	&6000	&100	&180 &385&	7.7$\times 10^{-4}$	&100&	(4)\\ \hline
AT~2025abao & $6-9$ & AGB & 0 & 4.1 & $7.2\times10^5$ & 5740 & -- & -- & -- & -- & -- & (5) \\ 
 &  &  & 0.19 & 10 & 1100 & 4000 & -- & -- & -- & -- & -- & (5) \\ 
 &  &  & 0.33 & 8.3 & 440 & 3600 & -- & -- & -- & -- & -- & (5) \\    \hline  
       AT~2018bwo&	12--16&	post-MS&	0	&25.8	&3.41$\times 10^6$	&3330	&220 &--		&--	& $<$3.1$\times 10^{-3}$	&--& (6,7)	\\
    &	&	&	6.1	&--	&1.48$\times 10^5$	&2600	&100	&140	&670	&2.1$\times 10^{-5}$	&--& (4)	\\ \hline
    AT~2021blu	&13--18	&post-MS	& 0 &	--	& 	1.698$\times 10^7$	&8800&	500 &--&	-- &	--&	22&	(6)\\
    	&	&	&3.1&	--	&	3.30$\times 10^5$	&2200&	100&	87&1100&	4.2$\times 10^{-5}$&	22&	(4)\\ \hline
    AT~2021biy&	17--24&	post-MS	&0&	7.34	&4.180$\times 10^7$	&	11\,450&	640&--&	--	&--&	--&(8)	\\
    
    &	&		&3.3&	--	&	4.20$\times 10^5$&	2400&	250&240&	700	&3.6$\times 10^{-5}$&	22& (4)	\\ \hline
    AT~2011kp&	20--30	&post-MS	& 0&	--&		2.612$\times 10^7$	& 16\,000	&450	&--&--	&--&	--&	(9)\\
    &		&	&12.5&	2.55&		41\,000	&3500	&--	&291&450	&1.8$\times 10^{-3}$&	133&	(10)\\ \hline
    AT~1997bs&	20$-$60	 & post-MS&	0&	--&		8.62$\times 10^6$&	--	& 585&--&	--&	--&	--& (9,10,11)\\
    &		& &	7&	3.8&		1.52$\times 10^5$&	4000	&--&301&	600&	1$\times10^{-3}$&	69&(10)\\ 
    \hline     
    \multicolumn{13}{l}{(1) \citealt{Tylenda2013}; (2) \citealt{Steinmetz2025}; (3) \citealt{Blagorodnova2020}; (4) \citealt{Karambelkar2026}; (5) \updated{\citealt{Reguitti2026b}};}\\ 
    \multicolumn{13}{l}{(6) \citealt{Pastorello2023}; (7) \citealt{Blagorodnova2021}; (8) \citealt{Cai2022AA};  (9) \citealt{Pastorello2019b}; (10) \citealt{Reguitti2026}; (11) \citealt{VanDyk2000PASP}. }\\
    \multicolumn{13}{l}{$^a$Post peak unless stated otherwise;}\\ 
    \multicolumn{13}{l}{$^b$Observational papers may refer to this quantity as effective temperature; here we refer to it as colour temperature, $T_{\rm colour}$, because the quantity is derived}\\
    \multicolumn{13}{l}{from the \gls{SED} as done by us for the simulations.}\\
    \multicolumn{13}{l}{$^c$ Time post outburst's start on MJD~52\,500 or August 14, 2002.}\\
    \multicolumn{13}{l}{$^d$This is assumed, but the value at 3-4~yrs after outburst start was 850~K.}\\
    \multicolumn{13}{l}{$^e$This luminosity is for a distance of 4.09~kpc. For the distance assumed by \citet{Tylenda2013}, 8.2~kpc, the luminosity would be 6600~\Lsun; }\\
    \multicolumn{13}{l}{$^f$Outer radius of the inner shell; outer shell radius is 9500~au. }
    
    \end{tabular}
    \caption{Observationally-derived parameters of red novae and luminous red novae (more massive stars) that can be compared with suitable simulations are presented.}
    \label{tab:LRNs}
\end{table*}

\section{Conclusions}
\label{sec:discussion}

We have calculated the lightcurves over 44 years for two simulations representing 1.7~\msun\ and 3.7~\msun\ AGB stars undergoing common envelope evolution with a 0.6~\msun\ companion, with dust nucleation. The lighturves are obtained in post-processing, and the simulations are adiabatic. Despite a series of shortcomings, we can draw a number of (primarily qualitative) conclusions.

\begin{enumerate}
\item The \updated{lightcurves} for both models (Figure~\ref{fig:lc:both}) have the following characteristics.
(a) The light rise \updated{takes place over 1-5 years (depending on model and viewport) from the start of the simulations, i.e., the start of Roche lobe overflow and} is driven by the expansion of the photosphere.
(b) The effective temperature plays a more minor role: besides a cooling tendency due to adiabatic expansion, fluctuations are due to optical depth effects and, on balance, the temperature remains in the range 6000-7000~K during the light rise (Figure~\ref{fig:Teff}).
(c) The peak of the lightcurve is at about $2-5\times 10^6$~\Lsun, depending on model and viewing angle.
(d) The broader peak for the more massive model is likely due to its lower mass ratio ($q=0.16$ vs. 0.35), resulting in a weaker interaction leading to a delayed dust nucleation.
(e) The bolometric luminosity peak subsides to a trough in 2--4~years (depending on model and viewport) as dust formation generates an optically thick, expanded, and cool photosphere. Subsequent minor variations—peaks and troughs—result from the complex, asymmetrical interplay between gas and dust layers, which dynamically exposes or hides deeper surface layers.
(f) After a trough, which has a temperature of $\sim 6000$~K, the luminosity settles on a value of $100 \, 000 - 200\,000$~\Lsun\ and a temperature that gradually decreases to $\sim 400$~K at 20~yrs. The luminosity is too high on account of the lack of cooling in our simulations. 

\item \updated{For context, the uncertain CE pre-inspiral time is approximately 1 year long and likely too short (driven by the stellar surface expansion due to stellar structure instability, and to the abrupt insertion of the companion in the computational domain - \citealt{GonzalezBolivar2022,Roepke2023}). On the other hand, the inspiral itself takes place over $\sim$6 years, a timescale generally considered reliable \citep{Roepke2023}, typically ending near the trough of the lightcurve. This contextualises the timing of lightcurve rise and fall: the initial rise (0--1 year) is likely too fast, as it keeps pace with a likely too fast early expansion. However, what follows, the final stage of the rise to the peak luminosity, the timing of the dust formation, and the subsequent light decline, are the most accurate outcomes of these simulations.}

\item The primary source of uncertainty stems from insufficient spatial resolution at the photosphere \updated{(quantified by the error bar shading in Figure~\ref{fig:lc:both})}. This issue is particularly severe at the onset of the simulation; consequently, we conclude that we lack meaningful information for the entire initial 1-2 year of both simulations (Figures~\ref{fig:lc:both} and~\ref{fig:lc:ABmag}). Adequately resolving the photosphere---specifically, ensuring that particles at the nominal photosphere have a resolution length, $h$, \updated{commensurate with} the photospheric \updated{scale height}---would require approximately \updated{1.7~billion} particles for the 1.7~M$_{\odot}$ simulation and \updated{16~billion} particles for the 3.7~M$_{\odot}$ model. Such high resolution is currently computationally prohibitive.
A possible future solution is to use the novel adaptive particle refinement method of \citet{Nealon2025}.

\item Our simulations are adiabatic and, as such, result in too bright a luminosity, particularly in the last $\sim$30 years of the simulation, when the bolometric luminosity is approximately constant at $1-2\times10^5$~\Lsun. During that time the object would radiate $\sim 4\times 10^{47}$~erg or $\sim$10 times the initial internal energy of the star. Such radiation, would contribute to a smaller cooler and less luminous photosphere. The solution is adopting a radiative cooling approach, e.g., \citet{Lau2025}.
\updated{The lack of radiative transport and dust destruction in our simulations also means that the dust shells are assumed to be adiabatically-expanding, instead of being heated up by the hot gas photosphere on the inside, while cooling into space. }

\item An optically thin dust shell emerges above the warmer, gaseous  photosphere between 3 and 5 years after the start of the simulations, at which point the \gls{SED} shows the emergence of a second, cooler peak (Figures~\ref{fig:spec:both} and \ref{fig:Teff}, where we see the temperature drops below 1000~K). The dust shell becomes fully optically thick by 20 years and it continues to cool and expand.
\updated{We predict the dust will disperse sufficiently to become optically thin to visible radiation between 2000 and 20\,000 years later (Figure~\ref{fig:tau-dust}), at which point an inner object would be revealed. If dust destruction were efficient, these times could be shorter.}

\item Equivalent models with no dust show a very different lightcurve with a gently increasing luminosity throughout the simulation---only the lower mass model shows a turn-over of the bolometric luminosity, while for the higher mass, lower $q$ model, the calculation was carried out for too short a time to see the turn-over.

\item \updated{Despite the shortcoming of our simulations, we took some of the results at face value and compared them to} the lightcurves and other observationally-derived quantities of OGLE-2002-BLG-360 and of AT~2025abao, two \glspl{LRN} with progenitor masses similar to our more massive model. Both likely derive from giant progenitors, the former from a $\sim$30~\Rsun, RGB star, the latter from a $\sim$350~\Rsun{} AGB star quite similar to our models. The timing of the peaks measured from the start of the light rise or the start of the simulations is similar for OGLE-2002 (1.6 vs. $1-6$~yrs for the simulations), but much shorter for AT~2025abao (36 days), with a photospheric size that, for OGLE-2002 starts smaller (at peak it is 1.1~\updated{au} vs. $3-6$~\updated{au} for the simulations), but becomes quite comparable at 8 years (50~\updated{au} for the dust photosphere), while for AT~2025abao it is within range of the simulations (4.1~au) at \updated{peak}. The peak luminosities are very discrepant for OGLE-2002 (8100~\Lsun\ vs. $2-4\times 10^6$~\Lsun\ for the simulations), but much closer for AT~2025abao (720\,000~\Lsun). The mass of the dust is comparable between our simulations and OGLE-2002, while AT~2025abao may not have had time to make any dust yet (at the time of writing, only 8 months have passed from the start of the outburst).   

\end{enumerate}

Bridging the gap between simulations and observations requires an ability to model the onset of CE phases in simulations as well as resolving two primary limitations identified in this study: low photospheric resolution, which chiefly degrades the light curve during the first year, and the adiabatic assumption, which overestimates the luminosity in later phases. The first issue could be mitigated using the adaptive particle refinement method developed by \citet{Nealon2025}, an approach currently under development. The second limitation can be addressed by implementing radiative cooling techniques, such as those demonstrated by \citet{Lau2025}, steps we reserve for future work.


\section*{Acknowledgements}

This research was supported by the Commonwealth through an Australian Government
Research Training Program Scholarship [DOI: 10.82133/C42F-K220].
CM acknowledge the use of the python package {\sc Sarracen} \citep{Harris2023a} used to post-process \phant{} data dumps.
OD, LS, MK and LB acknowledge support through the Australian Research Council discovery programme grant DP210101094.
LS is senior research associates from F.R.S.- FNRS (Belgium). ML acknowledges funding by the European Union (ERC, ExCEED, project number 101096243). NB acknowledges being funded by the European Union (ERC, CET-3PO, 101042610). Views and opinions expressed are, however, those of the author(s) only and do not necessarily reflect those of the European Union or the European Research Council Executive Agency. Neither the European Union nor the granting authority can be held responsible for them. The publication is part of the research project PID2024-155585NA-I00 funded by MICIU/AEI /10.13039/501100011033. NB acknowledges financial support from grant CEX2024-001451-M funded by MICIU/AEI/10.13039/501100011033.

This work was supported in part by Oracle Cloud credits and related resources provided by Oracle for Research.
This work was supported by the OzSTAR national facility at Swinburne University of Technology. OzSTAR is funded by Swinburne and the Australian Government’s National Collaborative Research Infrastructure Strategy (NCRIS).

\updated{We finally thank a very thorough referee who helped us improve this manuscript.}


\section{Data Availability}
\label{sec:data-availability}
\updated{
    The source code for {\sc clmuphantomlib} is available on GitHub (\url{https://github.com/chunliangmu/clmuphantomlib}) and Zenodo (\url{https://doi.org/10.5281/zenodo.22912863}).
    Scripts for data analysis and plot generation are also available on GitHub (\url{https://github.com/chunliangmu/RTinCEE-scripts-2024}) and Zenodo (\url{https://doi.org/10.5281/zenodo.22913417}).
    Movies versions of Figure~\ref{fig:profile:2md-single-ray},~\ref{fig:lum-image:2md} and~\ref{fig:lum-contr:2md} are available on Zenodo (\url{https://zenodo.org/records/18239846}).
}


\bibliographystyle{mnras}
\bibliography{bibliography}

\setglossarystyle{list}
\printglossary[type=\acronymtype]

\appendix
\section{Light Calculation Details}
\label{app:light-calculation}

\begin{figure}
    \centering
    \includegraphics[width=9cm, trim={0 0 0 80}, clip]{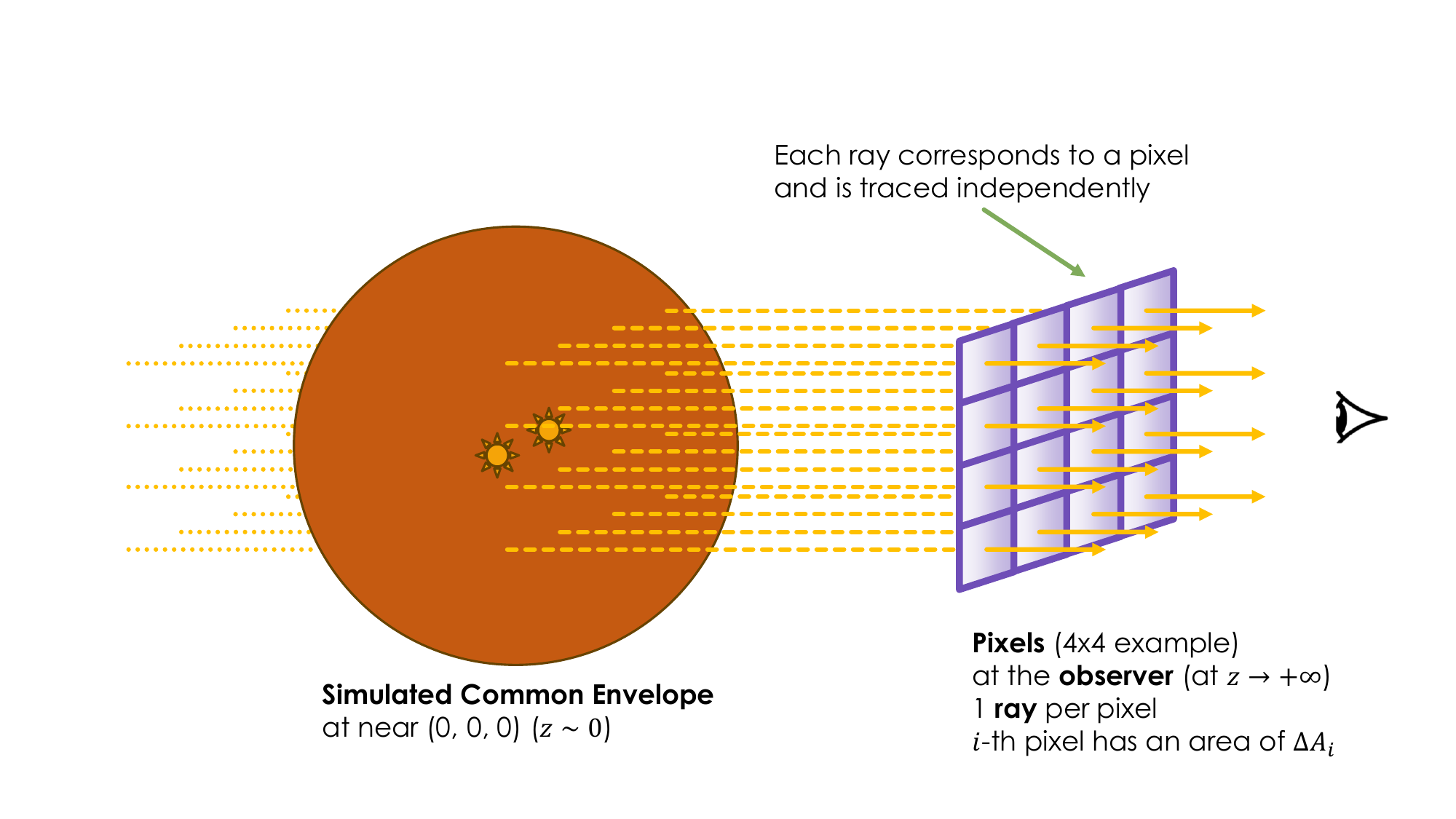}
    \caption{
    Diagram of the grid (purple) of rays (yellow). The area of each pixel is $\Delta A_i$. All pixels have the same area. }
    \label{fig:diagram:ray-grid}
\end{figure}

\begin{figure*}
    \centering
    \includegraphics[width=14cm, trim={0 0 0 100}, clip]{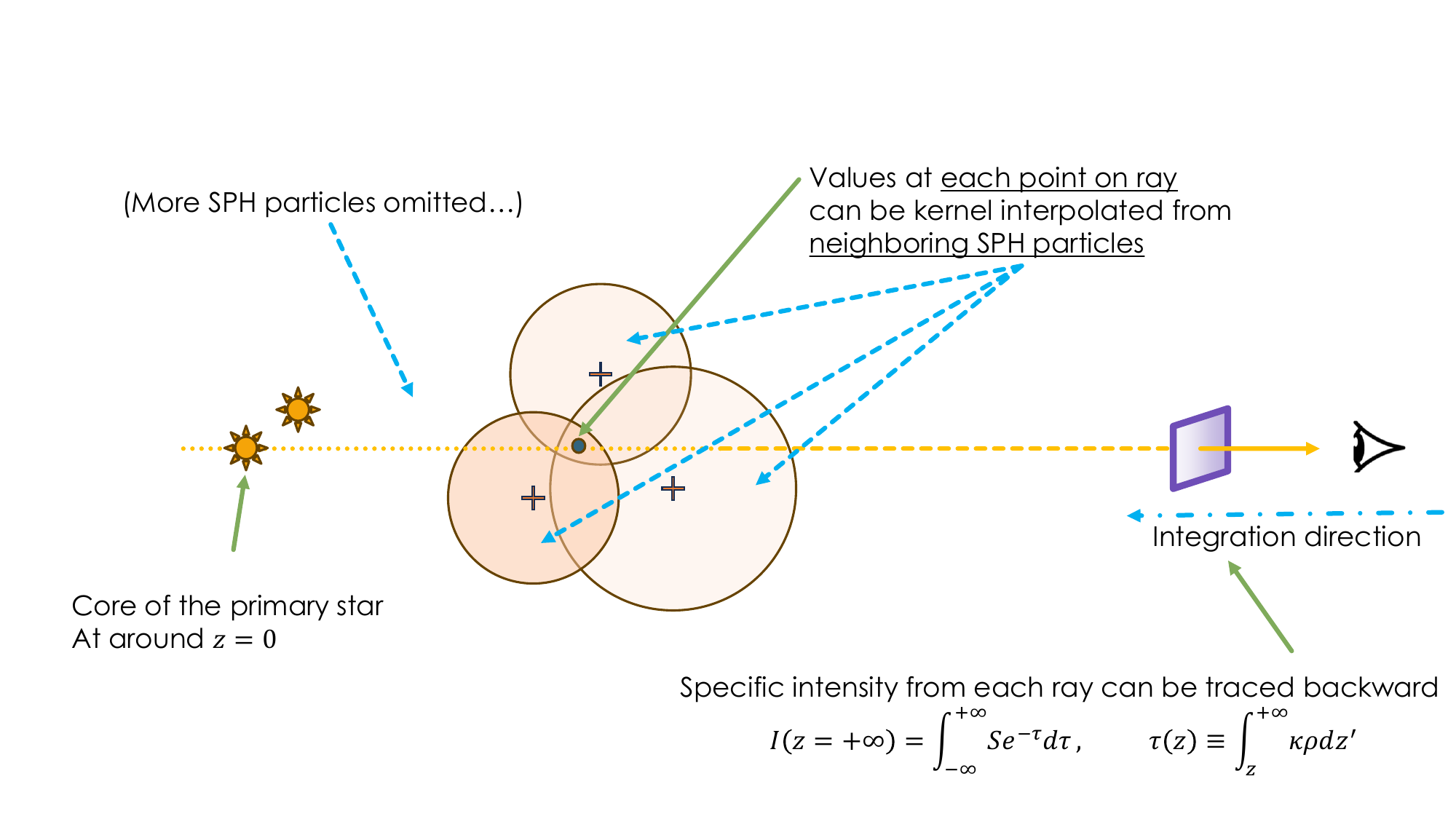}
    \caption{Diagram of how rays (yellow line) from $z=-\infty$ (far left) passing through the simulated \gls{CE}, which is represented by a collection of \gls{SPH} particles (plus signs with circles around), reach the observer (black eye) at $z=+\infty$. Some \gls{SPH} particles contribute to quantities at a specific point in the ray (dark green circle on the yellow ray) when they are closer to that point than $2h$, where $h$ is the smoothing length of the particle. Near the surface, the density, $\rho$, is low and thus the smoothing lengths of particles at that location is large, increasing the size of the particles, resulting in much more overlap with rays. This becomes a significant source of uncertainty if the particle has an  opacity, $\kappa_j$, large enough to single-handedly block out the ray (i.e., $\Delta \tau_j \gtrsim 1$, see Equation~\ref{eq:def:Delta_tau}),
    meaning that the value of $I_i$ for that tile is provided by a single (or few) particle(s), thus failing to capture the details of the structure near the surface.
    }
    \label{fig:diagram:ray-vs-particles}
\end{figure*}

In Figures~\ref{fig:diagram:ray-grid} and \ref{fig:diagram:ray-vs-particles} we give a graphical representation of the post-processing calculation. 
Following the description in Section~\ref{sec:post-processing},
the calculations of the light is parallelized across rays using the python package {\sc numba}'s {jit(parallel=True)} and {prange()} functions, with the following steps:
\begin{enumerate}
    \item The coordinate system was rotated so that the observer is always at $z \rightarrow +\infty$.
    Currently, this is achieved by simply re-arranging the axis labels, so we only have 3 view ports $+x, +y, +z$.
    \item A grid of parallel rays pointing towards the $+z$ direction were generated to sample the $xy$-plane.
    The size of the grid in the $x$ and $y$ directions includes all SPH particle positions $\pm$ the kernel radius ($2h$) along that axis. In other words, it ranges in $\left[
    \min{\{x_j - 2 h_j | \forall j\}},\;
    \max{\{x_j + 2 h_j | \forall j\}} \right]$, and same for the $y$ direction.
    \item \label{list:lcgen:para:s} We sort all SPH particles by their $z$ coordinates and go through them for each ray in a backward tracing fashion, i.e., from the ones with highest $z$ (observer's side) to the lowest $z$ (the opposite side).
    \item For each ray $i$, and each \gls{SPH} particle $j$, we first check whether that particle lies within the range of of the ray, in other words, whether the distance from the particle $j$ to ray $i$ is smaller than twice its smoothing length $2h_j$.
    The factor of 2 here comes from the smoothing kernel radius.
    We record a list of all relevant particles for each ray $i$.
    \item Then we compute the effective area $A_{\mathrm{eff}, ij}$ for each particle $j$ and ray $i$, using Equation~\ref{eq:def:Aeffj}.
    This is achieved by integrating along 1000 sample points (per particle per ray) across the segment of the ray that lies inside the $2h_j$ radius of the particle.
    The contribution to each sample points from the neighbouring particles were also included when calculating the optical depth $\tau_i(z)$ in Equation~\ref{eq:calc:tau_z}.
    \item If the optical depth becomes larger than the stop condition (see equation~\ref{eq:radiative-transfer-stop-cond} below), the calculation for the $i$'th ray is stopped, as we know that further particles no longer make a meaningful contribution to $I_i$.
    \item  \label{list:lcgen:para:e} The effective area $A_{\mathrm{eff}, j}$ for each particle $j$ is computed by summing up among the rays using Equation~\ref{eq:def:Aeffj}.
    \item The luminosity is then computed using Equation~\ref{eq:calc:luminosity}.  
\end{enumerate}
Step~\ref{list:lcgen:para:s} to~\ref{list:lcgen:para:e} are parallelized across the rays.
To save computational time, we implement a stop condition (expressed in Equation~\ref{eq:radiative-transfer-stop-cond}) once $\tau_i(z)$ becomes sufficiently large such that further particles' contributions to $I_i$ are smaller than the rounding error.
For simplicity,
we begin by approximating the solution to the radiative transfer equations (Equation~\ref{eq:radiative-transfer-solution-backwards}) with
\begin{equation}
    I_i = \sum_j S_j e^{-\tau_{ij}} (1 - e^{-\Delta \tau_{ij}})
    ,
    \label{eq:radiative-transfer-discretized-delta_taus}
\end{equation}
where
\begin{equation}
    \Delta \tau_{ij} \equiv \frac{m_j \kappa_j}{h_j^3} \int_{-\infty}^{+\infty} w(q_{ij}(z')) dz'.
    \label{eq:def:Delta_tau}
\end{equation}
is the optical depth contribution of particle $j$ for ray $i$, the integral in the equation above (the column kernel) is a pre-calculated, tabulated quantity,
and
\begin{equation}
    \tau_{ij} \equiv \sum_{n=0}^{j-1} \Delta \tau_{in}
    \label{eq:def:tau_max_i}
\end{equation}
is the optical depth at the location of particle $j$, which we are approximating by summing up the optical depth contribution from all particles in front of it.
In other words, by invoking the above approximation,
we are ignoring---for the purpose of deciding on the stop condition for the post-processing interpolation---how spread out the particles along the ray are;
instead, we assume particles in front of $j$ are fully in front (added to $\tau_{ij}$), and those behind are fully behind (ignored).
In other words, here we are treating \gls{SPH} particles as pre-integrated 2D circles instead of 3D spheres.
This should be reasonable in the context of deciding the stop condition (which requires less precision). 

To determine whether it is reasonable to stop at a given particle, $k$, we follow the method below. Given that the source function $S_j \geq 0$ and $\Delta \tau_{ij} \geq 0$ remains true for all particles, we can infer that
\begin{equation}
    I_{i, k} \leq I_i \leq \sum_{j=0}^{N_i} S_j,
\end{equation}
where
\begin{equation}
    I_{i, k} \equiv  \sum_{j=0}^k S_j e^{-\sum_{n=0}^{j-1} \Delta \tau_n} (1 - e^{-\Delta \tau_j})
\end{equation}
is the specific intensity for $i$-th ray integrated up to $k$-th particle.
Therefore, the absolute error of using $I_{i, k}$ to estimate $I_i$ is
\begin{equation}
    I_i - I_{i, k}
    = \sum_{j=k+1}^{N_i} e^{-\tau_j} \left( 1 - e^{-\Delta \tau_j} \right)  S_j
\end{equation}
for simulations with $N_i$ particles relevant to ray $i$,
and that
\begin{equation}
e^{-\tau_j} \left( 1 - e^{-\Delta \tau_j} \right)
= e^{-\tau_k} e^{-\sum_{n=k}^{j-1} \Delta \tau_n} \left( 1 - e^{-\Delta \tau_j} \right)
\leq e^{-\tau_k}.
\end{equation}
Since $j > k$ and $\Delta \tau_j > 0$,
we can deduct that
\begin{equation}
    I_i - I_{i, k}
    \leq e^{-\tau_k} \sum_{j=0}^{N_i} S_j.
\end{equation}
Hence, the relative error of $I_i$ is
\begin{equation}
    \frac{I_i - I_{i, k}}{I_i}
    \leq \frac{e^{-\tau_k} \sum_{j=0}^{N_i} S_j}{I_{i, k}}.
\end{equation}
When the right hand side is smaller than the machine precision $\delta_\mathrm{tol} \approx 10^{-16}$,
the relative error (represented by the left hand side of the equation) is smaller than the machine precision $\delta_\mathrm{tol}$,
therefore integrating any further will make no difference,
i.e., when
\begin{equation}
    \tau_k \geq \ln{\sum_{j=0}^{N_i} S_j} - \ln{\delta_\mathrm{tol}} - \ln{I_{i, k}},
    \label{eq:radiative-transfer-stop-cond}
\end{equation}
we can safely ignore all particles for ray $i$ behind this point.

\section{Integration order}
\label{app:uncertainties:integration-order}

We note that there are two ways of combining the quantities in Equation~\ref{eq:calc:kernel-interpolation:I_i} and~\ref{eq:calc:tau_z}: One first interpolates each individual component ($S$, $\kappa$, $\rho$) to the ray location and then multiplies them together; alternatively, one first multiplies the quantities for each particle ($S \kappa \rho$)  and then interpolates the product to the ray (the method of this paper).
These two approaches produce somewhat different results. For example, the lightcurve in Figure~\ref{fig:lc:both} produced by the multiply-then-interpolate approach is brighter by a factor of approximately 2, at time zero. In theory the two methods should approach each other when the resolution is higher, or, in other words when the SPH particles contributing to each location on the ray are closer together, with smaller smoothing lengths.
It is not clear to us which approach is more correct. Our method (Section~\ref{sec:grey_luminosity}) allows us to clearly identify the times when we have little knowledge of the light and those where we have relatively good knowledge. Where the shaded areas are dominant in Figure~\ref{fig:lc:both}, the lightcurve is virtually unknown, while where the shaded area disappears into the curve, i.e., where the upper and lower limits according to our definition converge, we have reasonable information on the light.

\section{On the source function}
\label{app:source}
\updated{The full frequency-dependent source function in the case of absorption and scattering combined can be expressed as
\begin{equation}
    S_\nu=\frac{\kappa_\mathrm{abs}B_\nu+\kappa_\mathrm{sca}J_\nu}{\kappa_\mathrm{abs}+\kappa_\mathrm{sca}},
\end{equation}
where $\kappa_\mathrm{abs}$ is the absorption opacity, $\kappa_\mathrm{sca}$ is the scattering opacity, $B_\nu$ is the Planck function and $J_\nu$ is the local mean intensity \citep{RybickiLightman1986}. Approximating the source function as Eq.~(\ref{eq:assumption:S_j}) is equivalent to setting $J_\nu=B_\nu$, which means that the local mean intensity can be well approximated with a blackbody spectrum of the local temperature. This is a good approximation in absorption-dominated optically thick regions.} 

\updated{However, this will break down in scattering-dominated regions where the local intensity is determined by photons created far away. If the thermalisation radius (where the photons are created) is located in higher temperature regions than the scattering photosphere, therefore $J_\nu\gtrsim B_\nu$. By taking the assumption in Eq.~(\ref{eq:assumption:S_j}), we are likely underestimating the source function in scattering-dominated regions. That said, if $J_\nu$ is zero or less than $B_\nu$ we would overestimate the source function. Ultimately a complete radiation transport calculation is needed.
}

\section{Location of the photosphere}
\label{app:photosphere_table}

In Table~\ref{tab:ray_values} we present the photospheric values that correspond to Figure~\ref{fig:profile:2md-single-ray}.
\updated{
    The reported values pertain to one point on the photosphere, corresponding to one ray, where a comparison of three rays ($x \rightarrow +\infty$, $y \rightarrow +\infty$, $z \rightarrow +\infty$) give a general idea about the photosphere's shape. The error bars on the photospheric radii listed in  Table~\ref{tab:ray_values}, derive {\it only} from the optical depth uncertainty.
    Systematic uncertainties, such as those due to the surface instability of the mapped 3D stellar structure at time zero (see Appendix~\ref{app:uncertainties:initial-years}), or those due to the adiabatic approximation affecting the light a few decades after the start of the simulation, or others discussed in Section~\ref{sec:uncertainties},  are not included.
}

\begin{table*}
    \centering
    \begin{tabular}{lcccccccc}
    \hline
    &\multicolumn{4}{c}{1.7~\Msun} & \multicolumn{4}{c}{3.7~\Msun}\\
    &0 &3.0&  12.1&  44.4 &0 &3.0&  12.1&  44.4\\
    & (yr) & (yr) & (yr) & (yr) & (yr) & (yr) & (yr) & (yr)\\
    \hline
    $\tau_\mathrm{ph, +z}$ & $0.67^{+30.22}_{-0.67}$ & $0.67^{+1.79}_{-0.54}$ & $0.67^{+2.96}_{-0.46}$ & $0.67^{+0.57}_{-0.35}$ & $0.67^{+494.09}_{-0.67}$ & $0.67^{+21.34}_{-0.58}$ & $0.67^{+3.25}_{-0.61}$ & $0.67^{+0.38}_{-0.26}$ \\[5pt]
    $\tau_\mathrm{ph, +y}$ & $0.67^{+15.49}_{-0.67}$ & $0.67^{+3.56}_{-0.54}$ & $0.67^{+1.05}_{-0.42}$ & $0.67^{+0.31}_{-0.20}$ & $0.67^{+725.51}_{-0.67}$ & $0.67^{+0.62}_{-0.41}$ & $0.67^{+3.20}_{-0.52}$ & $0.67^{+0.52}_{-0.33}$ \\[5pt]
    $\tau_\mathrm{ph, +x}$ & $0.67^{+5.94}_{-0.67}$ & $0.67^{+2.55}_{-0.62}$ & $0.67^{+1.27}_{-0.55}$ & $0.67^{+0.32}_{-0.19}$ & $0.67^{+106.15}_{-0.65}$ & $0.67^{+0.92}_{-0.37}$ & $0.67^{+2.75}_{-0.35}$ & $0.67^{+0.61}_{-0.35}$ \\\\
    
    $\kappa_\mathrm{ph, +z}$ / $\mathrm{cm^{2}\,g^{-1}}$ & $0.09^{+0.12}_{-0.06}$ & $3.4^{+2.4}_{-1.9}$ & $5.6^{+4.0}_{-1.6}$ & $5.3^{+0.6}_{-0.9}$ & $2.8^{+1.4}_{-0.0}$ & $0.02^{+0.29}_{-0.02}$ & $6.4^{+5.0}_{-3.3}$ & $4.4^{+0.3}_{-0.4}$ \\[5pt]
    $\kappa_\mathrm{ph, +y}$ / $\mathrm{cm^{2}\,g^{-1}}$ & $0.12^{+0.11}_{-0.00}$ & $7.3^{+2.5}_{-0.4}$ & $6.2^{+1.0}_{-0.8}$ & $4.0^{+0.6}_{-0.6}$ & $4.4^{+0.7}_{-0.0}$ & $4.2^{+0.0}_{-3.2}$ & $5.4^{+4.5}_{-2.4}$ & $4.5^{+0.6}_{-0.8}$ \\[5pt]
    $\kappa_\mathrm{ph, +x}$ / $\mathrm{cm^{2}\,g^{-1}}$ & $0.03^{+0.05}_{-0.02}$ & $9.9^{+0.2}_{-0.2}$ & $5.5^{+2.0}_{-2.1}$ & $3.9^{+0.8}_{-0.6}$ & $2.8^{+0.0}_{+0.0}$ & $0.01^{+0.01}_{-0.00}$ & $4.7^{+3.8}_{-1.0}$ & $4.9^{+0.7}_{-1.3}$ \\\\
    
    $T_\mathrm{ph, +z}$ / $\mathrm{K}$ & $6659^{+406}_{-437}$ & $911^{+400}_{-136}$ & $468^{+325}_{-124}$ & $426^{+45}_{-67}$ & $8675^{+169}_{-0}$ & $5528^{+1738}_{-1166}$ & $719^{+291}_{-272}$ & $351^{+26}_{-27}$ \\[5pt]
    $T_\mathrm{ph, +y}$ / $\mathrm{K}$ & $6858^{+251}_{-0}$ & $619^{+392}_{-43}$ & $509^{+81}_{-63}$ & $326^{+48}_{-50}$ & $8969^{+19}_{-2}$ & $959^{+1229}_{-130}$ & $517^{+288}_{-132}$ & $366^{+44}_{-55}$ \\[5pt]
    $T_\mathrm{ph, +x}$ / $\mathrm{K}$ & $6148^{+367}_{-345}$ & $846^{+176}_{-17}$ & $465^{+139}_{-183}$ & $319^{+60}_{-47}$ & $8660^{+5}_{-0}$ & $5600^{+286}_{-224}$ & $392^{+323}_{-62}$ & $401^{+55}_{-107}$ \\\\
    
    $R_{1, \mathrm{ph, +z}}$ / $\mathrm{AU}$ & $2.2 \pm 0.3$ & $28 \pm 6$ & $104 \pm 9$ & $525 \pm 12$ & $2.1 \pm 0.2$ & $5 \pm 1$ & $136 \pm 15$ & $530 \pm 22$ \\[5pt]
    $R_{1, \mathrm{ph, +y}}$ / $\mathrm{AU}$ & $2.1 \pm 0.2$ & $31 \pm 7$ & $142 \pm 11$ & $451 \pm 14$ & $2.1 \pm 0.2$ & $24 \pm 10$ & $151 \pm 16$ & $610 \pm 19$ \\[5pt]
    $R_{1, \mathrm{ph, +x}}$ / $\mathrm{AU}$ & $2.2 \pm 0.3$ & $36 \pm 5$ & $137 \pm 10$ & $461 \pm 15$ & $2.3 \pm 0.2$ & $10 \pm 1$ & $142 \pm 16$ & $585 \pm 18$ \\\\
    \hline
    \end{tabular}
    \caption{Values of
    opacity, temperature and radius at optical depth 2/3, along three rays traced along the $x$, $y$ and $z$ directions through the middle of the 1.7~\Msun{} and 3.7~\Msun\ simulations,
    as shown in Figure~\ref{fig:profile:2md-single-ray}.
    Uncertainties refers to the range of values within one smoothing length.
    }
    \label{tab:ray_values}
\end{table*}
\section{Convergence and resolution}

\subsection{Convergence with respect to ray-grid resolution}
\label{app:convergence:ray-grid}

\begin{figure}
    \centering
    \includegraphics[width=\linewidth]{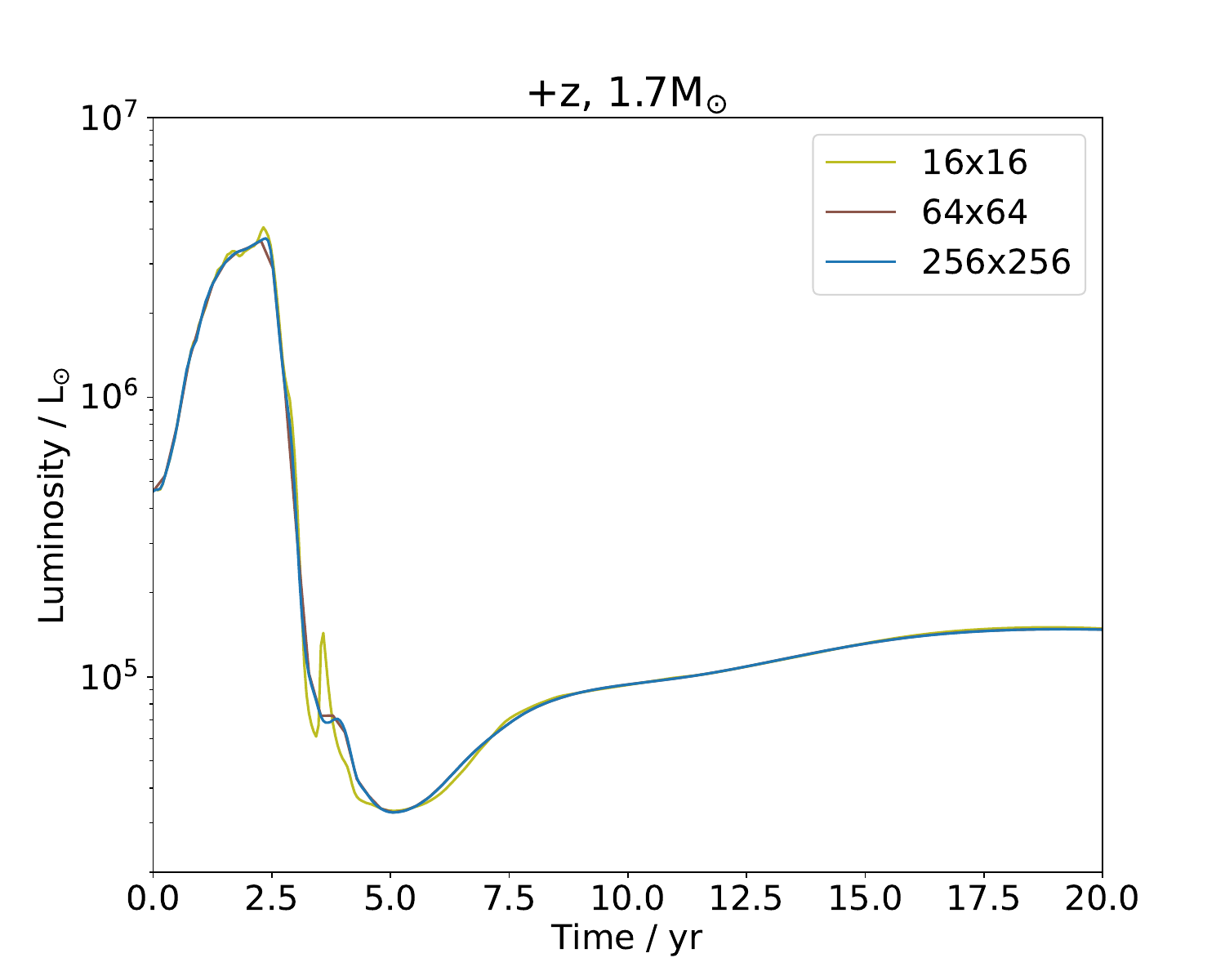}
    \caption{
    Convergence study for the bolometric lightcurve of the 1.7~\Msun{} simulation, as viewed from the polar ($z \rightarrow +\infty$) direction, using different ray-grid resolutions (yellow: $16 \times 16$ rays; brown: $64 \times 64$ rays; blue $256 \times 256$ rays). An additional resolution of $1024 \times 1024$ rays was tested, and the results are visually indistinguishable from $256 \times 256$ ray results.
    } \label{fig:lc-dusty-16x16}
\end{figure}

Here we give an account of the effect of changing the resolution in the grid used to construct the lightcurve. It is clear from Figure~\ref{fig:lc-dusty-16x16} that grid resolution does not greatly affect the lightcurve and that a resolution of $256\times 256$ cells is sufficient. Also, the size of the particles' smoothing lengths at the photosphere is two to ten times larger than the pixels' size, denoting that the grid resolution is sufficient.

\subsection{SPH resolution study}
\label{app:sph-resolution}

\begin{figure*}
    \centering
    \includegraphics[width=0.454\linewidth, trim={0 90 112 150}, clip]{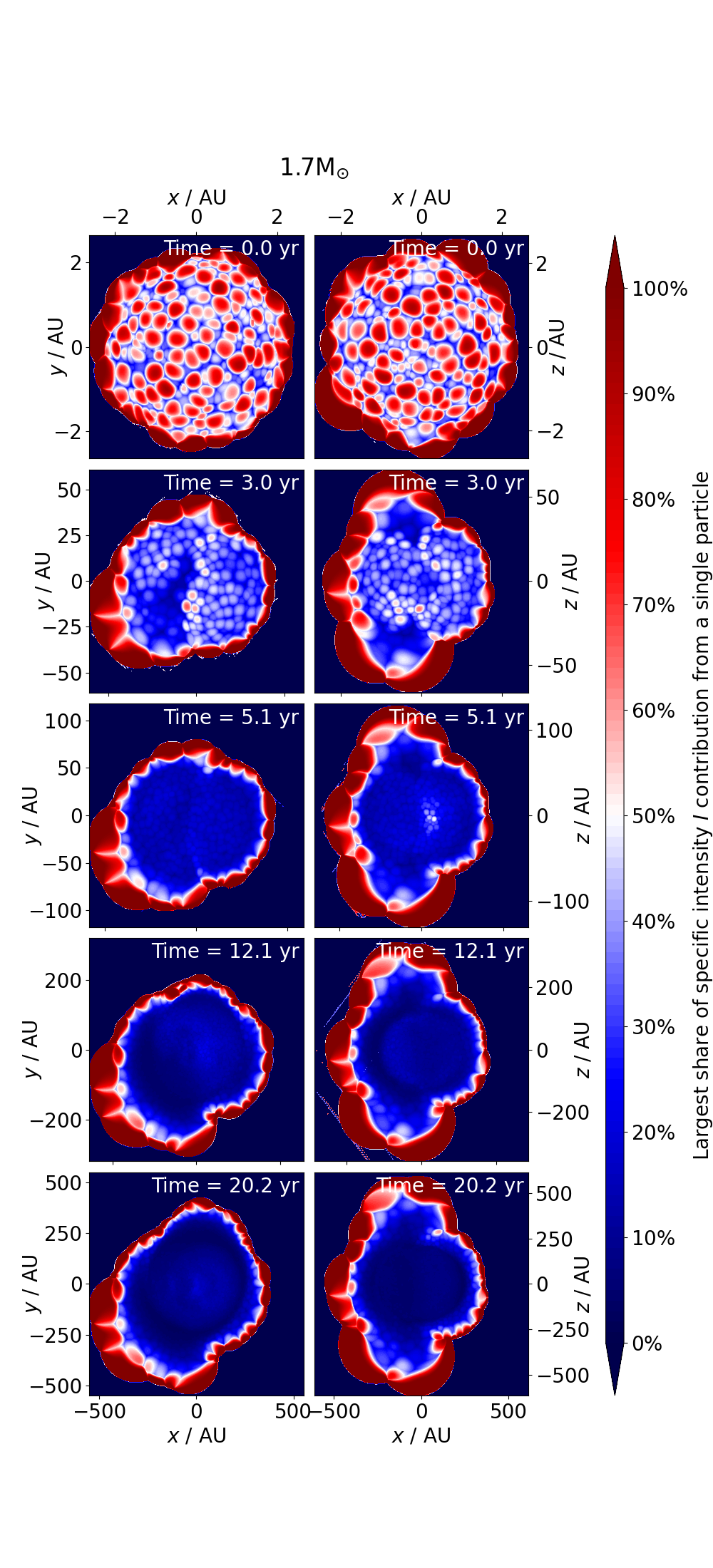}
    \includegraphics[width=0.454\linewidth, trim={0 90 112 150}, clip]{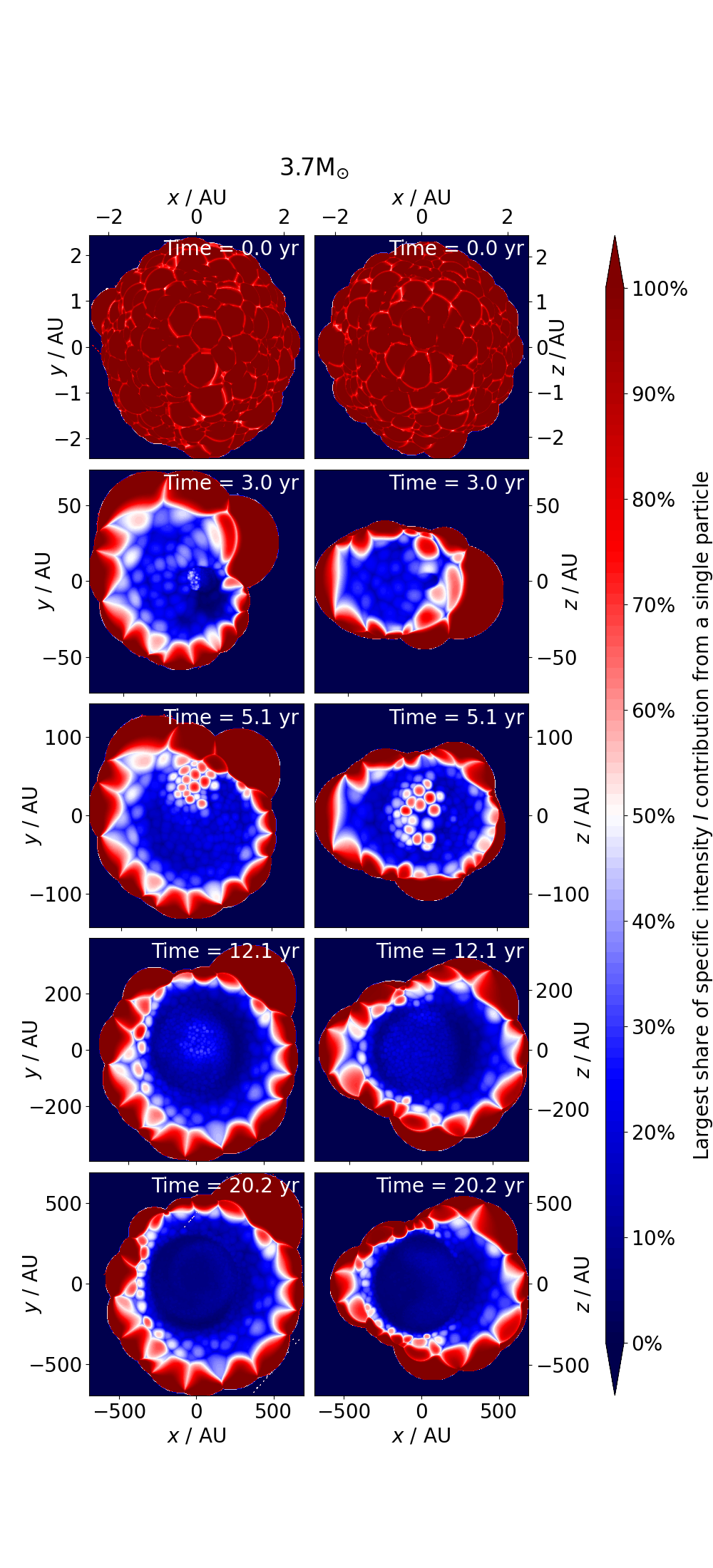}
    \includegraphics[width=0.082\linewidth, trim={610 90 0 150}, clip]{figures/contr_4md_256x256.png}
    \caption{The fractional contribution to the specific intensity of a pixel $i$, $I_i$, for the $j^{\rm th}$ SPH particle, that contributes the most (see Equation~\ref{eq:def:F_contr}),
    for the 1.7~\Msun{} (left two columns) and the 3.7~\Msun{} (right two columns) simulations. The redder colours on the top row show that one particle dominates the calculation of the light at that location, while the dark blue colours later on show that many more particles contribute to the calculation of the light at that location, with a resulting much smaller uncertainty.
    Similarly to Figure~\ref{fig:lum-image:2md}, data is viewed from the polar ($z \rightarrow +\infty$) direction (left panels) and the orbital ($y \rightarrow +\infty$) direction (right panels), at $t=0, 3, 12, 20$~yr for the top, second, third and bottom rows respectively.
    The origin is fixed at the primary's core particle's location.
    Movies of this image see \url{https://zenodo.org/records/18239846}
    for both 1.7~\Msun{} (2md) and 3.7~\Msun{} (4md) simulations, viewed from $z \rightarrow +\infty$ (xyz), $y \rightarrow +\infty$ (xzy) and $x \rightarrow +\infty$ (zyx).
    }
    \label{fig:lum-contr:2md}
\end{figure*}

In Section~\ref{sec:uncertainties:SPHres} we scrutinised the resolution of the simulation and its ability to accurately reproduce the light by following a different argument. There we calculate the percentage of the light emitted by pixel (ray) $i$ by the most contributing particle. If that fraction is close to 100\% it means only one particle contributes to the light of that pixel, denoting very low resolution. In Figure~\ref{fig:lum-contr:2md} we show polar ($z$) and side ($y$) maps of the quantity $f$ on the same physical scale as the brightness maps in Figure~\ref{fig:lum-image:2md}. This gives an immediate idea of where the resolution is poor and where it is better.

\section{1D vs 3D photosphere at time zero}
\label{app:uncertainties:initial-years}

\begin{figure*}
   \centering
    \includegraphics[width=0.49\linewidth]{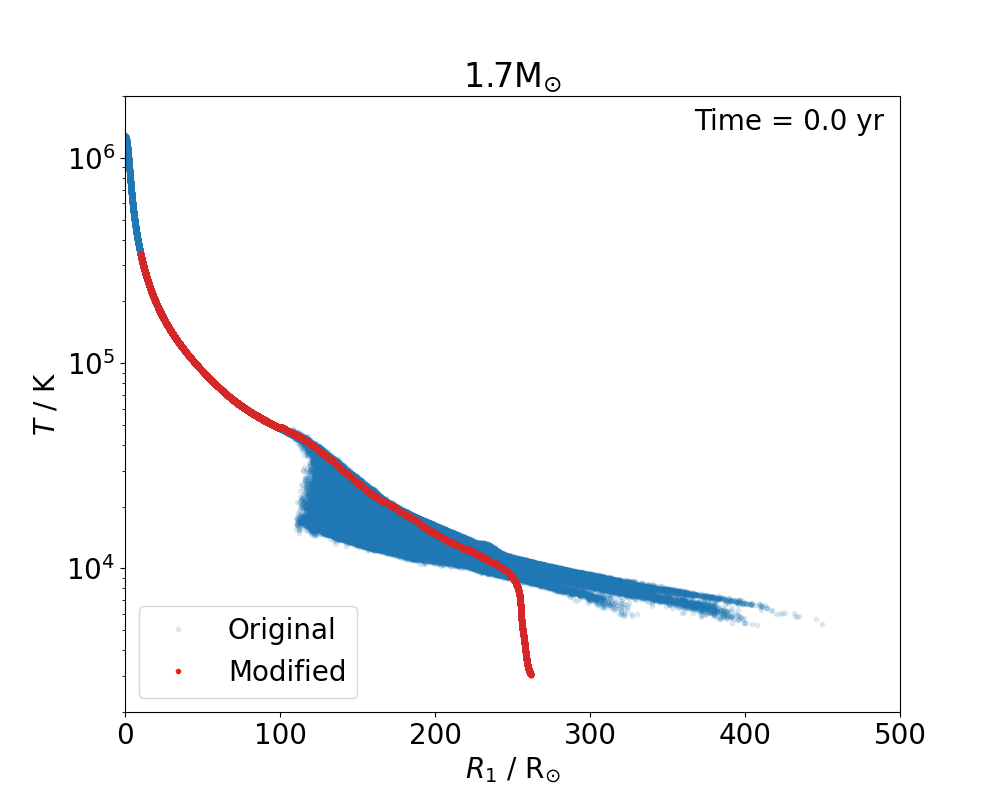}
    \includegraphics[width=0.49\linewidth]{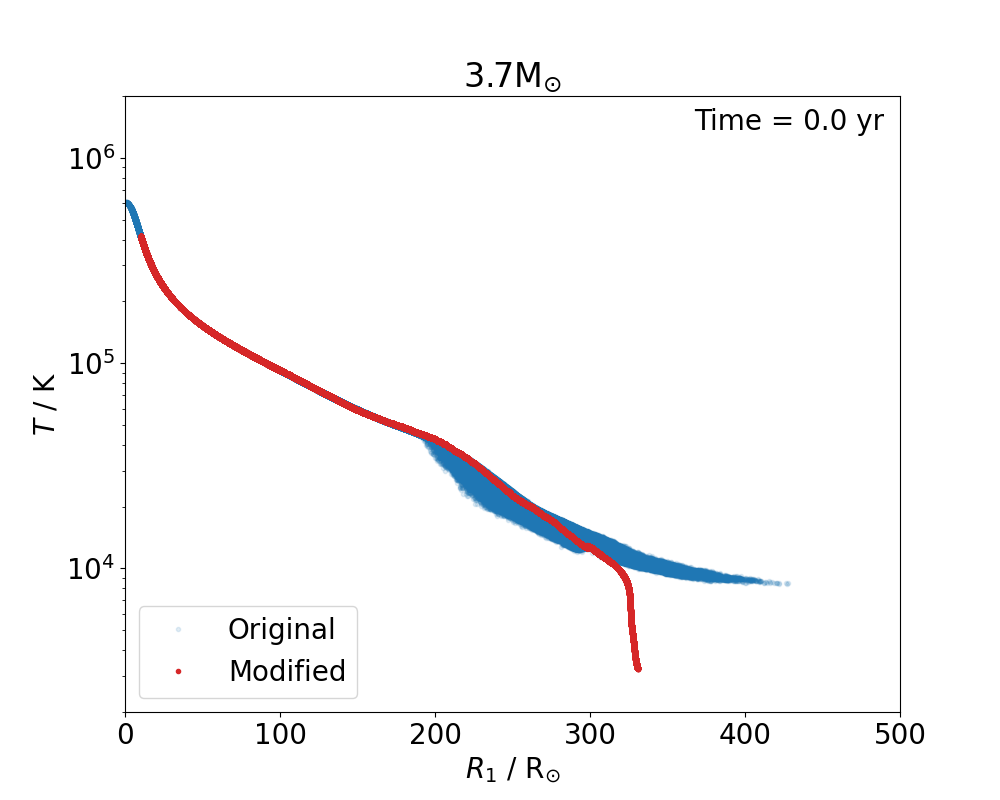}
\caption{Stellar temperature profile of the models, before (blue symbols) and after (red symbols/curve) forcing \phant{} initial dump to have \mesa-prescribed temperatures  (see text).  Left panel is for the 1.7~\Msun{} simulation, while the right panel is for 3.7~\Msun{}.   
\label{fig:profile:Xmdd}}
\end{figure*}

Another source of systematic lightcurve uncertainty comes from a mismatch of the outer layers of the initial temperature profile of the AGB star, between the 1D \mesa{} profile, and the profile mapped and relaxed in the 3D computational domain (Figure~\ref{fig:profile:Xmdd}). The methods used to carry out the mapping and relaxation have been thoroughly investigated \citep[e.g.,][]{Reichardt2019,GonzalezBolivar2022,Lau2022} with an eye to ensuring that the result of the common envelope simulations are not dependent on the imperfect reproduction of the 1D profile in the 3D domain. The lack of sensitivity of the hydrodynamic simulations to the specifics of the stellar surface layers is mostly due to the fact that the mass that expands past the 1D photosphere is small.

In our 3D AGB stars, the few low density and relatively low temperature surface particles that protrude outside the 1D photosphere are optically thick. As we can see from Figure~\ref{fig:profile:Xmdd} this results in a photosphere that is both hotter and larger than the original 1D photosphere, with a resulting larger luminosity as seen in Figure~\ref{fig:lc:both}, where the luminosities of the two models at time zero are $5\times10^5$ and $10^6$~\Lsun{} for the 1.7~\Msun{} and 3.7~\Msun{} models, respectively, compared to the value they should have as given by the {\sc mesa} calculation: $5\,180$~\Lsun{} for the 1.7~\Msun{} simulation and $11\,900$~\Lsun{} for the 3.7~\Msun{} one.


Here we evaluate the effect of these particles on the lightcurve {\it after} time zero. We re-run the first 5 years of the same 1.7~\Msun{} and 3.7~\Msun{} simulations, but the surface particles have been forced to have the same temperature, $T$, as prescribed by \mesa{}. By assigning the \mesa{} temperatures to the 3D star after relaxation (which has a different density distribution compared to the original \mesa{} profile), leads to a different entropy distribution compared to the 1D MESA model. We do this here to achieve the correct thermal properties for this purpose.

Since \phant{} finds the SPH particles' temperatures by interpolating the \mesa{} \gls{EoS} from the particles' densities, $\rho$, and their specific internal energy, $u$ (see also~\citealt{Bermudez2024a}),
we enforce the required $T$ 
for each particle by solving for the corresponding $u$ keeping $\rho$ fixed in the tabulated \mesa{} \gls{EoS} via root-finding, and then setting the particles to have that value of $u$.

Since the particles' temperatures only deviate from the \mesa{} prescription outside $\sim100$~\Rsun{} (see Figure~\ref{fig:profile:Xmdd}), only particles at radii larger than this value are treated in this way. 
We also removed all particles outside the \mesa{} photosphere surface (261.8~\Rsun{} for the 1.7~\Msun{} simulation and 330.9~\Rsun{} for the 3.7~\Msun{} simulation). The resulting $T$ profile at $t=0$ is illustrated in Figure~\ref{fig:profile:Xmdd}.

This new, adjusted star evolved in isolation to test its stability, shows a similar stability to the original model, as demonstrated by inspecting the ``particle" and ``density" radii as plotted by \citet[][]{GonzalezBolivar2022}, as well as their figure A1, second row, fourth column. This reassures us that the star's hydrodynamic behaviour would be comparable to that of the main model used here.

\begin{figure*}
    \centering
    \includegraphics[width=0.49\linewidth]{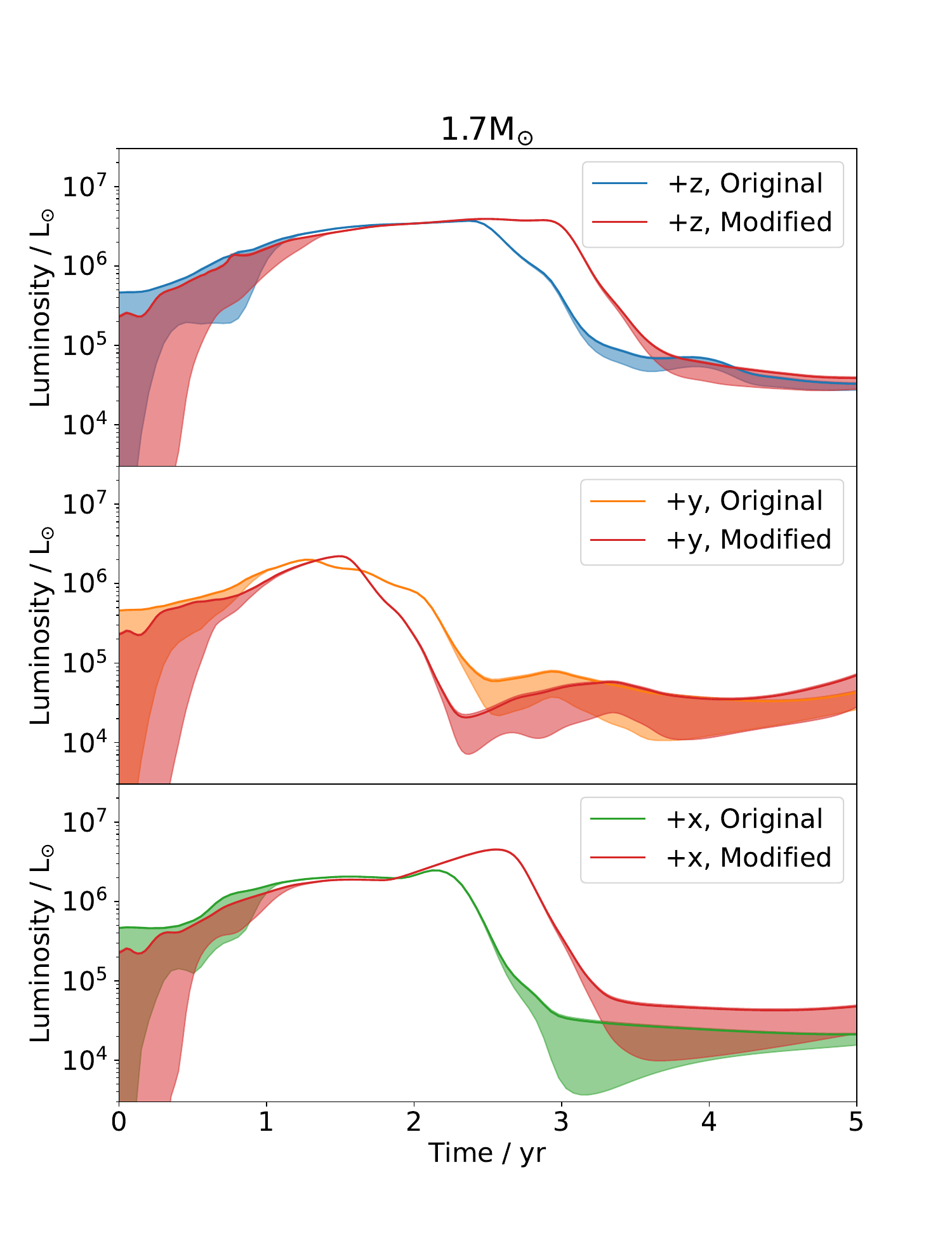}
    \includegraphics[width=0.49\linewidth]{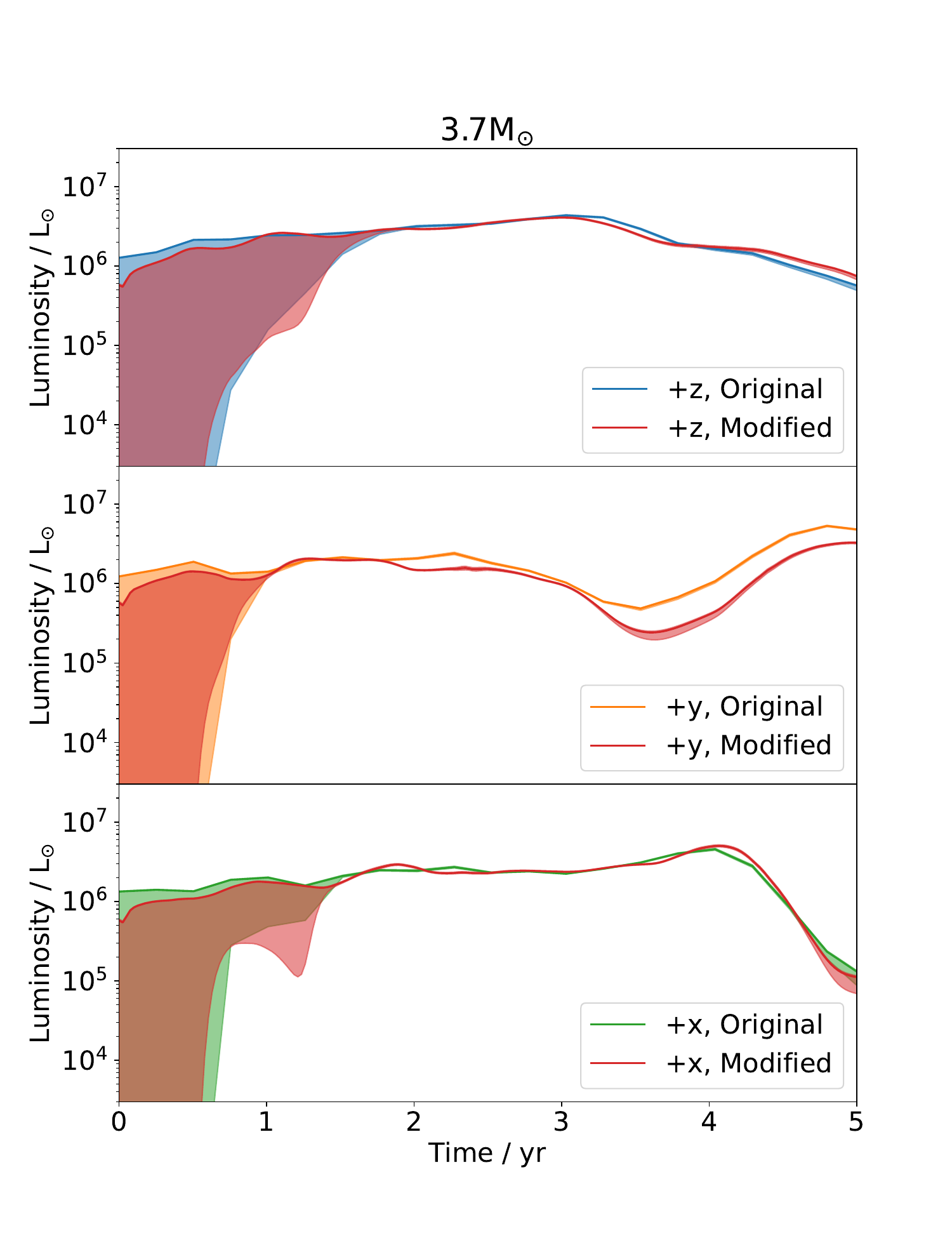}
    \caption{First 5 years of the lightcurves as viewed from three directions: +x, +y and +z, top, middle and bottom rows, respectively, for the 1.7~\Msun{} (left \updated{column}) and the 3.7~\Msun{} (right column) simulations.
    The blue/orange/green curves are for the original simulations as seen in Figure~\ref{fig:lc:both}, as reference; red curves are for simulations where the surface particles have their temperature reset to match the \mesa{} profile values, as described in this Appendix (see text). The shaded areas denote the error range as calculated in Section~\ref{sec:uncertainties:surface-particles} and Appendix~\ref{app:upper-lower-bounds}.
    }\label{fig:lc:xmdx}
\end{figure*}

The resulting lightcurve, calculated for only 5 years with the new star with an adjusted surface, is compared with the original case in Figure~\ref{fig:lc:xmdx}. 
The luminosity at time zero is a factor of two lower, but still much higher than the 1D, correct luminosity, due to the surface resolution issues detailed in Section~\ref{sec:uncertainties:surface-particles}.

Besides the difference in luminosity at time zero, the lightcurves (red curves) follow the same general trend as the original light curves, despite some differences in the duration of the peak (especially in the 1.7~\Msun{} case), where it can be broader or narrower by several months. \updated{The overall brightness of the entire lightcurve is surprisingly similar in the original and modified models, with at most a factor of two discrepancy and only for some moments in time of some of the viewports (e.g., in the $+x$ direction after 3.5 years for the 1.7~\msun\ model, or in the $+y$ direction after 3.5 years in the 3.7~\msun\ model)}. At the qualitative level, the mismatch of initial surface temperature profile is not problematic, but at the quantitative level, systematic uncertainties are very likely.

\section{Upper and lower lightcurve limits}
\label{app:upper-lower-bounds}

Here we provide details to quantify the uncertainty deriving from the lack of  resolution near the surface, as a direct result of too few \gls{SPH} particles sampling the steep increase of density, temperature and opacity as we enter the stellar surface. Effectively, surface particles are statistically distributed around the \mesa-prescribed stellar surface. However, these particles effectively protrude $2h$ outwards and our ray integration starts as soon as the ray enters the particle volume thus defined. Simply put, the integration of density, opacity, and source functions starts too soon, likely increasing the luminosity\footnote{Technically, since  the integration of the opacity also starts too early, the extra source function could be self absorbed completely and even obscure the emissivity from particles behind it (closer to the centre of the star), making the luminosity actually lower.}.
This is evidenced in Figure~\ref{fig:lc:both}, where our raw bolometric luminosity estimates are two orders of magnitude larger than the value suggested by \mesa{}.

We next calculate what must be a lower limit to the luminosity effectively eliminating the source function coming from the outer half of surface \gls{SPH} particles. In order to determine what classes as ``surface" \gls{SPH} particles, we divide the region travelled by each ray into two parts: the poorly resolved ``surface" region, which the ray meets first as it travels from the observer into the gas distribution, and the well-resolved region, farther inside the object.

What distinguishes the well-resolved from the poorly-resolved regions (as described in Section~\ref{sec:uncertainties:SPHres}) is that in the well resolved region the ray has already crossed approximately a quarter of the average number of neighbours per particle, $N_\mathrm{neigh}/4$ - by "cross" we intend that the ray has passed through that number of SPH particles, each with a radius of $2h$.
We choose $N_\mathrm{neigh}/4$, because it is expected that, in the well-resolved region, any point on the ray should have an average of $N_\mathrm{neigh}$ particles neighbouring it. This implies that well-resolved point on ray should expect $N_\mathrm{neigh}/2$ neighbour particles on the closer-to-observer side, while the other half is on the farther-from-observer side. At the boundary between an the well-resolved and the poorly-resolved regions, this is true as well. However, we acknowledge that one does not need the full $N_\mathrm{neigh}$ of neighbouring particles to have a reasonable idea of the various quantities (such as $\rho$, $\kappa$ and $S$) at that point through \gls{SPH} interpolation. Since an ideally less-resolved region would have zero neighbouring particles on the closer-to-observer side (instead of $N_\mathrm{neigh}/2$ neighbouring particles), we somewhat arbitrarily pick the midway point---$N_\mathrm{neigh}/4$ neighbouring particles on the closer-to-observer side---as our limit that separates the well-resolved region to the less-resolved region.


There is only one small addendum to this accounting. Each particle has a different impact on the values of quantities interpolated on the ray, that depends on the particle's (dimensionless) distance $q_{xy}$ to the ray location. It is therefore  important to weigh each particle's contribution by its distance to the ray, especially within the poorly-resolved region.
As such, instead of directly counting the neighbouring particles encountered by the ray, we weigh the counting of neighbouring particles, $j$, by their contribution to the smoothing kernel integration to the ray $i$:
\begin{equation}
    K_i(z)
    \equiv \sum_j \int_{z}^{+\infty} w(q_{ij}(z')) \frac{dz'}{h_j}.
    \label{eq:calc:K_z}
\end{equation}
In other words, $K_i(z)$ is a quantification of the dimensionless column kernel value - where the column kernel is itself a quantification of the particle's influence on the ray to point $z$. This variable is essentially the sum of the dimensionless column kernel ($w_\mathrm{col}(q_{xy, ij}) \equiv \int_{-\infty}^{+\infty} w(q_{ij}(z')) dq_z'$) associated with each particle $j$ up until the point $z$.
(The column kernel is proportional to each particles' influence on the ray: note how Equation~\ref{eq:calc:K_z} is similar to Equation~\ref{eq:calc:tau_z}, just without the factor of $m\kappa / h^2$). 

A point on ray $i$ in an {\it ideally} resolved region should have an average of $N_\mathrm{neigh}$ neighbouring particles. 
Each particle $j$, with its distance to the ray $i$ being $q_{xy, ij}$, contributes to the ray a dimensionless column kernel of 

\[ 
w_\mathrm{col,ij} = \int_{-R_\mathrm{kern}}^{+R_\mathrm{kern}} w\left(\sqrt{q^2_{xy,ij} + q^2_{z,j}}\right) dq_{z,j}, 
\]
where $R_\mathrm{kern} = 2$ is the kernel radius (Equation~\ref{eq:def:q_ij} and $w(q)=0$ when $q>R_\mathrm{kern}$).
Since these particles are expected to be randomly distributed, the probability of a neighbouring particle residing in the shell of $q_{xy} \rightarrow q_{xy} + dq_{xy} $ from the ray is
\[
\frac{2 \pi q_{xy} dq_{xy}}{\int_0^{R_\mathrm{kern}} 2 \pi q'_{xy} dq'_{xy}}
= \frac{2 \pi q_{xy} dq_{xy}}{\pi R^2_\mathrm{kern}}
. \]
Together, the above allows us to calculate a critical value of the dimensionless column kernel for location $z$, $K_{\rm crit}$ on ray $i$, that separates the poorly resolved outer layers from the well resolved inner layers
\[
\begin{split}
    K_{\rm crit} =
    N_\mathrm{neigh}
    \int_{0}^{R_\mathrm{kern}} dq_{xy} \,
    \left[ \frac{2 \pi q_{xy}}{\pi R^2_\mathrm{kern}}
        \int_{-R_\mathrm{kern}}^{+R_\mathrm{kern}} dq_z \,
        w\left(\sqrt{q^2_{xy} + q^2_z}\right)
    \right] \\
    = 4.608,
\end{split}
\]
Following the same logic, we suggest that requiring $K_i(z) > 4.608/4$ is the equivalent of requiring $N_\mathrm{neigh}/4 = 14.5$ neighbouring particles before declaring entry of the well-resolved region, i.e.,
\begin{equation}
    K_i(z) > 1.152
\end{equation}
becomes the condition for the region inside of location $z$ on the ray, to be considered well resolved.

Comparing with the approach of neglecting the first 14.5 particles for each ray, the above method accounts instead for the lower influence of particles farther away from the ray.

\bsp	
\label{lastpage}
\end{document}